\pdfoutput=1
\RequirePackage{ifpdf}
\ifpdf 
\documentclass[pdftex]{sigma}
\else
\documentclass{sigma}
\fi

\usepackage{mathrsfs}
\usepackage{amstext}

\newcommand{\blue}{\color{blue}}
\newcommand{\red}{\color{red}}

\usepackage{mathtools}
\usepackage{bm}
\usepackage{dsfont}
\usepackage{tikz}
\usetikzlibrary{patterns, calc, decorations.markings, arrows.meta}

\tikzset{->-/.style={line width=0.25mm,decoration={
			markings,
			mark=at position 0.53 with {\arrow{Stealth}}},postaction={decorate}}}
\tikzset{-<-/.style={line width=0.25mm,decoration={
			markings,
			mark=at position 0.47 with {\arrowreversed{Stealth}}},postaction={decorate}}}

\renewcommand{\imath}{\mathrm{i}}
\renewcommand{\Re}{\operatorname{Re}}
\renewcommand{\Im}{\operatorname{Im}}
\renewcommand{\phi}{\varphi}

\newcommand{\II}{\hbox{{1}\kern-.25em\hbox{l}}}

\usepackage{tocloft}

\numberwithin{equation}{section}

\begin{document}
\allowdisplaybreaks

\newcommand{\arXivNumber}{2507.09568}

\renewcommand{\PaperNumber}{107}

\FirstPageHeading

\ShortArticleName{A-Type Open ${\rm SL}(2,\mathbb{C})$ Spin Chain}

\ArticleName{A-Type Open $\boldsymbol{{\rm SL}(2,\mathbb{C})}$ Spin Chain}	

\Author{Pavel V.~ANTONENKO~$^{\rm ab}$, Sergey \'E.~DERKACHOV~$^{\rm a}$ and Pavel A.~VALINEVICH~$^{\rm a}$}

\AuthorNameForHeading{P.V.~Antonenko, S.\'E.~Derkachov and P.A.~Valinevich}

\Address{$^{\rm a)}$~Saint-Petersburg Department of Steklov Mathematical Institute of Russian Academy\\
\hphantom{$^{\rm a)}$}~of Sciences, Fontanka 27, 191023 St. Petersburg, Russia}
\EmailD{\mail{antonenko\_pavel@pdmi.ras.ru}, \mail{derkach@pdmi.ras.ru}, \mail{valinevich@pdmi.ras.ru}}

\Address{$^{\rm b)}$~Leonhard Euler International Mathematical Institute, Pesochnaya nab.~10,\\
\hphantom{$^{\rm b)}$}~197022 St.~Petersburg, Russia}

\ArticleDates{Received August 13, 2025, in final form December 08, 2025; Published online December 21, 2025}	

\Abstract{For the noncompact open ${\rm SL}(2, \mathbb{C})$ spin chain, the eigenfunctions of the special matrix element of monodromy matrix are constructed. The key ingredients of the whole construction are local Yang--Baxter $\mathcal{R}$-operators, $Q$-operator and raising operators obtained by reduction from the $Q$-operator. The calculation of various scalar products and the proof of orthogonality are based on the properties of $Q$-operator and demonstrate its hidden role. The symmetry of eigenfunctions with respect to reflection of the spin variable $s \to 1-s$ is established. The Mellin--Barnes representation for eigenfunctions is derived and equivalence with initial coordinate representation is proved. The transformation from one representation to another is grounded on the application of $A$-type Gustafson integral 	generalized to the complex field.}


\Classification{81R12; 17B80; 33C70}

\section{Introduction}

We consider the spin chain with infinite-dimensional unitary principal series representation of the group ${\rm SL}(2,\mathbb{C})$ in each site.
This integrable model appears in description of the high-energy behavior of
quantum chromodynamics, see~\cite{FK, L0,L1,L2} and
in connection to the two-dimensional version \cite{DKO, KO} of conformal
fishnet field theory \cite{BCF,BD,BD1,CK,GGKK,GKK,GKKNS,KG}.

The main objects of the quantum inverse scattering method \cite{F,KSk,Skl91,FST,FaTa} for the model under consideration are known.
The Hilbert space of the model $H = V_1\otimes V_2\otimes\cdots\otimes V_n$ is given by the direct product of the spaces $V_k=\mathrm{L}^2(\mathbb{C})$.
To each site, we associate the quantum $L$-operator
\begin{align*}
L(u) =
\begin{pmatrix}
u + S & S_{-} \\
S_{+} & u - S \end{pmatrix} ,
\end{align*}
where $S_{+}$, $S_{-}$ and $S$ are generators of the Lie algebra $\mathfrak{sl}(2,\mathbb{C})$ in representation labelled by spin~$s$.
The monodromy matrix is defined as a product of $L$ operators
\begin{equation} \label{monodromy}
T(u)=L_1(u)L_2(u)\cdots L_n(u) = \begin{pmatrix}
A(u)& B(u)\\
C(u)& D(u)
\end{pmatrix},
\end{equation}
where subscript $k$ shows that $L_k(u)$ acts nontrivially on the $k$-th space in the tensor product.
Monodromy matrix is a two by two matrix in the auxiliary two-dimensional space
$\mathbb{C}^2$ with entries that are operators on the quantum space $H$.
 The RTT-relations \cite{F,KSk,Skl91,FST,FaTa} imply the commutativity
$t(u) t(v) = t(v) t(u)$ of the transfer matrix -- the trace of the
monodromy matrix in auxiliary space
$t(u) = A(u)+D(u)$.
The commutativity relation means that $t(u)$ is the generating function for the set of $n$ commuting operators $t_k$, $k=1,\dots,n$,
\begin{align*}
t(u) = A(u)+D(u) = 2u^n + u^{n-1} t_1 + u^{n-2} t_2
+ \cdots + t_n.
\end{align*}
In the case of the closed (or periodic) spin chain, the main problem is
diagonalization of the transfer matrix $t(u)$ or, equivalently, the set
of $n$ commuting operators $t_k$ \cite{F,KSk,Skl91,FST,FaTa}.

The RTT-relations imply the commutation relations
\begin{align*}
A(u) A(v) = A(v) A(u), \qquad B(u) B(v) = B(v) B(u)
\end{align*}
so that operators $A(u)$ and $B(u)$ are also generating functions
for the sets of $n$ commuting operators.
It turns out that the systems of eigenfunctions of the operators
$B(u)$ and $A(u)$ are of independent interest.
Moreover, the diagonalization of the operator $B(u)$ is the important step in the diagonalization of the transfer matrix $t(u)$.

In the case of spin chain with infinite-dimensional unitary principal series representation of the group ${\rm SL}(2,\mathbb{C})$ in each site, the standard technique of
the algebraic Bethe ansatz \cite{F,KSk,FST} is not applicable.
The principal series representation has neither lowest nor highest weight vectors, and as a consequence there is no reference state for ABA.
The situation is similar to the quantum Toda chain and for the
solution of the model one uses the Sklyanin's method of
separation of variables \cite{KL1,KL2,KK, S1,S2}.
The main idea of the SoV method is very natural from the quantum
mechanical point of view and in our particular case consists in a performing
of quantum canonical transformation \cite{Fock2, Fock1} to the representation where operator $B(u)$ is diagonal.

In precise and most general terms \cite{KK, S1,S2}, quantum separation of variables
consists in explicit construction of a unitary map $\mathcal{U}$ between some
distinguished Hilbert space and the original Hilbert space on which the model is formulated. This map must greatly simplify the spectral problem associated with
the family of commuting operators.
The original spectral problem is essentially a multi-parameter spectral problem
in which each eigenvector is associated with a tower of eigenvalues of the
family of commuting operators
($t_k$ in the case of closed spin chain). The role of the map $\mathcal{U}$ is to implement unitary equivalence in such a way that the original multi-dimensional and multi-parameter spectral problem is reduced to a multi-parameter, but one-dimensional spectral problem.
In our particular example of the closed spin chain, Sklyanin showed
that the mapping $\mathcal{U}$ can be realized as an integral transformation,
where the integral kernel is given by an eigenfunction of the operator $B(u)$, and
the one-dimensional spectral problem corresponds to solving the so-called
$T-Q$ Baxter equation associated with the model.

Thus, the entire problem is divided into two separate tasks -- the explicit
construction of the mapping $\mathcal{U}$ and the solution of the $T-Q$ Baxter equation.
In turn, the construction of the mapping $\mathcal{U}$ is divided into its own separate tasks -- the construction of the corresponding eigenfunctions and the proof that this is a complete orthogonal set, which is equivalent to proving the unitarity of the mapping $\mathcal{U}$. In fact, the construction of $\mathcal{U}$ can be dealt with by making the best of the Yang--Baxter algebra underlying the integrability of the model.

The eigenfunctions of the operator $B(u)$ were constructed in \cite{DKM} and
their orthogonality was also proved there.
It turns out that the systems of eigenfunctions of the remaining operators~$A(u)$
and $D(u)$ and the corresponding integral transformations are of
independent interest~\mbox{\cite{BDM,DKO, L4}}. The corresponding eigenfunctions were
constructed in \cite{DM2} and completeness of all sets of
eigenfunctions (including eigenfunctions of $B(u)$) was proved quite recently \cite{M}.

In the present paper, we focus on the investigation of eigenfunctions of the operator
$A(u,z_0) = A(u)+z_0 B(u)$.
The variable $z_0$ plays the role of the auxiliary interpolating parameter --
the operator $A(u,z_0)$ coincides with operator $A(u)$ for $z_0=0$ and gives operator $B(u)$ for $z_0 \to \infty$. Correspondingly,
the eigenfunctions of operator $A(u,z_0)$ depend on $z_0$ and reproduce
eigenfunctions of operator $A(u)$ for $z_0=0$ and eigenfunctions of operator
$B(u)$ in appropriate limit~${z_0 \to \infty}$.
In~analogy with SoV representation, an integral transformation,
where the integral kernel is given by an eigenfunction of the
operator $A(u,z_0)$, is very useful.
First of all, it is unitary mapping to the representation where graph-building operator
for the two-dimensional Basso--Dixon diagram is diagonal \cite{DKO}.
Secondly, we use this transformation for construction of the Mellin--Barnes representation for the BC-type spin chain \cite{ADV2}.

An essential part of the works \cite{DKM,DM2}
is the use of the Feynman diagram technique.
This allows one to represent the kernels of the integral operators as Feynman diagrams
and to perform all the necessary operations at the graphical level.
It makes all the calculations visual, but it is sometimes difficult to reconstruct the simple algebraic meaning of the transformations.
The central aim of the present work is to demonstrate the new algebraic approach to calculations involving the eigenfunctions of the operator~$A(u,z_0)$.
The eigenfunctions of~$A(u)$ and~$A(u,z_0)$ differ by the shift of argument, therefore they are given by the same formula~\cite{DM2}, see Section~\ref{sect:eigen}.
By means of the algebraic method, the new proofs of known properties of these functions are developed.
For example, we give the new proof of orthogonality relation and the symmetry with respect to permutations of spectral variables (these properties were established in~\cite{DKO,DM2} with the help of the diagram technique)
and present the algebraic calculation of some scalar products of eigenfunctions including the kernel of Mellin--Barnes integral representation~\cite{Val20} and the matrix element of shift operator (the overlap integral)~\cite{DMV2}.
In addition, we obtain some new results like the symmetry relation between the eigenfunctions corresponding to spins~$s$ and~$1-s$, and the direct transformation from the Mellin--Barnes representation to the initial integral representation without using the completeness of the set of eigenfunctions.

Our motivation for developing the algebraic approach is twofold.
First of all, in recent years significant progress has been made in constructing SoV representations for higher rank finite-dimensional models, see~\cite{Cavaglia:2019pow,GromovRyan20,GromovSizov17,Gromov:2022waj,Levk,MailletNiccoli18,
MR3983970,MailletNiccoli19,Maillet:2020ykb,Ryan:2021duf,RyanVolin19,
Ryan:2020rfk}.
In the case of higher rank, the diagrammatical approach does not exist
so that development of an algebraic approach is the first needed step for the generalization of the SoV to the case of noncompact groups of the higher rank.
Secondly, this work is the first part in a series of two papers.
The second work~\cite{ADV2} is devoted to the open spin chain of the BC-type.
One of the goals of the present paper is to collect all formulas needed
for the second part and derive them in a unified way.
This allows to perform similar calculations in the case of the spin chain
of BC-type because in this case Feynman diagrams have significantly more
complicated form and algebraic approach becomes more suitable.\looseness=-1

The paper is organized as follows.
In Section~\ref{PRSL}, we present all necessary formulas concerning
principal series representations.
In Section~\ref{RQ}, we introduce the main building block --
$\mathcal{R}$-operator which gives solution of the defining
Yang--Baxter relation~\eqref{defR} and then use it for the
construction of the $Q$-operator.
In contrast to Baxter's $Q$-operator, it commutes not with the transfer matrix~$t(u)=A(u)+D(u)$, but with the operator $A(u,z_0)=A(u)+z_0 B(u)$.
We call this object the $Q$-operator since it has the same characteristic properties as the canonically defined Baxter $Q$-operator -- it forms the commutative family of operators (see~\eqref{comQQ}) and satisfies the Baxter equation~\eqref{Bax}.
In our case, the latter takes the form of the first order difference equation, in~contrast to the second order difference equation in the conventional case.
All these properties are proved algebraically without
using diagrammatic technique.

The common eigenfunctions of $A$- and $Q$-operators are discussed
in Section~\ref{sect:eigen}.
The eigenfunctions are constructed in an iterative way using raising
operator.
The $Q$-operator coincides with integral graph building operator from~\cite{DKO} which was used to compute Feynman diagrams of Basso--Dixon type in two-dimensional conformal fishnet theory.
The mentioned inductive construction of eigenfunctions of $Q$-operator was presented in that work on diagrammatic level. These functions play the key role in the exact calculation of Basso--Dixon diagrams.
In the present work, this construction is translated to the algebraic language.
This way, we show the connection between the graph building operator and the monodromy matrix~\eqref{monodromy} of~${\rm SL}(2,\mathbb{C})$ spin chain which was not discussed in~\cite{DKO}.

Raising operator is a closed relative of the $Q$-operator and
can be obtained by simple reduction.
In fact, all properties of eigenfunctions are defined by the commutation
relations between raising operators and $Q$-operators.
We derive all needed commutation relations from the commutativity
of $Q$-operators using reduction.
In comparison with diagrammatic consideration, this approach
clearly shows the necessary restructuring of the product of
$\mathcal{R}$-operators.
Principal series representations characterized by the
parameters $s$ and $1-s$ are equivalent and in the second part of this section
we consider the manifestation of this symmetry $s \rightleftarrows 1-s$ on the
level of eigenfunctions.

Section~\ref{sect:Basic} is devoted to the calculations of some
specific scalar products. We present an~iterative algebraic computation
that allows us to reveal the hidden role of the $Q$-operator.

In Section~\ref{sect:Ort}, we prove orthogonality of eigenfunctions and
calculate the corresponding Sklya\-nin measure.
The introduction of the special regularization allows to use the explicit
formulas derived in the previous section. We show that after removing the
regularization one obtains the needed symmetric $\delta$-function.

Section~\ref{sect:MB} is devoted to the derivation of the equivalent
representation for eigenfunction.
In analogy with the similar representation for the eigenfunctions of
A-type Toda chain \cite{KL1,KL2,KK}, we call it the Mellin--Barnes representation.
Using generalization to the complex field of A-type Gustafson integral,
we present the direct transformation from one representation to another.

In the last Section~\ref{con}, we collect the main formulas and
shortly discuss the possible applications.

In appendices, we collect some useful formulas.

\section[Principal series representations of SL(2,C)]{Principal series representations of $\boldsymbol{{\rm SL}(2,\mathbb{C})$}}
\label{PRSL}

To describe the model, we recall the main formulas for the \textit{principal series representation} of the group ${\rm SL}(2,\mathbb{C})$~\cite{GGV, GN}. It is defined on the space $\mathrm{L}^2(\mathbb{C})$ of square integrable functions on $\mathbb{C}$ with the scalar product
\begin{equation} \label{sp}
	\langle \Phi|\Psi\rangle = \int \mathrm{d}^2z \, \overline{\Phi(z,\bar{z})} \Psi(z, \bar{z}) .
\end{equation}
The integration measure is the Lebesque measure in $\mathbb{C}$: $\mathrm{d}^2 z = \mathrm{d} x \mathrm{d} y$, where as usual $z =x+ \imath y$
and $x =\Re(z)$, $y =\Im(z)$. The integration is over the whole complex plane.

To avoid misunderstanding, we denote complex conjugation of a number $a \in \mathbb{C}$ as~$a^*$ so that in generic situation $\bar{a}$ does not mean complex conjugated to $a$.
There is only one traditional exception for the variable $z$ so that~\mbox{$\bar{z} \equiv z^*$} and for the complex conjugation of functions \smash{$\overline{\Psi(z, \bar{z})}$}.

The representation is parametrized by the pair of parameters (spins) $(s, \bar{s})$ with restriction $2(s - \bar{s}) \in \mathbb{Z}$.
The group element
\begin{equation*}
	g =
	\begin{pmatrix}
		a & b \\
		c & d
	\end{pmatrix}, \qquad ad - bc = 1,
\end{equation*}
is represented by the operator $T^{(s,\bar{s})}(g)$ acting on functions $\Psi(z ,\bar{z})$ according to the formula
\begin{align*}
	\bigl[T^{(s,\bar{s})}(g) \Psi\bigr] (z ,\bar{z}) =
	(d-bz)^{-2s} (d^*-b^*\bar{z})^{-2\bar{s}}
\Psi \biggl(\frac{-c+az}{d-bz} ,\frac{-c^*+a^* \bar{z}}{d^*-b^*\bar{z}}\biggr).
\end{align*}
To avoid misunderstanding, we write everything explicitly but in the following
we will use compact notations so that this formula reduces to the form
\begin{equation*}
	\bigl[T^{(s,\bar{s})}(g) \Psi\bigr] (z) =
	[d-bz]^{-2s} \Psi \biggl(\frac{-c+az}{d-bz}\biggr).
\end{equation*}
First of all, for the sake of simplicity, we will display only argument $z$
so that~\mbox{$\Psi(z) \equiv \Psi(z, \bar{z})$}.
Secondly, here and in what follows for a pair $(a,\bar{a}) \in \mathbb{C}^2$ such that $a-\bar{a} \in \mathbb{Z}$ we denote the double power
\begin{align} \label{power}
	[z]^a \equiv z^a \bar{z}^{\bar{a}} = |z|^{a+\bar{a}} {e}^{\imath(a-\bar{a}) \arg z}.
\end{align}
The function~\eqref{power} depends on both ``holomorphic'' parameter $a$ and ``antiholomorphic'' one $\bar{a}$, but for brevity we display only the ``holomorphic'' exponent. The numbers $a$, $\bar{a}$ are not complex conjugate in general. We impose the condition $a-\bar{a} \in \mathbb{Z}$, so that the function $[z]^a$ is single-valued in the whole complex plane. In addition, for $\rho \in \mathbb{R}$ we denote
\begin{equation*}
	[z]^{\rho+a} \equiv z^{\rho+a} \bar{z}^{\rho+\bar{a}} .
\end{equation*}
The principal series representation is unitary if $s$ and $\bar{s}$ satisfy the condition
\begin{equation} \label{scond}
	s^\ast + \bar{s} = 1 ,
\end{equation}
where $s^\ast$ denotes the complex conjugation of $s$. Together with the condition $2(s - \bar{s}) \in \mathbb{Z}$, this gives the parametrization of spins
\begin{equation}\label{sparam}
	s = \frac{1+n_s}{2} + \imath\nu_s, \qquad
	\bar{s} = \frac{1-n_s}{2} + \imath\nu_s ,
	\qquad
	n_s \in \mathbb{Z}+\sigma, \quad \nu_s \in \mathbb{R},
\end{equation}
where for the rest of the paper we fix the parameter $\sigma$
\begin{equation} \label{sigma}
	\sigma \in \bigl\{0, \tfrac{1}{2}\bigr\}
\end{equation}
so that for $\sigma=0$ the parameter $n_s$ is integer and for
\smash{$\sigma=\tfrac{1}{2}$} the parameter $n_s$ is half-integer.

Generators of the representation are defined in the standard way as the coefficients of decomposition of the map $T^{(s,\bar{s})}(g)$ in the neighbourhood of unity.
They have the form
\begin{alignat}{4}
		&S = z\partial_z + s, \qquad&& S_- = -\partial_z, \qquad&&
		S_+ = z^2\partial_z+2sz,& \nonumber\\
		&
		\bar{S} = \bar{z}\partial_{\bar{z}} + \bar{s}, \qquad&& \bar{S}_- = -\partial_{\bar{z}}, \qquad&&
		\bar{S}_+ = \bar{z}^2\partial_{\bar{z}}+2\bar{s}\bar{z} .&\label{gen}
\end{alignat}
We call $S$, $S_{\pm}$ holomorphic generators and $\bar{S}$, $\bar{S}_\pm$ -- antiholomorphic ones. They obey the standard commutation relations of the Lie algebra $\mathfrak{sl}(2,\mathbb{C})$
\begin{align*}
	[S_+, S_-] = 2S, \qquad [S,S_\pm] = \pm S_\pm,
\end{align*}
and similarly for the antiholomorphic generators. Note that the holomorphic generators commute with the antiholomorphic ones. Two types of generators are conjugate to each other with respect to the scalar product~\eqref{sp},
\begin{equation} \label{Shc}
	S^\dagger = -\bar{S}, \qquad S_-^\dagger = -\bar{S}_-, \qquad
	S_+^\dagger = -\bar{S}_+
\end{equation}
due to relation~\eqref{scond}.

\subsection{Equivalent representations and intertwiner}

It is well known \cite{GGV, GN} that the representations characterized by the
parameters $(s, \bar s)$ and $(1-s, 1-\bar{s})$ are equivalent.
There exists an integral operator $W$ which intertwines
equivalent principal series representations $T^{(s,\bar{s})}$
and $T^{(1-s,1-\bar{s})}$ for generic complex $s$ and $\bar{s}$,
\begin{equation*}
W(s,\bar{s}) T^{(s,\bar{s})}(g) =
T^{(1-s,1-\bar{s})}(g) W(s,\bar{s}) .
\end{equation*}
Intertwining operator can be written in the following explicit form~\cite{GGV}
(for a justification of the taken normalization factor,
see \cite{DM1}):
\begin{align*}
 [ W(s,\bar{s}) \Phi ](z,\bar{z}) =
\frac{\imath^{-|2s-2\bar s|}}{\pi}
 \frac{\Gamma (2-s-\bar s+|s-\bar s| )}
{\Gamma (s+\bar s+|s-\bar s| -1 )}
\int \mathrm{d}^2 x \frac{\Phi(x,\bar{x})}{[z-x]^{2-2s}},
\end{align*}
where $\Gamma(x)$ is the standard Euler gamma function.
This is a well-defined operator for generic values of $s$ and $\bar{s}$.
Despite of the diverging integral for the discrete values $2s = -n$,
$2\bar{s} = -\bar{n}$, where $n$, $\bar{n} \in \mathbb{Z}_{\geq 0}$,
it remains well defined in this case too due the appropriate
normalizing factor.

The described intertwining operator has a meaning of the
pseudo-differential operator. Such an interpretation is reached
with the help of the following explicit Fourier transformation:
\begin{equation}\label{F}
A(\alpha,\bar\alpha) \int \mathrm{d}^2 z \frac{\mathrm{e}^{\imath pz+\imath\bar{p}\bar{z}}}
{z^{1+\alpha}\bar{z}^{1+\bar\alpha}} =
p^{\alpha}\bar{p}^{\bar\alpha} ,\qquad \alpha-\bar\alpha\in \mathbb{Z},
\end{equation}
where the normalization constant has the canonical form~\cite{GGV}
\begin{equation*}
A(\alpha,\bar\alpha):= \frac{\imath^{-|\alpha-\bar\alpha|}}{\pi}
 \frac{\Gamma\bigl(\frac{\alpha+\bar\alpha+|\alpha-\bar\alpha|+2}{2}\bigr)}
{\Gamma\bigl(\frac{-\alpha-\bar\alpha+|\alpha-\bar\alpha|}{2}\bigr)}.
\end{equation*}
For non-integer values of $\alpha$,
the form of this constant can be simplified
\begin{equation*}
A(\alpha,\bar\alpha) =
\frac{[\imath]^{-\alpha}}{\pi} \bm{\Gamma}(\alpha+1), \qquad
\bm{\Gamma}(\alpha):=\frac{\Gamma(\alpha)}{\Gamma(1-\bar\alpha)}, \qquad
[\imath]^{\alpha} \equiv \imath^{\alpha-\bar{\alpha}},
\end{equation*}
where $\bm{\Gamma}(\alpha)$ is the gamma function
associated with the field $\mathbb{C}$~\cite{GGR,N}.

Let us replace in formula~\eqref{F} the complex variables $p$ and $\bar p$
by the momentum operators, $p \to \hat{p} = -\imath\partial_x$ and
$\bar{p} \to \hat{\bar{p}} = -\imath\partial_{\bar{x}}$.
Then one can use the standard finite-difference operator
\begin{equation*}
\mathrm{e}^{\imath a \hat{p}+\imath\bar{a} \hat{\bar{p}}}f(x) =
\mathrm{e}^{a\partial_x+\bar{a}\partial_{\bar{x}}}f(x,\bar{x})=
f(x+a,\bar{x}+\bar{a})
\end{equation*}
in order to set by definition
\begin{align} \label{d}
\left[\hat{p}\right]^{\alpha}\Phi(z,\bar z) :=
\hat{p}^{\alpha}\hat{\bar{p}}^{\bar{\alpha}}\Phi(z,\bar z) =
c(\alpha)
\int \mathrm{d}^2 x
\frac{\Phi(x,\bar x)}{[z-x]^{1+\alpha}},
\end{align}
where
\begin{align}\label{c}
c(\alpha) = [\imath]^{2\alpha} A(\alpha,\bar\alpha) = \frac{[\imath]^{\alpha} \bm{\Gamma}(\alpha+1)}{\pi}.
\end{align}
It is this pseudo-differential operator that is used for fixing the normalizing
factor in the intertwining operator -- one simply sets
\begin{align}
\label{W1}
W(s,\bar{s}) := [\hat{p}]^{1-2s}.
\end{align}
From~\eqref{W1} and relation $[p]^\alpha [p]^{-\alpha} = \II$, it follows that
\begin{equation*}
	W^{-1}(s,\bar{s}) = W(1-s,1-\bar{s}).
\end{equation*}
The rules of conjugation for the momentum
operators have the standard form $\hat{p}^{\dagger} = \hat{\bar{p}}$,
$\hat{\bar{p}}^{\dagger} = \hat{p}$ so that
\begin{align}\label{conj}
([\hat{p}]^{\alpha})^{\dagger} =
\bigl(\hat{p}^{\alpha}\hat{\bar{p}}^{\bar{\alpha}}\bigr)^{\dagger} =
\hat{\bar{p}}^{\alpha^*}\hat{p}^{\bar{\alpha}^*} =
[\hat{p}]^{\bar{\alpha}^*}.
\end{align}
In our parametrization~\eqref{sparam}, $\bar{s}^* = 1-s$ and therefore
\begin{equation*}
	W^{\dagger}(s,\bar{s}) = W(1-s,1-\bar{s}) = W^{-1}(s,\bar{s})
\end{equation*}
so that $W(s,\bar{s})$ is unitary operator
$W^{\dagger}(s,\bar{s}) W(s,\bar{s}) = \II$.

\section[SL(2,C) spin chain]{$\boldsymbol{{\rm SL}(2,\mathbb{C})}$ spin chain}
\label{RQ}

In this section, we define the integrable model under
consideration and introduce Yang--Baxter $\mathcal{R}$-operators
and corresponding $Q$-operators.

\subsection[L operators and monodromy matrices]{$\boldsymbol{L}$ operators and monodromy matrices}

The Hilbert space of the model is given by the direct product of the $\mathrm{L}^2(\mathbb{C})$ spaces,
\begin{equation}\label{HN}
H=V_1\otimes V_2\otimes\cdots\otimes V_n, \qquad V_k=\mathrm{L}^2(\mathbb{C}) ,\qquad
k=1,\dots,n.
\end{equation}
The dynamical variables are given by two sets of spin operators: holomorphic
\smash{$\bigl(S^{(k)}_{\pm} ,S^{(k)}\bigr)$} and anti-holomorphic
\smash{$\bigl(\bar S^{(k)}_{\pm} ,\bar S^{(k)}\bigr)$},
$k=1,\dots,n$.\footnote{It is assumed that the
generators with index $k$ act non-trivially
only on $k-$th space in the tensor product, $V_k$.}
In what follows, we will consider only homogeneous chains, $s_k=s$, $\bar s_k=\bar s$, for all $k$.
To each site $k$, we associate the quantum $L$-operators
with subscript $k$, acting nontrivially on the $k$-th space in the tensor product~(\ref{HN})
\begin{equation*}
L_k(u) =
\begin{pmatrix}
u + S^{(k)} & S_{-}^{(k)} \\
S_{+}^{(k)} & u - S^{(k)} \end{pmatrix}, \qquad
\bar L(\bar u) = \begin{pmatrix}
\bar u + \bar S^{(k)} & \bar S_{-}^{(k)}\\
\bar S_{+}^{(k)}& \bar u - \bar S^{(k)}
\end{pmatrix}.
\end{equation*}
The monodromy matrix is defined as a product of $L$ operators
\begin{equation*}
T(u)=L_1(u)L_2(u)\cdots L_n(u), \qquad
 \bar T(\bar u)=\bar L_1(\bar u)\bar L_2(\bar u)\cdots \bar L_n(\bar u) .
\end{equation*}
It is a two by two matrix in the auxiliary two-dimensional space
$\mathbb{C}^2$ with entries that are operators on the quantum space $H$
\begin{equation*}
T(u)
=\begin{pmatrix}
A(u)& B(u)\\
C(u)& D(u)
\end{pmatrix},\qquad
\bar T(\bar u)
 =\begin{pmatrix}
\bar A(\bar u)& \bar B(\bar u)\\
\bar C(\bar u)& \bar D(\bar u)
\end{pmatrix} .
\end{equation*}
The monodromy matrix satisfies the $RTT$-relation~\cite{F,KSk,Skl91}
\begin{align}\label{RTT}
R(u-v) \bigl(T(u)\otimes \bm{1}\bigr) \bigl(\bm{1} \otimes T(v)\bigr)
= \bigl(\bm{1} \otimes T(v)\bigr) \bigl(T(u)\otimes \bm{1}\bigr) R(u-v).
\end{align}
Here $R(u)$ is the Yang's $R$-matrix that acts in the tensor product $\mathbb{C}^2\otimes \mathbb{C}^2$
\begin{equation*}
	R(u) = u + P, \qquad P a \otimes b = b \otimes a.
\end{equation*}
From~\eqref{RTT}, one obtains in a standard way that operators $A(u)$ and $B(u)$
form commutative families
\begin{align}\label{AA}
[A(u), A(v)]=0, \qquad [B(u), B(v)]=0.
\end{align}
Another consequence of~\eqref{RTT} are commutation relations
\begin{align}\label{S-}
[S_- , B(u)] = 0 ,\qquad [S_- , A(u)] = B(u)
\end{align}
with global generators
\begin{align*}
S_- = S^{(1)}_-+\dots+S^{(n)}_-.
\end{align*}
Due to~(\ref{S-}), we have
\begin{align}\label{Az0B}
A(u,z_0) = A(u)+z_0 B(u) = {\rm e}^{z_0 S_{-}} A(u) {\rm e}^{-z_0 S_{-}}
\end{align}
so that the operators $A(u,z_0)$ and $A(u)$ are connected
by similarity transformation.
Then the relation~\eqref{AA} implies commutativity of
$A(u,z_0)$ also
\begin{align*}
[A(u,z_0) ,A(v,z_0)] = 0.
\end{align*}
By construction, the operator $A(u,z_0)$ is polynomial of degree $n$ in $u$
\begin{align}\label{ABCDexp}
A(u,z_0)=u^n+u^{n-1} \left(S+z_{0} S_{-}\right) +\sum_{k=2}^{n} u^{n-k} I_k ,
\end{align}
where operators $I_k$ are homogeneous polynomials of degree $k$
from generators $S^{(j)}_{\pm}$ and $S^{(j)}$, $j=1,\dots,n$, i.e., differential operators of $k$-th order in the variables $z_1,\dots,z_n$.
The construction for anti-holomorphic
sector is essentially the same and we will omit the corresponding similar expressions as a rule.

Let $\Psi(\bm z_n) = \Psi(z_1,\bar z_1,\dots, z_n,\bar z_n)$
be a common eigenfunction of the operators
$A(u,z_0)$ and $\bar{A}(\bar{u},\bar{z}_0) = \bar{A}(\bar{u})+\bar{z}_0 \bar{B}(\bar{u})$.
By virtue of~\eqref{ABCDexp}, the corresponding eigenvalues are
polynomials of degree $n$ in $u$ and $\bar u$, respectively and
eigenfunction can be labelled by zeroes of these polynomials
\begin{align}
&A(u,z_0)
\Psi_{\bm x_n }(\bm z_n)= (u-x_1)\cdots(u-x_n)
\Psi_{\bm x_n }(\bm z_n),
\nonumber\\
&\bar{A}(\bar{u},\bar{z}_0) \Psi_{\bm x_n }(\bm z_n)=
(\bar u-\bar x_1)\cdots(\bar u-\bar x_n)
\Psi_{\bm x_n }(\bm z_n).\label{AbAN}
\end{align}
To avoid cumbersome formulas, we will use the following
notations for tuples of $n$ variables:
\begin{align*}
\bm z_n=(z_1,\bar{z}_1, \dots, z_n, \bar{z}_n) , \qquad
\bm x_n=(x_1,\bar{x}_1, \dots, x_n, \bar{x}_n).
\end{align*}
By analogy with spin variables $s$, $\bar{s}$, we
impose for $x_k$, $\bar x_k$ the same restriction
\begin{equation*}
x_k-\bar x_k \in \mathbb{Z}+\sigma,
\end{equation*}
where the parameter $\sigma$ was introduced in~\eqref{sigma}.
Due to~\eqref{ABCDexp} and rules of conjugation for generators~\eqref{Shc}, we have $x_k^*=-\bar{x}_k$ so that together it result
in the following parametrization~\cite{DKM}:%
\begin{equation}\label{x-n-nu}
x_k=\frac{n_k}{2}+\imath\nu_k,\qquad \bar x_k=-\frac{n_k}{2}+\imath\nu_k ,
\end{equation}
where $\nu_k$ is real and $n_k \in \mathbb{Z}+\sigma$.
The set of relations $x_k^*=-\bar{x}_k$ for $k=1 ,\dots , n $
means that operator $A(u,z_0)$ is an hermitian adjoint
of \smash{$\bar A(\bar u,\bar{z}_0)$} (up to overall sign factor)
\begin{equation*}
A^\dagger(u,z_0) = (-1)^n \bar A(\bar u, \bar{z}_0)
\end{equation*}
provided that $u^*=-\bar u$. It is useful to
parameterize the spectral parameters $u$ and $\bar{u}$ in the same~way
\begin{equation*}
u=\frac{n_u}{2}+\imath\nu_u,\qquad \bar u = -\frac{n_u}{2}+\imath\nu_u ,
\end{equation*}
where $\nu_u$ is real and $n_u \in \mathbb{Z}+\sigma$.

\subsection[R-operators and Q-operators]{$\boldsymbol{R}$-operators and $\boldsymbol{Q}$-operators}

We have two free parameters in the L-operator:
spin $s$ and spectral parameter $u$, but equivalently
one can use another pair of parameters
\begin{align*}
u_1 = u+s-1,\qquad u_2 = u - s,
\end{align*}
which appear in a natural way in factorized expression below.
In representation~\eqref{gen} the $L$-operator can be
expressed in various useful factorized forms
$\bigl(\partial_k = \frac{\partial}{\partial z_k}\bigr)$
\begin{align}
\nonumber
L_k(u) & =
\begin{pmatrix}
u + S^{(k)} & S_{-}^{(k)} \\
S_{+}^{(k)} & u - S^{(k)} \end{pmatrix} =
\begin{pmatrix} u+s+z_k \partial_k & -\partial_k\\
z_k^2 \partial_k + 2s z_k & u-s-z_k
\partial_k
\end{pmatrix}
\\
\label{Lfac}
&=
\begin{pmatrix}
 1 & 0 \\
 z_k & 1 \\
\end{pmatrix}
\begin{pmatrix}
 u_1 & -\partial_{k} \\
 0 & \hphantom{-} u_2 \\
\end{pmatrix}
\begin{pmatrix}
 \hphantom{-}1 & 0 \\
 -z_k & 1 \\
\end{pmatrix} =
\begin{pmatrix}
 1 & 0 \\
 z_k & u_2 \\
\end{pmatrix}
\begin{pmatrix}
 1 & -\partial_{k} \\
 0 & \hphantom{-}1 \\
\end{pmatrix}
\begin{pmatrix}
 \hphantom{-}u_1 & 0 \\
 -z_k & 1 \\
\end{pmatrix}.
\end{align}
We will use the following notations for
the Lax operator:
\begin{equation*}
L_k(u) = L_k(u_1 ,u_2) =
L_k(u+s-1 ,u - s) .
\end{equation*}
Let us introduce the integral operator acting on functions $\Psi(z_k, z_j)$
\begin{align} \label{R}
 [\mathcal{R}_{k j}(x,\bar{x}) \Psi ](z_k, z_j) =
	c(x-s)
	\int \mathrm{d}^2 w
	\frac{[z_k-z_j]^{1-2s}}{[z_k-w]^{1-s+x} [w-z_j]^{1-s-x}}
	\Psi(w,z_j)
\end{align}
where $x = \frac{m}{2}+\imath\nu$, $\bar{x} = -\frac{m}{2}+\imath\nu$ and
$c(x-s)$ is given by~\eqref{c} for $\alpha=x-s$.
In the simpler form~\eqref{d}, this operator
can be represented in two equivalent ways ($z_{kj} = z_k-z_j$)
\begin{align}\label{R1}
\mathcal{R}_{k j}(x,\bar{x}) =
[z_{kj}]^{1-2s} [\hat{p}_k]^{x-s} [z_{kj}]^{s+x-1} =
[\hat{p}_k]^{x+s-1} [z_{kj}]^{x-s}
 [\hat{p}_k]^{1-2s}
\end{align}
The equivalence of these two representation
is essentially the statement of the star-triangle
relation~\eqref{star-tr}.
Note that initial variables $z_k$, $\bar{z}_k$ and spectral variables
$x_k$, $\bar{x}_k$ enter almost everywhere in pairs.
For the sake of simplicity, we will omit dependence on $\bar{z}_k$
and $\bar{x}_k$ in explicit notations. For example, we will
use simpler notation $\mathcal{R}_{k j}(x)$ instead of
$\mathcal{R}_{k j}(x,\bar{x})$.

$R$-operator is defined as solution of the following key relation which is similar to the usual RLL-relation~\cite{KRS}:
\begin{align}\label{defR}
\mathcal{R}_{1 2}(x) L_1(u_1, u_2) L_2(u_1, u-x) =
L_1(u_1, u-x) L_2(u_1, u_2) \mathcal{R}_{1 2}(x)
\end{align}
so that it interchanges the arguments in the product of
two holomorphic Lax matrices in a~very specific way.
The same relation holds for the antiholomorphic Lax matrices
but we skip it for simplicity.
Operator $\mathcal{R}_{i j}(x)$ is related to the general ${\rm SL}(2,\mathbb{C})$-invariant solution of the Yang--Baxter equation \cite{D1,DM1} which is generalization of \cite{KRS}.

The composite operator
\begin{align}\label{Q}
Q_n(x) =
\mathcal{R}_{12}(x) \mathcal{R}_{23}(x)\cdots
\mathcal{R}_{n-1 n}(x) \mathcal{R}_{n 0}(x)
\end{align}
has all characteristic properties of the Baxter $Q$-operator:
\begin{itemize}\itemsep=0pt
\item
it commutes with operators $A(u,z_0)$ and
$\bar{A}(\bar{u},\bar{z}_0)$
\begin{align}\label{comQA}
Q_n(x) A(u,z_0) = A(u,z_0) Q_n(x) ,\qquad
Q_n(x) \bar{A}(\bar{u},\bar{z}_0) =
\bar{A}(\bar{u},\bar{z}_0) Q_n(x),
\end{align}
\item it forms commutative family
\begin{align}\label{comQQ}
Q_n(x) Q_n(y) = Q_n(y) Q_n(x),
\end{align}
\item it obeys the Baxter relations
\begin{align}\label{Bax}
&A(u,z_0) Q_n(u,\bar{u}) = \imath^n Q_n(u+1,\bar{u}),
\\
\label{Baxbar}
&\bar{A}(\bar{u},\bar{z}_0) Q_n(u,\bar{u}) = \imath^n Q_n(u,\bar{u}+1).
\end{align}
\end{itemize}
In fact, all these relations are consequence of the defining
relation~(\ref{defR}) for the $\mathcal{R}$-operator.
We will call~$Q_n(x)$~\eqref{Q} the $Q$-operator.
Note however that unlike~\eqref{Q}, the conventionally defined Baxter $Q$-operator commutes not with~$A(u,z_0)$, but with the transfer matrix~$t(u)=A(u)+D(u)$,
and satisfies the second order difference equation (Baxter equation) instead of the first order difference equation~\eqref{Bax}.
The operator formula~\eqref{Q} can be written in a more explicit form using~\eqref{R1}
\begin{align}\label{Qexp}
Q_n( x) = S_n
[\hat{p}_1]^{x-s} [z_{12}]^{s+x-1}
[\hat{p}_2]^{x-s} [z_{23}]^{s+x-1}
\cdots
[\hat{p}_n]^{x-s} [z_{n0}]^{s+x-1},
\end{align}
where
\begin{equation} \label{Sn}
S_n = [z_{12}]^{1-2s} [z_{23}]^{1-2s}
\cdots [z_{n0}]^{1-2s} .
\end{equation}
The explicit expression~\eqref{Qexp} for the $Q$-operator
clearly shows that there are special values of the spectral
parameter $x=s$ and $x=1-s$, where $Q$-operator admits
reductions to the simpler form
\begin{itemize}\itemsep=0pt
\item For the point $x=s$, we have reduction $[\hat{p}_k]^{x-s} \to \II$ and finally
\begin{align*}
Q_n(s) = \II.
\end{align*}
\item
At the point $x=1-s$, we have reduction $[z_{k k+1}]^{s+x-1}\to \II$ so that
\begin{align}\label{Q_1-s}
Q_n(1-s) = S_n
[\hat{p}_1]^{1-2s}
[\hat{p}_2]^{1-2s}
\cdots
[\hat{p}_n]^{1-2s} =
S_n W_1(s) \cdots W_n(s).
\end{align}
\end{itemize}
Let us start the proofs of relations
\eqref{comQA},~\eqref{comQQ} and~\eqref{Bax} from the proof of
the commutativity of $Q$- and $A$-operators.
By construction~(\ref{Q}), operator
$Q_n(x)$ has the following commutation relation with the
product of $L$-operators:
\begin{align}
& \nonumber
Q_n(x)
L_{1}(u_{1},u_{2})\cdots
L_{n-1}(u_{1},u_{2})
L_{n}(u_{1},u_{2})
L_{0}(u_{1},u-x) \\
& \qquad
= L_{1}(u_{1},u-x)
L_{2}(u_{1},u_{2})\cdots
L_{n}(u_{1},u_{2})
L_{0}(u_{1},u_{2}) Q_n(x).\label{main0}
\end{align}
Note that parameter $u-x$ enters the second row of the matrix $\mathrm{L}_{1}\left(u_{1},u-x\right)$ only
so that the replacement $\mathrm{L}_{1}\left(u_{1},u-x\right)\to \mathrm{L}_{1}\left(u_{1},u_{2}\right)$ does not change the first
row of the matrix in the right-hand side of equality~(\ref{main0}).
Extracting from the whole matrix equality the relations for two elements in the first row and using the definition of the matrix elements of the monodromy matrix, we obtain
\begin{equation*}
Q_n(x)
\begin{pmatrix}
A(u) & B(u)
\end{pmatrix}
L_{0}(u_{1},u-x) = \begin{pmatrix}
A(u) & B(u)
\end{pmatrix}
L_{0}(u_{1},u_{2}) Q_n(x).
\end{equation*}
Next, we use the factorized representation~\eqref{Lfac} for the matrices
$L_0$
\begin{align*}
&
Q_n(x)
\begin{pmatrix}
A(u) & B(u)
\end{pmatrix}
\begin{pmatrix}
 1 & 0 \\
 z_0 & 1 \\
\end{pmatrix}
\begin{pmatrix}
 1 & -\partial_{0} \\
 0 & u-x \\
\end{pmatrix}
\begin{pmatrix}
 \hphantom{-}u_1 & 0 \\
 -z_0 & 1 \\
\end{pmatrix} \\
& \qquad{}
=
\begin{pmatrix}
A(u) & B(u)
\end{pmatrix}
L_{0}(u_{1},u_{2})
\begin{pmatrix}
 1 & 0 \\
 z_0 & 1 \\
\end{pmatrix}
\begin{pmatrix}
 1 & -\partial_{0} \\
 0 & u-s \\
\end{pmatrix}
\begin{pmatrix}
 \hphantom{-}u_1 & 0 \\
 -z_0 & 1 \\
\end{pmatrix} Q_n(x)
\end{align*}
and evident commutativity of the operator $Q_n( x)$
with $z_0$ which allows to cancel the most right matrix
in both sides of our relation.
In this way, we transform it to the simpler form
\begin{align*}
Q_n(x)
\begin{pmatrix}
A(u)+z_0 B(u) & B(u)
\end{pmatrix}
\begin{pmatrix}
 1 & -\partial_{0} \\
 0 & u-x \\
\end{pmatrix}
 = \begin{pmatrix}
A(u)+z_0 B(u) & B(u)
\end{pmatrix}
\begin{pmatrix}
 1 & -\partial_{0} \\
 0 & u-s \\
\end{pmatrix}
Q_n(x)
\end{align*}
and as a consequence one immediately obtains
\begin{align*}
Q_n(x)
(A(u)+z_0 B(u))
= (A(u)+z_0 B(u))
Q_n(x) .
\end{align*}
To derive Baxter equation, we start from the
defining relation~(\ref{defR}) and using the factorized representation for $L$-operator and explicit expression~\eqref{R1} for the operator $R_{12}( x)$
transform it to the form
\begin{align*}
&
Z_1^{-1}
\mathcal{R}_{12}(x) L_1(u_1,u_2) Z_2 \\
&\qquad{}
=
\begin{pmatrix}
\imath\mathcal{R}_{12}(x+1,\bar{x}) +
(u-x)\mathcal{R}_{12}( x) & -\mathcal{R}_{12}( x)\partial_{1} \\
-(u-x) z_{12} \mathcal{R}_{12}( x) & \imath(x+s-1)(x-s)
\mathcal{R}_{12}(x-1,\bar{x}) \\
\end{pmatrix},
\end{align*}
where \smash{$Z_k =
\bigl(\begin{smallmatrix}
 1 & 0 \\
 z_k & 1 \\
\end{smallmatrix}\bigr)$}.
Note the important property: for $x=u$ matrix is triangular.
Let us put ${x=u}$, ${\bar{x}=\bar{u}}$ and write down analog of
this relation for the site $k$
\begin{align*}
& Z^{-1}_k \mathcal{R}_{k k+1}( u)
 L_k(u_1,u_2) Z_{k+1} \\
& \qquad{} =
\begin{pmatrix}
\imath\mathcal{R}_{k k+1}(u+1,\bar{u}) & -\mathcal{R}_{k k+1}( u)\partial_{k} \\
 0 & \imath(u+s-1)(u-s) \mathcal{R}_{k k+1}(u-1,\bar{u}) \\
\end{pmatrix}.
\end{align*}
All matrices $Z_k$ and \smash{$Z^{-1}_k$} with $k=2,3,\dots,n$ cancel
each other in the product over all sites so that one obtains
\begin{align*}
&
Z_1^{-1} \mathcal{R}_{12}( u)\cdots
\mathcal{R}_{n0}( u)
L_1(u_1 ,u_2)\cdots L_n(u_1 ,u_2) Z_0 \\
&
\qquad{} =
\begin{pmatrix}
\imath\mathcal{R}_{12}(u+1,\bar{u}) & \dots \\
 0 & \dots \\
\end{pmatrix}
\cdots
\begin{pmatrix}
\imath\mathcal{R}_{n0}(u+1,\bar{u}) & \dots \\
 0 & \dots \\
\end{pmatrix} .
\end{align*}
The first Baxter equation~(\ref{Bax})
\begin{equation*}
Q_n(u,\bar{u}) A(u ,z_0) = \imath^n Q_n(u+1,\bar{u})
\end{equation*}
is direct consequence of this matrix relation.
The second relation~(\ref{Baxbar}) can be proved in the same way.
Operator proof of the commutation relation
\begin{align}\label{com}
Q_n( x) Q_n( y) = Q_n(y) Q_n(x)
\end{align}
is based on the following relations for $R$-operators:
\begin{align}\label{YB}
\mathcal{R}_{12}(x,y) \mathcal{R}_{23}(x) \mathcal{R}_{12}(y) =
\mathcal{R}_{23}(y) \mathcal{R}_{12}(x) \mathcal{R}_{23}(x,y) ,
\end{align}
where
\begin{align}\label{Rxy}
\mathcal{R}_{12}(x,y) = [z_{12}]^{1-s-y} [\hat{p}_1]^{x-y} [z_{12}]^{s+x-1} = \mathcal{R}_{12}(x) \mathcal{R}^{-1}_{12}(y) =
\mathcal{R}^{-1}_{12}(y) \mathcal{R}_{12}(x) .
\end{align}
The relations~\eqref{YB} and~\eqref{Rxy} can be checked using the
star-triangle relation in operator form~\eqref{star-tr}.
To show how it works, we will use representative example
and hope that generalization will be evident
\begin{align*}
Q_3( x) Q_3( y) & =
\mathcal{R}_{12}(x) \mathcal{R}_{23}(x) \mathcal{R}_{30}(x)
\mathcal{R}_{12}(y) \mathcal{R}_{23}(y) \mathcal{R}_{30}(y) \\
& =
\mathcal{R}_{12}(x) \mathcal{R}_{23}(x)
\mathcal{R}_{12}(y) \mathcal{R}_{30}(x)
\mathcal{R}_{23}(y) \mathcal{R}_{30}(y) \\
& =
\mathcal{R}_{12}(y) {\blue \mathcal{R}_{12}(x,y)} \mathcal{R}_{23}(x)
\mathcal{R}_{12}(y) \mathcal{R}_{30}(x)
\mathcal{R}_{23}(y) \mathcal{R}_{30}(y) \\
& =
\mathcal{R}_{12}(y) \mathcal{R}_{23}(y)
\mathcal{R}_{12}(x) {\blue \mathcal{R}_{23}(x,y)} \mathcal{R}_{30}(x)
\mathcal{R}_{23}(y) \mathcal{R}_{30}(y) \\
& =
\mathcal{R}_{12}(y) \mathcal{R}_{23}(y)
\mathcal{R}_{12}(x) \mathcal{R}_{30}(y)
\mathcal{R}_{23}(x) {\blue \mathcal{R}_{30}(x,y)} \mathcal{R}_{30}(y) \\
& =
\mathcal{R}_{12}(y) \mathcal{R}_{23}(y)
\mathcal{R}_{12}(x) \mathcal{R}_{30}(y)
\mathcal{R}_{23}(x) \mathcal{R}_{30}(x) = Q_3( y) Q_3( x) .
\end{align*}
The alternative proof of the commutation relation~\eqref{com}
consists in diagrammatic proof of the corresponding relation
for the integral kernels.

The operator $Q_n(x)$ is an integral operator.
The action on a function $\Phi(\bm z_n)$ is defined in a~standard way
\begin{align*}
[Q_n( x) \Phi](\bm z_n) = \int \mathrm{d}^{2} \bm w_n
 Q(\bm z_n,\bm w_n;x) \Phi(\bm w_n) ,
\end{align*}
where $\mathrm{d}^{2}\bm w_n = \prod_{k=1}^{n} \mathrm{d}^2 w_k$.
The integral kernel $Q(\bm z_n,\bm w_n;x)$
\begin{align*}
Q(\bm z_n,\bm w_n;x) = c^n(x-s)
\prod_{k=1}^{n}
[z_k-z_{k+1}]^{1-2s}
[z_k-w_k]^{s-x-1}[w_k-z_{k+1}]^{s+x-1} .
\end{align*}
can be easily obtained
from the explicit representation~\eqref{Qexp} for operator $Q_n(x)$

We write all products uniformly assuming identification $z_{n+1}=z_0$.
Note that operator $Q_n( x)$ maps the function of the $n$ variables
$z_1 ,\dots ,z_n$ to the function of $n$ variables and
distinguished variable $z_0$ plays the role of auxiliary parameter.

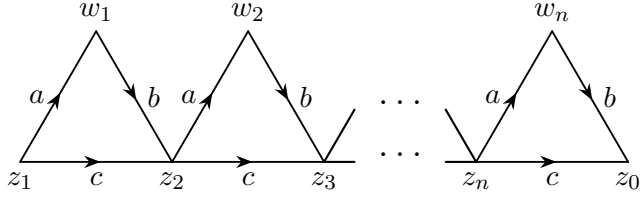
\begin{figure}[t]
	\centering
	\begin{tikzpicture}[thick, line cap = round, scale=2]
		\def\si{0.866} 
		\def\cut{0.4} 
		\foreach \k in {0, 1, 3}
		{
			\draw[->-] (\k,0) to node[midway,left=1]{$a$} (0.5+\k, \si);
			\draw[->-] (0.5+\k, \si) to node[midway,right=1]{$b$} (1+\k,0);
			\draw[->-] (\k,0) to node[midway,below]{$c$} (1+\k,0);
		}
		\draw (2+0.5*\cut,\si*\cut) -- (2,0) -- (2+0.5*\cut,0);
		\draw (3-0.5*\cut, \si*\cut) -- (3,0) -- (3-0.5*\cut, 0);
		\foreach \k in {0, 1} {\node at (2.5,\k*\si*\cut+0.05) {\Large{$\dots$}};}
		\foreach \k in {1, 2, 3} {\node[below] at (\k-1, 0) {$z_{\k}$};}
		\node[below] at (3, 0) {$z_{n}$}; \node[below] at (4, 0) {$z_0$};
		\foreach \k in {1, 2} {\node[above] at (\k-0.5,\si) {$w_{\k}$};}
		\node[above] at (3.5, \si) {$w_{n}$};
	\end{tikzpicture}
	\caption{The diagrammatic representation for
		$Q(\bm z_n,\bm w_n;x) $.
		The arrow with index $\alpha$ from $z$ to $w$ stands for $[z-w]^{-\alpha}$.
		The indices are given by the following expressions: $a=1-s+x$, $b=1-s-x$, $c=2s-1$.}
	\label{QDiag}
\end{figure}

The kernel $Q(\bm z_n,\bm w_n;x)$ can be represented in the diagrammatic form using the technique described in Appendix~\ref{App-DiagTech}. The corresponding diagram is shown in Figure~\ref{QDiag}.
Note that usually diagram represents nontrivial part of considered expression and for simplicity we do not show needed constants (in the present example it is the coefficient $c^n(x-s)$).
The commutativity of $Q$-operators~\eqref{com} can be alternatively proven diagrammatically by the use of transformations depicted in Figure~\ref{Rules}, see \cite{DKO, DKM}.

Later, we will need the inverse and conjugated $Q$-operators also.
The representation~\eqref{Qexp} for $Q$-operator allows to derive
the similar representation for the inverse operator
\begin{align}\label{Qinvexp}
Q^{-1}_n(x) = [z_{n0}]^{1-s-x} [\hat{p}_n]^{s-x}
\cdots
[z_{23}]^{1-s-x} [\hat{p}_2]^{s-x}
[z_{12}]^{1-s-x} [\hat{p}_1]^{s-x} S_n^{-1}
\end{align}
and conjugated operator
\begin{align}\label{Q-dagger}
Q^{\dagger}_n(x) = [z_{n0}]^{-s+\bar{x}^*} [\hat{p}_n]^{-1+s+\bar{x}^*}
\cdots
[z_{23}]^{-s+\bar{x}^*} [\hat{p}_2]^{-1+s+\bar{x}^*}
[z_{12}]^{-s+\bar{x}^*} [\hat{p}_1]^{-1+s+\bar{x}^*} S_n^{-1},
\end{align}
where we used the following conjugation rules:
\begin{equation*}
([z]^{\alpha})^{\dagger} =
[z]^{\bar{\alpha}^*} =
z^{\bar{\alpha}^*} \bar{z}^{\alpha^*},\qquad ([\hat{p}]^{\alpha})^{\dagger} =
[\hat{p}]^{\bar{\alpha}^*} = \hat{p}^{\bar{\alpha}^*}
\hat{\bar{p}}^{\alpha^*}.
\end{equation*}
Due to~\eqref{sparam}, we have $\bar{s}^* = 1-s$.
Note that we do not impose restriction $\bar{x}^* = -x$~\eqref{x-n-nu}
for the spectral parameter in the formula~\eqref{Q-dagger}
and write everything in general situation.
Comparison of representations
\eqref{Qinvexp} and~\eqref{Q-dagger} shows that
\begin{align} \label{QdagQinv}
Q^{\dagger}_n(x) = Q^{-1}_n(1-\bar{x}^*) .
\end{align}
To avoid misunderstanding, we collect explicit expressions for the operator $Q(u)$
at the special point $u = 1-s$
\begin{align*}
&
Q_n(1-s) = S_n
[\hat{p}_1]^{1-2s}
[\hat{p}_2]^{1-2s} \cdots [\hat{p}_n]^{1-2s} , \\
&
Q^{-1}_n(1-s) =
[\hat{p}_1]^{2s-1}
[\hat{p}_2]^{2s-1} \cdots [\hat{p}_n]^{2s-1} S^{-1}_n , \\
&
Q^{\dagger}_n(1-s) =
[\hat{p}_1]^{2s-1}
[\hat{p}_2]^{2s-1} \cdots [\hat{p}_n]^{2s-1} S^{-1}_n
= Q^{-1}_n(1-s) ,
\end{align*}
and in this case
\begin{align*}
Q^{\dagger}_n(1-s) = Q^{-1}_n(1-s) .
\end{align*}
We will need these formulas later.

\section{Eigenfunctions}\label{sect:eigen}

In this section, we will present iterative construction of
eigenfunctions~\eqref{AbAN}. The eigenfunction~$\Psi_{\bm x_n}(\bm z_n)$
is obtained from the eigenfunction~$\Psi_{\bm x_{n-1}}(\bm z_{n-1})$
from the previous level by application of
the raising operator $\Lambda_n(x_n)$
\begin{align*}
\Psi_{\bm x_n}(\bm z_n) = \Lambda_n(x_n) \Psi_{\bm x_{n-1}}(\bm z_{n-1}) ,
\end{align*}
and now we are going to the explicit construction of
the raising operators $\Lambda_n(x_n)$.
The main building blocks are operators $\mathcal{R}_{ik}( x)$
and on the first step we construct raising operator
for eigenfunctions of operator $A(u)$. Then using relation~\eqref{Az0B},
we construct raising operator for eigenfunctions of $A(u,z_0)$.

The composite operator
$
\mathcal{R}_{12}( x)\mathcal{R}_{23}( x)\cdots
\mathcal{R}_{n-1 n}( x)
$
has the following commutation relation with the
product of $L$-operators:
\begin{align}
& \nonumber
\mathcal{R}_{12}( x)\cdots
\mathcal{R}_{n-1 n}( x)
L_{1}(u_{1},u_{2})\cdots
L_{n-1}(u_{1},u_{2})
L_{n}(u_{1},u-x) \\
& \qquad
= L_{1}(u_{1},u-x)
L_{2}(u_{1},u_{2})\cdots
L_{n}(u_{1},u_{2}) \mathcal{R}_{12}( x)\cdots
\mathcal{R}_{n-1 n}(x) .\label{mL}
\end{align}
It is matrix equality and we will concentrate on
the relations for two elements in the first row
\begin{align*}
&
\mathcal{R}_{12}( x)\cdots
\mathcal{R}_{n-1 n}( x)
\begin{pmatrix}
A_{n-1}(u) & B_{n-1}(u)
\end{pmatrix}
\begin{pmatrix} u+s+z_n \partial_n & -\partial_n\\
z_n(z_n\partial_n + s+x) & u-x-z_n\partial_n
\end{pmatrix} \\
&
\qquad= \begin{pmatrix}
A_{n}(u) & B_{n}(u)
\end{pmatrix}
\mathcal{R}_{12}( x)\cdots
\mathcal{R}_{n-1 n}( x) .
\end{align*}
Note again that the replacement $L_{1}(u_{1},u-x)\to L_{1}(u_{1},u_{2})$ does not introduce any change
in the first row of the matrix in the right-hand side of equality~\eqref{mL}.
We are going to construct eigenfunction of $A$-operator so that
let us concentrate on the first relation
\begin{align*}
&
A_{n}(u) \mathcal{R}_{12}( x)\cdots
\mathcal{R}_{n-1 n}( x) \\
&
\qquad=
\mathcal{R}_{12}( x)\cdots
\mathcal{R}_{n-1 n}( x)
\bigl(A_{n-1}(u) (u+s+z_n \partial_n)
+ B_{n-1}(u) z_n (z_n \partial_n +
s+x)\bigl) .
\end{align*}
In the next step, we multiply obtained relation by the function $[z_n]^{-s-x}$
and using simple rule of conjugation
\begin{align*}
[z]^{\alpha} z\partial_{z} [z]^{-\alpha} = z\partial_{z} -\alpha
\end{align*}
derive the key commutation relation
\begin{align}\label{main}
A_{n}(u) \Lambda_n( x) =
\Lambda_n( x) \bigl(A_{n-1}(u) (u-x+z_n \partial_n)
+ B_{n-1}(u) z_n^2 \partial_n \bigl)
\end{align}
for the main building block in iterative construction of eigenfunctions -- composite operator
\begin{align*}
\Lambda_n( x) = \mathcal{R}_{12}( x)\mathcal{R}_{23}( x)\cdots
\mathcal{R}_{n-1 n}( x) [z_n]^{-s-x} .
\end{align*}
Note that this operator does not depend on
momentum operators $\hat{p}_n = -\imath\partial_{z_n}$ and
$\hat{\bar{p}}_n = -\imath\partial_{\bar{z}_n}$ and commutes with $z_n$ and $\bar{z}_n$.
The recurrent procedure for construction of eigenfunctions
is based on the relation~\eqref{main}.
Let us apply operators in lef- and right-hand sides of~\eqref{main} to the function
$\Psi(\bm z_{n-1})$ which does not depend on the variable $z_n$
\begin{align*}
A_{n}(u) \Lambda_n( x)
\Psi(\bm z_{n-1}) =
(u-x) \Lambda_n( x)
A_{n-1}(u)\Psi(\bm z_{n-1}) .
\end{align*}
In the next step, we use the same formula on the previous level
\begin{align*}
A_{n-1}(u) \Lambda_{n-1}( x)
\Psi(\bm z_{n-2}) =
(u-x) \Lambda_{n-1}( x)
A_{n-2}(u)\Psi(\bm z_{n-2})
\end{align*}
and obtain
\begin{align*}
&A_{n}(u) \Lambda_n( x_n)\Lambda_{n-1}( x_{n-1})
\Psi(\bm z_{n-2})=
(u-x_n) \Lambda_n( x_n)
A_{n-1}(u) \Lambda_{n-1}( x_{n-1})\Psi(\bm z_{n-2}) \\
&\qquad =
(u-x_n) (u-x_{n-1}) \Lambda_n( x_n)
\Lambda_{n-1}( x_{n-1})
A_{n-2}(u) \Psi(\bm z_{n-2}).
\end{align*}
Of course, it is possible to continue up to the last step
\begin{align*}
&
A_{n}(u) \Lambda_n( x_n)
\Lambda_{n-1}( x_{n-1})\cdots
\Lambda_{2}( x_{2}) \Psi(z_1)
\\
& \qquad
= (u-x_n)(u-x_{n-1})\cdots(u-x_2)
\Lambda_n( x_n)
\Lambda_{n-1}( x_{n-1})\cdots
\Lambda_{2}( x_{2}) A_{1}(u) \Psi(z_1) .
\end{align*}
The eigenfunction of the last operator
$A_{1}(u) = u+s+z_1\partial_1$ is $[z_1]^{-s-x_1}$.
Repeating all the same for the anti-holomorphic
sector, we construct eigenfunction of two commuting
operators $A_n(u)$ and $\bar{A}_n(\bar{u})$
\begin{align}
&A_n(u)
\Psi_{\bm{x}_n}(\bm z_n)= (u-x_1)\cdots(u-x_n)
\Psi_{\bm{x}_n}(\bm z_n),
\nonumber\\
&\bar{A}_n(\bar{u})
\Psi_{\bm{x}_n}(\bm z_n)= (\bar u-\bar x_1)\cdots(\bar u-\bar x_n)
 \Psi_{\bm{x}_n}(\bm z_n)\label{APsi}
\end{align}
in the following explicit form:
\begin{align*}
\Psi_{\bm{x}_n}(\bm z_n) =
\Lambda_n( x_n)
\Lambda_{n-1}( x_{n-1})\cdots
\Lambda_{2}( x_{2}) [z_1]^{-s-x_1} .
\end{align*}
Of course, the eigenfunction is defined up to overall
normalization. We will fix the useful and natural normalization later.
Note that due to relation~(\ref{Az0B})
\begin{align*}
A_n(u,z_0) = A_n(u)+z_0 B_n(u) = {\rm e}^{z_0 S_{-}} A_n(u) {\rm e}^{-z_0 S_{-}},
\end{align*}
eigenfunction of the
operator $A_n(u,z_0)$ can be obtained from the eigenfunctions
of the operator~$A_n(u)$ by transformation
\smash{$\Psi_{\bm{x}_n}(\bm z_n) \to
{\rm e}^{z_0 S_{-}+\bar{z}_0\bar{S}_-} \Psi_{\bm{x}_n}(\bm z_n)$}
which is equivalent to the shift of all variables $z_k\to z_k-z_0$.
In the following, we will work with eigenfunctions of the operator~$A_n(u,z_0)$
exclusively so that for the sake of simplicity we
will use the same notation as in the case of the operator $A_n(u)$
but with needed changes due to the shift $z_k\to z_k-z_0$.

\begin{figure}[t]
	\centering
	\begin{tikzpicture}[thick, line cap = round, scale=2]
		\def\si{0.866} 
		\def\cut{0.4} 
		\foreach \m in {0, 1, 3}
		{
			\draw[->-] (\m,0) to node[midway,left=1]{$a$} (0.5+\m, \si);
			\draw[-<-] (1+\m,0) to node[midway,right=1]{$b$} (0.5+\m, \si);
			\draw[->-] (\m,0) to node[midway,below]{$c$} (1+\m,0);
		}
		\draw (2+0.5*\cut,\si*\cut) -- (2,0) -- (2+0.5*\cut,0);
		\draw (3-0.5*\cut, \si*\cut) -- (3,0) -- (3-0.5*\cut, 0);
		\foreach \m in {0, 1} {\node at (2.5,\m*\si*\cut+0.05) {\Large{$\dots$}};}
		\draw[->-] (4,0) to node[midway,right]{$1-b$} (4.5,\si);
		\foreach \m in {1, 2, 3} {\node[below] at (\m-1, 0) {$z_{\m}$};}
		\node[below] at (3, 0) {$z_{k-1}$}; \node[below] at (4, 0) {$z_k$};
		\node[above] at (4.5, \si) {$z_0$};
		\foreach \m in {1, 2} {\node[above] at (\m-0.5,\si) {$w_{\m}$};}
		\node[above] at (3.5, \si) {$w_{k-1}$};
	\end{tikzpicture}
	\caption{The diagrammatic representation for
		$\Lambda(\bm z_k,\bm w_{k-1};x)$.
		The indices are given by the following expressions: $a=1-s+x$, $b=1-s-x$, $c=2s-1$.}
	\label{LambdaDiag}
\end{figure}
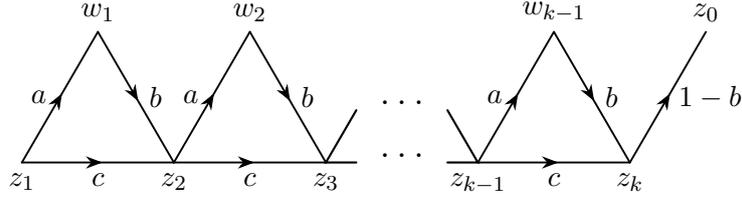

Finally, the solution of our problem~\eqref{AbAN} is given by
the formula
\begin{align}\label{reprPsi}
\Psi_{\bm{x}_n}(\bm z_n) =
\Lambda_n( x_n)
\Lambda_{n-1}( x_{n-1})\cdots
\Lambda_{2}( x_{2}) [z_{10}]^{-s-x_1} ,
\end{align}
where
\begin{align} \label{Lambda}
\Lambda_k( x) & = \mathcal{R}_{12}(x)\mathcal{R}_{23}(x)\cdots
\mathcal{R}_{k-1 k}(x) [z_{k0}]^{-s-x} \\
\label{LambdaExpl}
& = S_k [z_{k0}]^{s-1-x}
[\hat{p}_1]^{x-s} [z_{12}]^{s+x-1}
[\hat{p}_2]^{x-s} [z_{23}]^{s+x-1}\cdots
[\hat{p}_{k-1}]^{x-s} [z_{k-1 k}]^{s+x-1}
\end{align}
and
\begin{equation*}
S_k = [z_{12}]^{1-2s} [z_{23}]^{1-2s}
\cdots [z_{k0}]^{1-2s} .
\end{equation*}
From the explicit formula~\eqref{LambdaExpl} and the definition of operator $[\hat{p}]^\alpha$~\eqref{d}, it follows that
$\Lambda_k(x)$ is an integral operator of the form
\begin{equation*}
	[\Lambda_k(x) \Phi](\bm z_k) = \int \mathrm{d}^{2} \bm w_{k-1}
\Lambda(\bm z_k,\bm w_{k-1};x) \Phi(\bm w_{k-1}) ,
\end{equation*}
where
\begin{align*}
\Lambda(\bm z_n,\bm w_{n-1};x) = & \; c^{n-1}(x-s) [z_{n}-z_{0}]^{-s-x} \\
& \times
\prod\limits_{k=1}^{n-1}
[z_{k}-z_{k+1}]^{1-2s}[z_k-w_k]^{s-x-1}[w_k -z_{k+1}]^{s+x-1} .
\end{align*}
Diagrammatic representation for the kernel
$\Lambda(\bm z_n,\bm w_{n-1};x)$ is introduced in the Figure~\ref{LambdaDiag}.

\subsection[Eigenfunctions of Q-operator]{Eigenfunctions of $\boldsymbol{Q}$-operator}

The operators $A_n(u,z_0)$ and $Q_n(v)$ commute and
therefore have the common system of eigenfunctions.
The explicit proof of the fact that
$\Psi_{\bm x_n}(\bm z_n)$ is eigenfunction
of the operator $Q_n(u)$ and calculation of the corresponding eigenvalue
are grounded on the following exchange relation valid for $n=1 ,2 ,\dots$:
\begin{align}
\label{exchQA}
Q_n(u) \Lambda_n(x) \Psi(\bm z_{n-1})
= q(u ,x) \Lambda_n(x) Q_{n-1}(u) \Psi(\bm z_{n-1}) ,
\end{align}
where
\begin{equation*}
	q(u ,x) = [\imath]^{s-u} \frac{\mathbf{\Gamma}(u-x)}{\mathbf{\Gamma}(s-x)} .
\end{equation*}
Using this exchange relation in the expression
\begin{equation*}
	Q_n(u) \Psi_{\bm x_n}(\bm z_n) =
	Q_n(u) \Lambda_n(x_n)
	\Lambda_{n-1}(x_{n-1})\cdots
	\Lambda_{2}(x_{2}) [z_{10}]^{-s-x_1},
\end{equation*}
we move the $Q$-operator to the right and obtain
\begin{align}
	\label{QPsi}
	Q_n(u) \Psi_{\bm x_n}(\bm z_n)
	= q(u ,\bm x_{n})
	 \Psi_{\bm x_n}(\bm z_n) ,
\end{align}
where
\begin{align}\label{qn}
q(u ,\bm x_{n}) = \prod_{k=1}^n q(u ,x_{k}) =
[\imath]^{(s-u)n}
\frac{\mathbf{\Gamma}(u ,\bm{x}_n)}{\mathbf{\Gamma}(s ,\bm{x}_n)}
\end{align}
and we introduced the compact notation
\begin{align*}
\mathbf{\Gamma}(y ,\bm{x}_n) = \prod_{k=1}^n \mathbf{\Gamma}(y-x_k) .
\end{align*}
On the last step, we used the basic relation
\begin{equation}\label{Q1}
	Q_1(u) [z_{10}]^{-s-x_1} =
\mathcal{R}_{10}(u) [z_{10}]^{-s-x_1}
	= q(u,x_1) [z_{10}]^{-s-x_1} ,
\end{equation}
which is equivalent to the chain relation~\eqref{Chain1}.
Indeed, explicit expression~\eqref{R1} for the
operator~$\mathcal{R}_{10}(u)$
\begin{align*}
Q_1( u) = \mathcal{R}_{10}(u) = [z_{10}]^{1-2s}
[\hat{p}_1]^{u-s} [z_{10}]^{s+u-1}
\end{align*}
allows us to rewrite~\eqref{Q1} in an equivalent form~\eqref{Chain1}
\begin{align*}
[\hat{p}_1]^{u-s} [z_{10}]^{u-x_1-1} = [\imath]^{s-u} \frac{\mathbf{\Gamma}(u-x_1)}{\mathbf{\Gamma}(s-x_1)}
 [z_{10}]^{s-x_1-1} .
\end{align*}
To avoid misunderstanding, we should note that
it is not the operator identity like star-triangle relation~\eqref{star-tr} but it is result of application of the operator
$[\hat{p}_1]^{u-s}$ to the particular function of variable $z_1$.
We will see that relation~\eqref{Q1} plays the key role in
procedure of reduction from $Q$-operator to $\Lambda$-operator.

Let us return to the proof of the exchange relation~\eqref{exchQA}.
It can be proven diagrammatically~\cite{DKO, DKM,DM2} in the same
way as relation~\eqref{com} but in fact it is simple
reduction of~\eqref{com}.
The operators $Q_n(x)$ and $\Lambda_n(x)$ are closed relativies and comparison of the operator expressions%
\begin{align*}
&Q_n( x) = \mathcal{R}_{12}(x)\mathcal{R}_{23}(x)\cdots
\mathcal{R}_{n-1 n}(x) \mathcal{R}_{n 0}(x) , \\
&\Lambda_n( x) = \mathcal{R}_{12}(x)\mathcal{R}_{23}(x)\cdots
\mathcal{R}_{n-1 n}(x) [z_{n0}]^{-s-x}
\end{align*}
immediately shows that
\begin{align}\label{QL}
Q_n( x) = \Lambda_n(x) [z_{n0}]^{s+x} \mathcal{R}_{n 0}(x) .
\end{align}
This relation suggests the natural reduction from $Q$-operator to $\Lambda$-operator.
Action of the operator $Q_n( x)$ on the function
$[z_{n 0}]^{-s-\alpha} \Psi(\bm z_{n-1})$ with special
dependence on variable $z_n$ induces action
of the operator $\Lambda_n( x)$ on
the function $\Psi(\bm z_{n-1})$
\begin{align}\label{QL1}
Q_n( x) [z_{n 0}]^{-s-\alpha} \Psi(\bm z_{n-1}) =
q(x,\alpha) [z_{n0}]^{x-\alpha}
\Lambda_n( x) \Psi(\bm z_{n-1}),
\end{align}
where we used~\eqref{QL} and~\eqref{Q1} with evident changes
$z_1 \to z_n$, $u \to x$, $x_1 \to \alpha$.
After sending $\alpha \to x$, one obtains
\begin{align}\label{RedQL}
\Lambda_n( x) \Psi(\bm z_{n-1}) =
\lim\limits_{\alpha\to x} q^{-1}(x,\alpha) Q_n( x) [z_{n 0}]^{-s-\alpha} \Psi(\bm z_{n-1})
\end{align}
and next we will use this reduction to derive commutation
relation for $Q$- and $\Lambda$-operators~\eqref{exchQA} from
the commutation relation for $Q$-operators~\eqref{com}.
First of all, we have from~\eqref{QL1}
\begin{align*}
&Q_n( x) Q_n( y) [z_{n 0}]^{-s-\alpha} \Psi(\bm z_{n-1}) =
q(y,\alpha) Q_n( x) [z_{n 0}]^{y-\alpha} \Lambda_n(y) \Psi(\bm z_{n-1}) ,\\
&Q_n( y) Q_n( x) [z_{n 0}]^{-s-\alpha} \Psi(\bm z_{n-1}) =
q(x,\alpha) Q_n( y) [z_{n 0}]^{x-\alpha} \Lambda_n(x) \Psi(\bm z_{n-1}) ,
\end{align*}
and then commutativity relation~\eqref{com} implies
\begin{align*}
q(y,\alpha) Q_n( x) [z_{n 0}]^{y-\alpha} \Lambda_n(y) \Psi(\bm z_{n-1}) =
q(x,\alpha) Q_n( y) [z_{n 0}]^{x-\alpha} \Lambda_n(x) \Psi(\bm z_{n-1}) ,
\end{align*}
so that in the limit $\alpha \to y$ one obtains
\begin{align}\label{QL2}
 Q_n( x) \Lambda_{n}(y) \Psi(\bm z_{n-1}) = q(x,y)
\lim\limits_{\alpha\to y}
q^{-1}(y,\alpha) Q_n(y)
[z_{n 0}]^{x-\alpha} \Lambda_{n}(x) \Psi(\bm z_{n-1}) .
\end{align}
It remains to calculate the corresponding limit
in the right-hand side. Using representation~\eqref{QL} for $Q_n(y)$ and
then rearranging $\mathcal{R}$-operators, we derive
\begin{align*}
&
Q_n(y)
[z_{n 0}]^{x-\alpha} \Lambda_{n}(x) \Psi(\bm z_{n-1}) \\
&\qquad{} =
\Lambda_n(y) \mathcal{R}_{12}(x)\mathcal{R}_{23}(x)\cdots
\mathcal{R}_{n-2 n-1}(x)
[z_{n0}]^{s+y} \mathcal{R}_{n 0}(y) [z_{n0}]^{-s-\alpha}
\mathcal{R}_{n-1 n}(x) \Psi(\bm z_{n-1}) .
\end{align*}
The last step is the use of the identity~\eqref{del}
\begin{align*}
\lim\limits_{\alpha\to y}
q^{-1}(y,\alpha)
[z_{n0}]^{s+y} \mathcal{R}_{n 0}(y) [z_{n0}]^{-s-\alpha} \Phi(z_n) =
\Phi(z_0),
\end{align*}
which gives in our case
\begin{align*}
\lim\limits_{\alpha\to y}
q^{-1}(y,\alpha)
[z_{n0}]^{s+y} \mathcal{R}_{n 0}(y) [z_{n0}]^{-s-\alpha}
\mathcal{R}_{n-1 n}(x) \Psi(\bm z_{n-1}) =
\mathcal{R}_{n-1 0}(x) \Psi(\bm z_{n-1}) .
\end{align*}
After all, we obtain
\begin{align*}
&
\lim\limits_{\alpha\to y}
q^{-1}(y,\alpha) Q_n(y)
[z_{n 0}]^{x-\alpha} \Lambda_{n}(x) \Psi(\bm z_{n-1}) \\
&
\qquad{}= \Lambda_n(y) \mathcal{R}_{12}(x)\mathcal{R}_{23}(x)\cdots
\mathcal{R}_{n-2 n-1}(x) \mathcal{R}_{n-1 0}(x) \Psi(\bm z_{n-1}) =
\Lambda_n(y) Q_{n-1}(x) \Psi(\bm z_{n-1}) ,
\end{align*}
so that~\eqref{QL2} reduces to the~\eqref{exchQA}.

\subsection{Symmetry} \label{sect:PermSym}

The eigenvalues of operators $A_n$ and $\bar{A}_n$ in the
definition of eigenfunctions~\eqref{APsi} are invariant under
permutations of $x_1, \dots, x_n$.
We expect that the spectrum is simple so that for any
permutation $\tau \in \mathfrak{S}_n$ the eigenfunctions
$\Psi_{\tau \bm x_n}$ and $\Psi_{\bm x_n}$ can differ by
multiplication on the constant~only
\begin{align*}
\Psi_{\tau \bm x_n}(\bm z_n) =
c(\tau,\bm x_n) \Psi_{\bm x_n}(\bm z_n) ,
\end{align*}
where $\tau \bm x_n = \bigl(x_{\tau(1)},\bar{x}_{\tau(1)}, \dots , x_{\tau(n)},\bar{x}_{\tau(n)}\bigr)$.

It is indeed the case and the behaviour under permutations of $x_1, \dots, x_n$ is governed by the following
commutation relation between $\Lambda$-operators ($n=2 ,3 ,\dots$)
\begin{align} \label{exchLL}
[\imath]^{s-y} \mathbf{\Gamma}(s+y)
\Lambda_n(x) \Lambda_{n-1}(y) \Psi(\bm z_{n-2}) =
[\imath]^{s-x} \mathbf{\Gamma}(s+x)
\Lambda_n(y) \Lambda_{n-1}(x) \Psi(\bm z_{n-2}) .
\end{align}
We will use reduction~\eqref{RedQL} to derive commutation relation for $\Lambda$-operators~\eqref{exchLL} from the
commutation relation for $Q$- and $\Lambda$-operators~\eqref{exchQA}
\begin{align*}
Q_n( x) \Lambda_{n}( y) \Psi(\bm z_{n-1}) =
q(x,y) \Lambda_{n}( y) Q_{n-1}( x) \Psi(\bm z_{n-1}) .
\end{align*}
Using~\eqref{QL1}, we derive for the right-hand side ($z_{n-1 0} = z_{n-1} - z_0$)
\begin{align*}
\Lambda_n( y) Q_{n-1}( x) [z_{n-1 0}]^{-s-\alpha} \Psi(\bm z_{n-2}) =
q(x,\alpha) \Lambda_n( y) [z_{n-1 0}]^{x-\alpha} \Lambda_{n-1}(x) \Psi(\bm z_{n-2})
\end{align*}
so that in the corresponding limit $\alpha \to x$ one obtains the following relation:
\begin{align*}
\lim\limits_{\alpha\to x}
 q^{-1}(x,\alpha) Q_n(x) \Lambda_n(y)
[z_{n-1 0}]^{-s-\alpha} \Psi(\bm z_{n-2}) =
q(x,y) \Lambda_{n}(y) \Lambda_{n-1}(x) \Psi(\bm z_{n-2}) .
\end{align*}
It remains to calculate expression in the left-hand side.
We rearrange
$\mathcal{R}$-operators
\begin{align*}
\begin{split}
&Q_n(x) \Lambda_n(y)
[z_{n-1 0}]^{-s-\alpha} \Psi(\bm z_{n-2}) =
\Lambda_n(x) \mathcal{R}_{12}(y)\mathcal{R}_{23}(y)\cdots
\mathcal{R}_{n-2 n-1}(y) \\
&\qquad{} \times
[z_{n0}]^{s+x} \mathcal{R}_{n 0}(x) [z_{n0}]^{-s-y}
\mathcal{R}_{n-1 n}(y) [z_{n-1 0}]^{-s-\alpha} \Psi(\bm z_{n-2})
\end{split}
\end{align*}
then use the identity~\eqref{del1}
\begin{align*}
&
\lim\limits_{\alpha\to x}
q^{-1}(x,\alpha)
[z_{n0}]^{s+x} \mathcal{R}_{n 0}(x) [z_{n0}]^{-s-y}
\mathcal{R}_{n-1 n}(y) [z_{n-1 0}]^{-s-\alpha} \\
&
\qquad =
[\imath]^{s-y} \frac{\mathbf{\Gamma}(x-y,s+y)}
{\mathbf{\Gamma}(s-y,s+x)} [z_{n-1 0}]^{-s-y}
\end{align*}
and obtain
\begin{align*}
&
\lim\limits_{\alpha\to x}
q^{-1}(x,\alpha) Q_n(x) \Lambda_n(y)
[z_{n-1 0}]^{-s-\alpha} \Psi(\bm z_{n-2}) \\
&
\qquad =
[\imath]^{s-y} \frac{\mathbf{\Gamma}(x-y,s+y)}
{\mathbf{\Gamma}(s-y,s+x)} \Lambda_n(x) \mathcal{R}_{12}(y)\mathcal{R}_{23}(y)\cdots
\mathcal{R}_{n-2 n-1}(y) [z_{n-1 0}]^{-s-y} \Psi(\bm z_{n-2}) \\
&
\qquad =
[\imath]^{s-y} \frac{\mathbf{\Gamma}(x-y,s+y)}
{\mathbf{\Gamma}(s-y,s+x)}
\Lambda_n(x) \Lambda_{n-1}(y) \Psi(\bm z_{n-2}) .
\end{align*}
Collecting everything together, we have
\begin{align*}
[\imath]^{s-y} \frac{\mathbf{\Gamma}(x-y,s+y)}
{\mathbf{\Gamma}(s-y,s+x)}
\Lambda_n(x) \Lambda_{n-1}(y) \Psi(\bm z_{n-2}) = q(x,y) \Lambda_{n}(y) \Lambda_{n-1}(x) \Psi(\bm z_{n-2}) .
\end{align*}
This relation is evidently equivalent to~\eqref{exchLL}.

\subsection[The symmetry s to 1-s]{The symmetry $\boldsymbol{s \to 1-s}$} \label{sect:SpinSym}

Representations characterized by the
parameters $(s, \bar s)$ and $(1-s, 1-\bar{s})$ are equivalent \cite{GGV, GN}.
The integral operator~\eqref{W1}
\begin{align*}
W(s,\bar{s}) = [\hat{p}]^{1-2s}
\end{align*}
intertwines such equivalent principal series representations $T^{(s,\bar{s})}$
and $T^{(1-s,1-\bar{s})}$ for generic complex $s$ and $\bar{s}$
\begin{equation}
\label{splet1}
W(s,\bar{s}) T^{(s,\bar{s})}(g) W^{-1}(s,\bar{s}) =
T^{(1-s,1-\bar{s})}(g) .
\end{equation}
In this subsection, we consider the manifestation of this
symmetry $s \rightleftarrows 1-s$ on the level of eigenfunctions.
We should note that this symmetry is very similar to the
coupling constant reflection symmetry in Ruijsenaars hyperbolic
system~\cite{BDKK4}.

We are going to investigate the dependence of eigenfunctions
on parameters $(s, \bar s)$ so that we have to restore
parameter $s$ and use more detailed notation for eigenfunctions
$\Psi_{\bm x_n}(\bm z_n;s)$ and similar notations for operators.
For brevity, we omit $\bar{s}$ almost everywhere.

We are going to prove that transition $s \rightleftarrows 1-s$
can be performed by similarity transformation and is described by formulas
\begin{align}\label{AS}
&S^{-1}_n A_n(u;s) S_n = A_n(u;1-s) , \\
\nonumber
&S_n Q_n(x;1-s) S^{-1}_{n} = Q_n(x;s) Q^{-1}_{n}(1-s;s) , \\
\label{SL}
&S_n \Lambda_n(x;1-s) S^{-1}_{n-1} =
\Lambda_n(x;s) Q^{-1}_{n-1}(1-s;s) ,
\end{align}
where $S_n$~\eqref{Sn} is the operator of multiplication on
the function
\begin{align*}
S_n = [z_{12}]^{1-2s} [z_{23}]^{1-2s}
\cdots [z_{n0}]^{1-2s} .
\end{align*}
The corresponding transformation formula for eigenfunctions has the form
\begin{align*}
\Psi_{\bm x_n}(\bm z_n;1-s) = c(\bm x_{n-1})
S^{-1}_n \Psi_{\bm x_n}(\bm z_n;s) ,
\end{align*}
where
\begin{align}\label{cx}
c(\bm x_{n-1}) =
\prod_{k=1}^{n-1} q^{-1}(1-s ,\bm x_{k};s) =
 [\imath]^{\frac{n(n-1)(1-2s)}{2}}
\prod_{j=1}^{n-1}\frac{\mathbf{\Gamma}^{n-j}(s-x_j)}
{\mathbf{\Gamma}^{n-j}(1-s-x_j)}
\end{align}
and $q(u ,\bm x_{k};s)$ is eigenvalue
of operator $Q_k(u;s)$~\eqref{qn}
\begin{align*}
Q_k(u;s) \Psi_{\bm x_k}(\bm z_k;s)
= q(u ,\bm x_{k};s) \Psi_{\bm x_k}(\bm z_k;s) .
\end{align*}
Let us start from the transformation formula~\eqref{AS} for
the operator $A_n(u;s)$.
One of the ways of the derivation of~\eqref{AS} is
explicit transformation of all $L$-matrices
as is shown in \cite{BDM}. Here we use different approach.
First of all, note that operator $A_n(u;s)$ is a polynomial function of spin $s$
generators \smash{$S^{(k)}$}, \smash{$S^{(k)}_{\pm}$}~\eqref{gen}, where $k=1,\dots,n$.
In accordance with~\eqref{splet1}, $W_k(s)$ intertwines representations of spins $(s,\bar{s})$ and $(1-s,1-\bar{s})$ in the site $k$ and therefore
for the product of all~$W_k(s)$ we have
\begin{equation*}
W_1(s)\cdots W_n(s)
A_n(u;s) W^{-1}_1(s)\cdots W^{-1}_n(s) = A_n(u;1-s) .
\end{equation*}
Due to~\eqref{AS}, the same transformation can be performed using the operator of
multiplication by the function $S_n$
\begin{equation*}
S^{-1}_n A_n(u;s) S_n = A_n(u;1-s) .
\end{equation*}
It is consequence of the fact~\eqref{Q_1-s} that operator
$Q_n(x; s)$ at the point $x=1-s$ is reduced to the
following form
\begin{equation*}
	Q_n(1-s; s) = S_n W_1(s) \cdots W_n(s) .
\end{equation*}
Using this relation and commutativity of operators
$A_n(u;s)$ and $Q_n(x; s)$, we obtain
\begin{align*}
S^{-1}_n A_n(u;s) S_n
& =
W_1(s)\cdots W_n(s) Q^{-1}_n(1-s; s)
A_n(u,s) Q_n(1-s; s)
W^{-1}_1(s)\cdots W^{-1}_n(s) \\
& =
W_1(s)\cdots W_n(s)
A_n(u,s) W^{-1}_1(s)\cdots W^{-1}_n(s) = A_n(u;1-s) .
\end{align*}

Relation~\eqref{AS} shows that eigenfunction $\Psi_{\bm x_n}(\bm z_n;1-s)$
of the operator $A_n(u,1-s)$ should coincide with
$S^{-1}_n\Psi_{\bm x_n}(\bm z_n;s)$ up to overall constant $c(\bm x_{n-1})$
\begin{equation*}
\Psi_{\bm x_n}(\bm z_n;1-s) = c(\bm x_{n-1})
S^{-1}_n \Psi_{\bm x_n}(\bm z_n;s) .
\end{equation*}
The calculation of the constant $c(\bm x_{n-1})$ and explicit proof
of the previous formula is based on the relation~\eqref{SL}
\begin{align*}
S_k \Lambda_k(x;1-s) S^{-1}_{k-1} =
\Lambda_k(x;s) Q^{-1}_{k-1}(1-s;s) .
\end{align*}
Using this relation, we obtain
\begin{align*}
&
S_n \Psi_{\bm{x}_n}(\bm z_n;1-s ) =
S_n \Lambda_{n-1}( x_{n-1};1-s)\cdots
\Lambda_{2}( x_{2};1-s) [z_{10}]^{s-1-x_1} \\
& =\!
S_n \Lambda_n(x_n;1-s) S^{-1}_{n-1}
S_{n-1}\Lambda_{n-1}( x_{n-1};1-s) S^{-1}_{n-2}\cdots
S_{2} \Lambda_{2}( x_{2};1-s) S^{-1}_{1} S_{1} [z_{10}]^{s-1-x_1} \\
& =\!
\Lambda_n(x_n;s) Q^{-1}_{n-1}(1-s;s)
\Lambda_{n-1}( x_{n-1};s) Q^{-1}_{n-2}(1-s;s)\cdots
\Lambda_{2}( x_{2};s) Q^{-1}_{1}(1-s;s) [z_{10}]^{-s-x_1} \\
& =\!
c(\bm x_{n-1})
\Lambda_{n-1}( x_{n-1};s)\cdots
\Lambda_{2}( x_{2};s) [z_1-z_0]^{-s-x_1} =
c(\bm x_{n-1}) \Psi_{\bm{x}_n}(\bm z_n; s) ,
\end{align*}
where coefficient $c(\bm x_{n-1})$ is expressed in terms of
eigenvalues of operators $Q_k(u;s)$ as is shown in~\eqref{cx}.

Let us return to the proof of~\eqref{SL}.
Due to relation~\eqref{Q_1-s},
\begin{equation*}
Q_{n-1}(1-s; s) = S_{n-1} W_1(s) \cdots W_{n-1}(s) =
S_{n-1} [\hat{p}_1]^{1-2s}\cdots[\hat{p}_{n-1}]^{1-2s},
\end{equation*}
formula~\eqref{SL} can be rewritten in an equivalent form
 \begin{align}\label{SLp}
S_n \Lambda_n(x;1-s)
[\hat{p}_1]^{1-2s}\cdots[\hat{p}_{n-1}]^{1-2s} =
\Lambda_n(x;s) .
\end{align}
It remains to prove~\eqref{SLp}.
Everything is based on the local relation
\begin{align}\label{Rk}
[z_{k k+1}]^{1-2s} \mathcal{R}_{k k+1}(x;1-s) [\hat{p}_k]^{1-2s} = \mathcal{R}_{k k+1}(x;s) ,
\end{align}
and we immediately obtain
\begin{align*}
&
S_n \Lambda_n(x;1-s)
[\hat{p}_1]^{1-2s}\cdots[\hat{p}_{n-1}]^{1-2s} \\
& \qquad =
[z_{12}]^{1-2s} \mathcal{R}_{12}(x;1-s) [\hat{p}_1]^{1-2s}
[z_{23}]^{1-2s} \mathcal{R}_{23}(x;1-s) [\hat{p}_2]^{1-2s} \cdots \\
& \qquad\quad \times
[z_{n-1 n}]^{1-2s} \mathcal{R}_{n-1 n}(x;1-s)
[\hat{p}_{n-1}]^{1-2s} [z_{n0}]^{1-2s} [z_{n0}]^{s-1-x} \\
& \qquad =
\mathcal{R}_{12}(x;s) \mathcal{R}_{23}(x;s) \cdots
\mathcal{R}_{n-1 n}(x;s) [z_{n0}]^{-s-x} = \Lambda_n(x;s) .
\end{align*}
In fact, the relation~\eqref{Rk} is equivalent to the star-triangle
relation in operator form~\eqref{star-tr}
\begin{align*}
[z_{12}]^{1-2s} \mathcal{R}_{12}(x;1-s) [\hat{p}_1]^{1-2s} &=
[z_{12}]^{1-2s} [z_{12}]^{2s-1} [\hat{p}_1]^{x+s-1} [z_{12}]^{x-s}
 [\hat{p}_1]^{1-2s} \\
& =
[\hat{p}_1]^{x+s-1} [z_{12}]^{x-s}
 [\hat{p}_1]^{1-2s} \\
 &= [z_{12}]^{1-2s} [\hat{p}_1]^{x-s} [z_{12}]^{x+s-1} = \mathcal{R}_{12}(x;s) .
\end{align*}
In exactly the same way, one obtains the
relation for $Q$-operator
\begin{align}\label{SQp}
S_n Q_n(x;1-s) W_1(s) \cdots W_{n}(s) =
Q_n(x;s).
\end{align}
Indeed, we have
\begin{align*}
&S_n Q_n(x;1-s)
[\hat{p}_1]^{1-2s}\cdots[\hat{p}_{n}]^{1-2s}=
[z_{12}]^{1-2s} \mathcal{R}_{12}(x;1-s) [\hat{p}_1]^{1-2s} \\
& \qquad =
[z_{23}]^{1-2s} \mathcal{R}_{23}(x;1-s) [\hat{p}_2]^{1-2s} \cdots
[z_{n 0}]^{1-2s} \mathcal{R}_{n 0}(x;1-s)
[\hat{p}_{n}]^{1-2s} \\
& \qquad =
\mathcal{R}_{12}(x;s) \mathcal{R}_{23}(x;s) \cdots
\mathcal{R}_{n 0}(s) = Q_n(x;s) .
\end{align*}
Using~\eqref{Q_1-s}, it is easy to rewrite~\eqref{SQp} in the form
\begin{align*}
S_n Q_n(x;1-s) S^{-1}_{n} =
Q_n(x;s) Q^{-1}_{n}(1-s;s) .
\end{align*}
We have finished the proof
of relations~\eqref{AS} and \eqref{SL}.

The similarity transformation with operator $S_n$ allows us to
transform expression for eigenfunction to different but equivalent form.
It is possible to use the operator
\begin{equation} \label{Lambda'}
\Lambda'_n(x;s) = S_n \Lambda_n(x;1-s) S_{n-1}^{-1}
\end{equation}
as the main building block in construction of eigenfunction so that
there are two ways to construct eigenfunctions of $A_{n}(u;s)$ from eigenfunctions $\Psi_{\bm {x}_{n-1}}(\bm z_{n-1})$ of $A_{n-1}(u;s)$
\begin{equation} \label{PsiPsi'}
\Psi_{\bm {x}_{n}}(\bm z_{n}) = \Lambda_n(x_n;s)
\Psi_{\bm {x}_{n-1}}(\bm z_{n-1})
	\qquad \text{or} \qquad
\Psi^{\prime}_{\bm {x}_{n}}(\bm z_{n}) = \Lambda'_n(x_n;s)
\Psi_{\bm {x}_{n-1}}(\bm z_{n-1}) .
\end{equation}
Due to relation~\eqref{SL},
\begin{align*}
\Lambda^{\prime}_n(x;s) =
\Lambda_n(x;s) Q^{-1}_{n-1}(1-s;s), \qquad
\Lambda^{\prime}_1(x;s) =
\Lambda_1(x;s) = [z_1-z_0]^{-s-x} ,
\end{align*}
these two functions differ only in the normalization factor
which is eigenvalue of the operator~$Q_{n-1}(1-s;s)$
\begin{equation*}
\Psi_{\bm {x}_{n}}(\bm z_{n})
= q(1-s;\bm x_{n-1}) \Psi^{\prime}_{\bm {x}_{n}}(\bm z_{n}) .
\end{equation*}
The following expression for $\Lambda'$-operator
\begin{align} \label{Lambda'Expl}
	\Lambda'_n( x)
	= [z_{n0}]^{-s-x}
	[\hat{p}_1]^{x+s-1} [z_{12}]^{x-s}
	[\hat{p}_2]^{x+s-1} [z_{23}]^{x-s}\cdots
	[\hat{p}_{n-1}]^{x+s-1} [z_{n-1 n}]^{x-s}
	S_{n-1}^{-1}
\end{align}
is consequence of the definition~\eqref{Lambda'} and
formula~\eqref{LambdaExpl} for $\Lambda$-operator.
From~\eqref{Lambda'Expl}, we see that~$\Lambda^{\prime}_n$ is integral operator
\begin{equation*}
[\Lambda^{\prime}_n(x) \Phi](\bm z_n) =
\int \mathrm{d}^{2} \bm w_{n-1}
\Lambda^{\prime}(\bm z_n,\bm w_{n-1};x_n)
\Phi(\bm w_{n-1})
\end{equation*}
and integral kernel is given by the following expression:
\begin{align*}
\Lambda^{\prime}(\bm z_n,\bm w_{n-1}; x_n) ={}& c^{n-1}(x+s-1) [z_{n 0}]^{-x-s} \\
& {\times}\,
\prod\limits_{k=1}^{n-1} [z_k-w_k]^{-x-s} [w_k-z_{k+1}]^{x-s}
[w_k-w_{k+1}]^{2s-1} .
\end{align*}
We write all product uniformly assuming identification $w_{n} = z_0$.

\begin{figure}[t]
	\centering
	\begin{tikzpicture}[thick, line cap = round, scale=2]
		\def\si{0.866} 
		\def\cut{0.4} 
		\def\sa{0.1}
		\def\sb{0.85}
		\foreach \k in {0, 1, 3}
		{
			\draw[->-] (1+\k,0) to (1.5+\k, \si);
			\draw[-<-] (1+\k,0) to (0.5+\k, \si);
			\draw[->-] (0.5+\k,\si) to node[midway,above]{$-c$} (1.5+\k,\si);
		}
		\draw[->-] (0, 0) to (0.5, \si);
		\draw (2.5+0.5*\cut,\si-\si*\cut) -- (2.5,\si) -- (2.5+0.5*\cut,\si);
		\draw (3.5-0.5*\cut,\si-\si*\cut) -- (3.5,\si) -- (3.5-0.5*\cut,\si);
		\foreach \k in {0, 1} {\node at (3,\si-\k*\si*\cut-0.05) {\Large{$\dots$}};}
		\foreach \k in {1, 2, 4}
		{
			\node[above=16] at (\k-\sa*0.5, \sa*\si) {\small $1-a$};
			\node[below=19] at (\k+\sb*0.5, \sb*\si) {\small $1-b$};
		}
		\node[above=18] at (0,0) {\small $1-b$};
		\foreach \k in {1, 2, 3} {\node[below] at (\k-1, 0) {$z_{\k}$};}
		\node[below] at (4, 0) {$z_n$};
		\node[above] at (4.5, \si) {$z_0$};
		\foreach \k in {1, 2, 3} {\node[above] at (\k-0.5,\si) {$w_{\k}$};}
		\node[above] at (3.5, \si) {$w_{n-1}$};
	\end{tikzpicture}
	\caption{The diagrammatic representation for
		$\Lambda'(\bm{z}_n,\bm{w}_{n-1};x)$.
		The indices are given by the following expressions: $a=1-s+x$, $b=1-s-x$, $c=2s-1$.}
	\label{Lam_prime}
\end{figure}
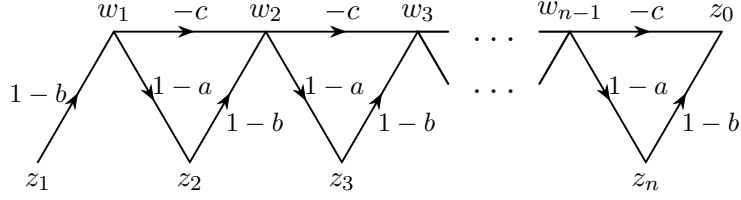

\begin{figure}[t]
	\centering
	\begin{tikzpicture}[thick, line cap = round, scale=1.4]
		\def\si{0.866} 
		\foreach \k in {1, 2}
		{
			\draw[->-] (0,-\k+1) to node[midway,above]{$a_3$} (\si, 0.5-\k);
			\draw[-<-] (0,-\k) to node[midway,above=1]{$b_3$} (\si, 0.5-\k);
			\draw[->-] (0,-\k+1) to node[midway,left]{$c$} (0, -\k);
			\node[left] at (0,-\k+1) {$z_{\k}$};
			\node[right] at (\si,0.5-\k) {$w_{\k}$};
		}
		\node[left] at (0,-2) {$z_3$};
		\draw[->-] (0,-2) to node[midway,above]{$\quad\; 1-b_3$} (\si,-2.5);
		\node[right] at (\si,-2.5) {$z_0$};
		\node at (0.5*\si, -2.8) {(1)};
	\end{tikzpicture}
	\begin{tikzpicture}[thick, line cap = round, scale=1.4]
		\def\si{0.866} 
		\draw[->-] (0,0) to node[midway,above]{$a_2$} (\si, -0.5);
		\draw[-<-] (0,-1) to node[midway,above=1]{$b_2$} (\si, -0.5);
		\draw[->-] (0,0) to node[midway,left]{$c$} (0, -1);
		\node[left] at (0,0) {$z_{1}$};
		\node[right] at (\si, -0.5) {$w_{1}$};
		\node[left] at (0,-1) {$z_2$};
		\draw[->-] (0,-1) to node[midway,above]{$\quad\; 1-b_2$} (\si,-1.5);
		\node[right] at (\si,-1.5) {$z_0$};
		\node at (0.5*\si, -1.8) {(2)};
	\end{tikzpicture}
	\begin{tikzpicture}[thick, line cap = round, scale=1.4]
		\def\si{0.866} 
		\node[left] at (0,0) {$z_1$};
		\draw[->-] (0,0) to node[midway,above]{$\quad\; 1-b_1$} (\si,-0.5);
		\node[right] at (\si,-0.5) {$z_0$};
		\node at (0.5*\si, -0.8) {(3)};
	\end{tikzpicture}
	\\
	\begin{tikzpicture}[thick, line cap = round, scale=1.4]
		\def\si{0.866} 
		\foreach \k in {1, 2}
		{
			\draw[->-] (0,-\k+1) to node[midway,above]{\footnotesize$\quad 1-b_3$} (\si, 0.5-\k);
			\draw[-<-] (0,-\k) to node[midway,above=2]{\footnotesize$1-a_3\quad\;$} (\si, 0.5-\k);
			\draw[->-] (\si,-\k+0.5) to node[midway,right]{$-c$} (\si, -\k-0.5);
			\node[left] at (0,-\k+1) {$z_{\k}$};
			\node[right] at (\si,0.5-\k) {$w_{\k}$};
		}
		\node[left] at (0,-2) {$z_3$};
		\draw[->-] (0,-2) to node[midway,above]{\footnotesize$\quad 1-b_3$} (\si,-2.5);
		\node[right] at (\si,-2.5) {$z_0$};
		\node at (0.5*\si, -2.8) {(4)};
	\end{tikzpicture}
	\begin{tikzpicture}[thick, line cap = round, scale=1.4]
		\def\si{0.866} 
		\draw[->-] (0,0) to node[midway,above]{\footnotesize$\quad 1-b_2$} (\si, -0.5);
		\draw[-<-] (0,-1) to node[midway,above=2]{\footnotesize$1-a_2\quad\;$} (\si, -0.5);
		\draw[->-] (\si,-0.5) to node[midway,right]{$-c$} (\si, -1.5);
		\node[left] at (0,0) {$z_{1}$};
		\node[right] at (\si, -0.5) {$w_{1}$};
		\node[left] at (0,-1) {$z_3$};
		\draw[->-] (0,-1) to node[midway,above]{\footnotesize$\quad 1-b_2$} (\si,-1.5);
		\node[right] at (\si,-1.5) {$z_0$};
		\node at (0.5*\si, -1.8) {(5)};
	\end{tikzpicture}
	\begin{tikzpicture}[thick, line cap = round, scale=1.4]
		\def\si{0.866} 
		\node[left] at (0,0) {$z_1$};
		\draw[->-] (0,0) to node[midway,above=1]{$\quad 1-b_1$} (\si,-0.5);
		\node[right] at (\si,-0.5) {$z_0$};
		\node at (0.5*\si, -0.8) {(6)};
	\end{tikzpicture}
	\\
	\begin{tikzpicture}[thick, line cap = round, scale=1.4]
		\def\si{0.866} 
		\def\rad{0.05}
		\foreach \k in {1, 2}
		{
			\draw[->-] (0,-\k+1) to node[midway,above]{$a_3$} (\si, 0.5-\k);
			\draw[-<-] (0,-\k) to node[midway,above=1]{$b_3$} (\si, 0.5-\k);
			\draw[->-] (0,-\k+1) to node[midway,left]{$c$} (0, -\k);
			\node[left] at (0,-\k+1) {$z_{\k}$};
		}
		\node[left] at (0,-2) {$z_3$};
		\draw[->-] (0,-2) to node[midway,above]{$\quad\; 1-b_3$} (\si,-2.5);
		\node[right] at (\si,-2.5) {$z_0$};
		
		\draw[->-] (\si,-0.5) to node[midway,above]{$a_2$} (\si+\si, -0.5-0.5);
		\draw[-<-] (\si,-1-0.5) to node[midway,above=1]{$b_2$} (\si+\si, -0.5-0.5);
		\draw[fill = black] (\si, -0.5) circle (\rad);
		\draw[fill = black] (\si, -1.5) circle (\rad);
		\draw[->-] (\si,-0.5) to node[midway,left]{$c$} (\si, -1-0.5);
		\draw[->-] (\si,-1-0.5) to node[midway,above]{$\quad\; 1-b_2$} (\si+\si,-1.5-0.5);
		\node[right] at (\si+\si,-1.5-0.5) {$z_0$};
		
		\draw[->-] (2*\si,-1) to node[midway,above]{$\quad\; 1-b_1$} (3*\si,-1.5);
		\draw[fill = black] (2*\si, -1) circle (\rad);
		\node[right] at (3*\si,-1.5) {$z_0$};
		
		\node at (1.5*\si, -3) {(7)};
	\end{tikzpicture}
	\begin{tikzpicture}[thick, line cap = round, scale=1.4]
		\def\si{0.866} 
		\def\rad{0.05}
		
		\foreach \k in {1, 2}
		{
			\draw[->-] (0,-\k+1) to node[midway,above]{$\quad\; 1-b_3$} (\si, 0.5-\k);
			\draw[-<-] (0,-\k) to node[midway,above=2]{$1-a_3\quad\;$} (\si, 0.5-\k);
			\node[left] at (0,-\k+1) {$z_{\k}$};
		}
		\node[left] at (0,-2) {$z_3$};
		\draw[->-] (0,-2) to node[midway,above]{$\quad\; 1-b_3$} (\si,-2.5);
		\node[right] at (\si,-2.5) {$z_0$};
		
		\draw[->-] (\si,-0.5) to node[midway,above]{$a_2$} (\si+\si, -1);
		\draw[-<-] (\si,-1.5) to node[midway,above=1]{$b_2$} (\si+\si, -1);
		\draw[->-] (\si,-1.5) to node[midway,above]{$a_2$} (\si+\si,-2);
		\draw[fill = black] (\si, -0.5) circle (\rad);
		\draw[fill = black] (\si, -1.5) circle (\rad);
		\node[right] at (\si+\si,-2) {$z_0$};
		
		\draw[->-] (2*\si,-1) to node[midway,above]{$\quad\; 1-b_1$} (3*\si,-1.5);
		\draw[fill = black] (2*\si, -1) circle (\rad);
		\node[right] at (3*\si,-1.5) {$z_0$};
		
		\node at (1.5*\si, -3) {(8)};
	\end{tikzpicture}
	\caption{Diagrammatic representations for kernels of $\Lambda$- and $\Lambda'$-operators:
		(1)~for~$\Lambda(\bm{z}_3, \bm{w}_2;x_3)$,
		(2)~for~$\Lambda(\bm{z}_2, w_1;x_2)$,
		(3)~for~$\Lambda(z_1;x_1) = [z_{10}]^{-s-x_1}$,
		(4)~for~$\Lambda'(\bm{z}_3, \bm{w}_2;x_3)$,
		(5)~for~$\Lambda'(\bm{z}_2, w_1;x_2)$,
		(6)~for~$\Lambda'(z_1;x_1) = [z_{10}]^{-s-x_1}$.
		(7)~Graphical representation for $\Psi_{\bm{x}_3}(\bm{z}_3) = \Lambda_3(x_3) \Lambda_2(x_2) \Lambda_1(x_1)$.
		(8)~Diagrammatic representation for $\Psi'_{\bm{x}_3}(\bm{z}_3) = \Lambda'_3(x_3) \Lambda_2(x_2) \Lambda'_1(x_1)$.
		The indices are given by the following expressions: $a_k=1-s+x_k$, $b_k=1-s-x_k$, $c=2s-1$.}
	\label{KerLam123}
\end{figure}
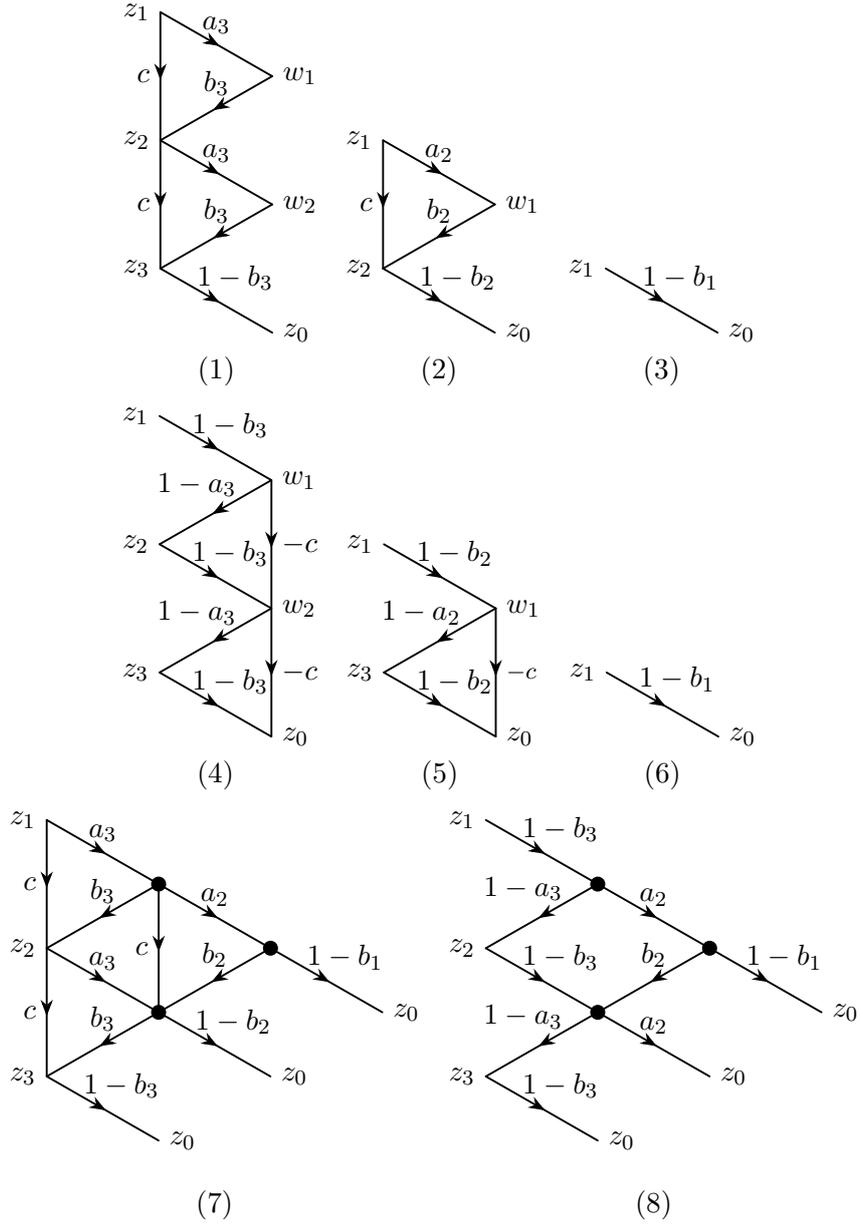

Diagrammatic representation for the kernel
$\Lambda^{\prime}(\bm z_n,\bm w_{n-1};x_n)$ is shown in the Figure~\ref{Lam_prime}.
The diagrammatic representation of eigenfunctions can be simplified by alternating left and right induction steps from~\eqref{PsiPsi'}.
Here we discuss the case $n=3$ for example and the generalization to arbitrary $n$ is straightforward.
The diagrams for integral kernels of operators $\Lambda_k(x)$, $\Lambda'_k(x)$, $k=1,2,3$ are shown in the first six pictures on the Figure~\ref{KerLam123}.
The graphical representation for eigenfunction $\Psi_{\bm{x}_3}(\bm{z}_3) = \Lambda_3(x_3) \Lambda_2(x_2) \Lambda_1(x_1)$ is displayed in seventh picture
on the same Figure~\ref{KerLam123}.
One can construct another eigenfunction by alternating $\Lambda'$- and $\Lambda$-operators: $\Psi'_{\bm{x}_3}(\bm{z}_3) = \Lambda'_3(x_3) \Lambda_2(x_2) \Lambda'_1(x_1)$ and its diagrammatic
representation is shown in the last picture on Figure~\ref{KerLam123}.
We repeat that $\Psi_{\bm{x}_3}$ and $\Psi'_{\bm{x}_3}$ differ only
in the normalization factor but in the last picture,
in contrast to previous ones, all vertical lines with index $2s-1$ disappear.
Similarly, when one constructs the eigenfunction for general $n$
by interlacing $\Lambda'$- and $\Lambda$-operators
\begin{equation*}
\Lambda'_n(x_n) \Lambda_{n-1}(x_{n-1}) \Lambda_{n-2}'(x_{n-2})
\Lambda_{n-3}(x_{n-2})\cdots
\end{equation*}
the lines with index $2s-1$ disappear due to cancellation of propagators
with indices $1-2s$ and~${2s-1}$ in every pair
$\Lambda'_k(x_k)\Lambda_{k-1}(x_{k-1})$.

\subsection{Symmetric normalization}\label{sect:SymmNorm}

In Section~\ref{sect:PermSym}, we discussed the symmetry of eigenfunctions under permutations of spectral parameters $\bm x_n$
\begin{equation*}
	\Psi_{\tau \bm x_n}(\bm z_n) =
	c(\tau,\bm x_n) \Psi_{\bm x_n}(\bm z_n) ,
\end{equation*}
where $\tau \bm x_n = \bigl(x_{\tau(1)},\bar{x}_{\tau(1)}, \dots , x_{\tau(n)},\bar{x}_{\tau(n)}\bigr)$, and $c(\tau,\bm x_n)$ is some constant. It is natural to choose the normalization of eigenfunctions, such that $c(\tau,\bm x_n)=1$.
To do so, hereinafter we use the new normalization for $\Lambda$-operators~\eqref{Lambda}
\begin{equation*}
	\Lambda_k(x) = \lambda_k(x) \mathcal{R}_{12}(x)\mathcal{R}_{23}(x)\cdots
	\mathcal{R}_{k-1 k}(x) [z_{k0}]^{-s-x} ,
\end{equation*}
where
\begin{equation} \label{lambda}
	\lambda_k(x) =
	\left(\frac{\pi [\imath]^{x-s}}{\mathbf{\Gamma}(s+x)}\right)^{k-1} .
\end{equation}
That is, we multiplied the operator $\Lambda_k(x)$~\eqref{Lambda} by the normalization factor $\lambda_k(x)$. The choice of this constant is dictated by the commutation relation~\eqref{exchLL} between $\Lambda$-operators. In new normalization, \eqref{exchLL} takes the form
\begin{equation*}
	\Lambda_n(x) \Lambda_{n-1}(y) =
	\Lambda_n(y) \Lambda_{n-1}(x) .
\end{equation*}
Consequently, from the definition~\eqref{reprPsi}
\begin{equation*}
	\Psi_{\bm{x}_n}(\bm z_n) =
	\Lambda_n( x_n)
	\Lambda_{n-1}( x_{n-1})\cdots
	\Lambda_{2}( x_{2}) [z_1-z_0]^{-s-x_1}
\end{equation*}
follows the complete symmetry of $\Psi_{\bm{x}_n}$ under permutations of $x_1, \dots, x_n$
\begin{equation*}
	\Psi_{\tau \bm x_n}(\bm z_n) =
	\Psi_{\bm x_n}(\bm z_n) .
\end{equation*}

\section{Basic integrals}
\label{sect:Basic}

In this section, we are going to calculate the following integrals:
\begin{align}
& \label{In}
I(\bm{y}_n,\bm{x}_{n};x) =
\int \mathrm{d}^{2}\bm{z}_n
\overline{\Psi_{\bm{y}_n}(\bm{z}_n)}
[z_{n0}]^{x}
\Psi_{\bm{x}_{n}}(\bm{z}_n) =
\langle\Psi_{\bm{y}_n}|[z_{n0}]^{x} \Psi_{\bm{x}_{n}}\rangle ,\\
& \label{Jn}
J(\bm{y}_n,\bm{x}_{n-1};x) =
\int \mathrm{d}^{2}\bm{z}_n
\overline{\Psi_{\bm{y}_n}(\bm{z}_n)}
[z_{n0}]^{x}
\Psi_{\bm{x}_{n-1}}(\bm{z}_{n-1}) =
\langle\Psi_{\bm{y}_n}|[z_{n0}]^{x} \Psi_{\bm{x}_{n-1}}\rangle ,
\end{align}
where in the second form we used the scalar product
\begin{equation*}
	\langle\Phi|\Psi\rangle =
	\int \mathrm{d}^{2}\bm{z}_n \overline{\Phi(\bm{z}_n)} \Psi(\bm{z}_n) .
\end{equation*}
The integral~\eqref{Jn} was calculated in \cite{Val20} with the help of diagram technique described in Appendix~\ref{App-DiagTech}.
Here we give some alternative way of its computation.
The iterative construction of eigenfunctions allows to derive the recurrence
relations for considered integrals. The derivation is heavily based on the fact that integrals~\eqref{In} and~\eqref{Jn} contain $Q$-operator in a hidden form. There are the following recurrence formulas connecting basic integrals:
\begin{align}
\label{In1}
&I(\bm{y}_n,\bm{x}_{n};x) = \frac{[\imath]^{x_n-s-1} \mathbf{\Gamma}(s+x-x_n) \lambda_n(x_n)}
{\mathbf{\Gamma}(x) q^*(x_n+1;\bm{y}_n)}
J(\bm{y}_n,\bm{x}_{n-1};x-x_n-s) ,\\
& \nonumber
J(\bm{y}_n,\bm{x}_{n-1};x) = \pi \mathbf{B}(s-y_n,s+y_n+x) \\
& \label{Jn1}
\hphantom{J(\bm{y}_n,\bm{x}_{n-1};x) =}{}
\times\frac{\lambda_n^\ast(y_n) q^{*}(1-s;\bm{y}_{n-1}) q(1-s;\bm{x}_{n-1})}
{q(y_{n}+1;\bm{x}_{n-1})}
 I(\bm{y}_{n-1},\bm{x}_{n-1};x+y_{n}+s) ,
\end{align}
where $q(u ,\bm x_{n})$ is the eigenvalue~\eqref{qn} of $Q$-operator
\begin{align*}
q(u ,\bm x_{n}) = [\imath]^{(s-u)n}
\frac{\mathbf{\Gamma}(u ,\bm{x}_n)}{\mathbf{\Gamma}(s ,\bm{x}_n)}
\end{align*}
and $\mathbf{B}(a,b)$ is the generalization of the Euler beta function for the complex field~\cite{GGR,N}
\begin{align*}
\mathbf{B}(a,b) =
\frac{\mathbf{\Gamma}(a) \mathbf{\Gamma}(b)}{\mathbf{\Gamma}(a+b)} .
\end{align*}
This system has iterative structure and can be solved using the initial condition
\begin{align*}
\begin{split}
I(y_1,x_{1};x) & =
\int \mathrm{d}^{2} z
\overline{\Psi_{y_1}(z)}
[z-z_0]^{x} \Psi_{x_{1}}(z) =
\int \mathrm{d}^{2} z
[z-z_0]^{x+y_1-x_1 -1} \\
& = 2\pi^2\delta^{(2)}(x+y_1-x_1).
\end{split}
\end{align*}
The main building blocks constructed from the eigenvalues of $Q$-operator
have the following explicit form:
\begin{align*}
q^{*}(1-s;\bm{y}_{n}) q(1-s;\bm{x}_{n}) =
[-1]^{\underline{\bm{x}_{n}}-\underline{\bm{y}_{n}}}
\prod_{k=1}^{n}\frac{\mathbf{\Gamma}(s - y_k)\mathbf{\Gamma}(s + y_k)}{\mathbf{\Gamma}(s - x_k)\mathbf{\Gamma}(s + x_k)}
\end{align*}
and
\begin{align*}
\frac{1}{q^{*}(x_n+1;\bm{y}_{n}) q(y_n+1;\bm{x}_{n-1})} ={}&
[-1]^{\underline{\bm{x}_{n-1}}-(n-1)x_n}
[\imath]^{(n-1)(x_n-y_n)+s-x_n} \\
& {\times}\,
\frac{\mathbf{\Gamma}(x_n-y_n)}{\mathbf{\Gamma}(s-y_n)}
\prod_{k=1}^{n-1}\frac{\mathbf{\Gamma}(s -x_k)}{\mathbf{\Gamma}(s -y_k)}
\prod_{k=1}^{n-1}\mathbf{\Gamma}(x_n-y_k) \mathbf{\Gamma}(x_k-y_n) .
\end{align*}
The general solution of the system reads
\begin{align*}
I(\bm{y}_n,\bm{x}_{n};x) = 2\pi^{n^2+1} [-1]^{(n-1)(ns+\underline{\bm{y}_n})} \frac{\mathbf{\Gamma}^{n-1}(s ,-\bm{y}_n) \mathbf{\Gamma}(\bm{x}_n,\bm{y}_n)}
{\mathbf{\Gamma}^{n-1}(s ,-\bm{x}_n) \mathbf{\Gamma}(x)}
 \delta^{(2)}\bigl(x-\underline{\bm{x}_n}+\underline{\bm{y}_n}\bigr)
\end{align*}
and
\begin{align}
\nonumber
J(\bm{y}_n,\bm{x}_{n-1};x) ={}&
2\pi^{2+n(n-1)}[-1]^{(n-1)(ns+\underline{\bm{y}_n})} \\
& \label{Jn2}
{\times}\,
\frac{\mathbf{\Gamma}^{n-1}(s ,-\bm{y}_n) \mathbf{\Gamma}(\bm{x}_{n-1},\bm{y}_n)
\mathbf{\Gamma}(s ,\bm{y}_n)}
{\mathbf{\Gamma}^{n-1}(s ,-\bm{x}_{n-1}) \mathbf{\Gamma}(2s+x)}
 \delta^{(2)}\bigl(x+s-\underline{\bm{x}_{n-1}}+\underline{\bm{y}_n}\bigr) ,
\end{align}
where we used the following compact notations:
\begin{align*}
	&
	\underline{\bm{y}_n} = \sum\limits_{j=1}^n y_j ,\qquad
	\mathbf{\Gamma}(\bm{x}_{m},\bm{y}_n) =
	\prod\limits_{i=1}^{m}\prod\limits_{j=1}^n\mathbf{\Gamma}(x_i-y_j) , \\
	&
	\mathbf{\Gamma}(s ,\bm{y}_n) = \prod\limits_{k=1}^n\mathbf{\Gamma}(s - y_k) ,\qquad \mathbf{\Gamma}(s ,-\bm{y}_n) = \prod\limits_{k=1}^n\mathbf{\Gamma}(s + y_k) .
\end{align*}
The argument of the delta function \smash{$\delta^{(2)}(v)$} must be of the form
\begin{equation*}
	v = \frac{m}{2}+\imath\nu , \qquad \bar{v} = -\frac{m}{2}+\imath\nu ,
	\qquad m\in\mathbb{Z}, \quad \nu\in\mathbb{R} ,
\end{equation*}
and \smash{$\delta^{(2)}(v)$} is defined in a natural way as the product of the Kronecker symbol
$\delta_{m,0}$ with respect to discrete variable and delta function
$\delta(\nu)$ with respect to continuous variable
\begin{equation} \label{delta_sum}
	\delta^{(2)}(v) = \delta_{m,0} \delta(\nu) .
\end{equation}
To avoid any misunderstanding, we demonstrate the derivation of the recurrence
relations starting from the simplest example. We have
\begin{align*}
\Psi_{\bm{y}_2}(z_1,z_2) =
\lambda_2(y_2) [z_{20}]^{-s-y_2} \mathcal{R}_{12}(y_2) \Psi_{y_1}
\end{align*}
so that
\begin{align*}
J(\bm{y}_2,x_{1};x) & =
\langle \Psi_{y_1 y_2}|[z_{20}]^{x} \Psi_{x_{1}} \rangle =
\langle \lambda_2(y_2) [z_{20}]^{-s-y_2} \mathcal{R}_{12}(y_2) \Psi_{y_1}|[z_{20}]^{x} \Psi_{x_{1}} \rangle \\
& =
\lambda_2^\ast(y_2) \bigl\langle \Psi_{y_1}|
\mathcal{R}^{\dagger}_{12}(y_2) [z_{20}]^{s-1+y_2+x} \Psi_{x_{1}} \bigr\rangle \\
& =
\lambda_2^\ast(y_2)
\bigl\langle \Psi_{y_1}|
[\hat{p}_1]^{2s-1} [z_{12}]^{s-y_2-1} [z_{20}]^{s-1+y_2+x}
[\hat{p}_1]^{-s-y_2}
\Psi_{x_{1}} \bigr\rangle ,
\end{align*}
or explicitly
\begin{gather*}
J(\bm{y}_2,x_{1};x)
=
\lambda_2^\ast(y_2)
\int
\mathrm{d}^{2} z_1
\mathrm{d}^{2} z_2
\overline{\Psi_{y_1}(z_1)}
[\hat{p}_1]^{2s-1} [z_{12}]^{s-y_2-1} [z_{20}]^{s-1+y_2+x}
[\hat{p}_1]^{-s-y_2}
\Psi_{x_{1}}(z_1) .
\end{gather*}
The integral over $z_2$ can be calculated with the help of the
chain rule~\eqref{Chain}
\begin{align*}
\int \mathrm{d}^2 z_2 [z_{12}]^{s-y_2-1} [z_{20}]^{s+y_2+x-1} =
\pi \mathbf{B}(s-y_2,s+y_2+x) [z_{10}]^{2s+x-1}
\end{align*}
and everything is reduced to the single integral
\begin{align*}
&
J(\bm{y}_2,x_{1};x) \\
&\qquad =
\pi \mathbf{B}(s-y_2,s+y_2+x) \lambda_2^\ast(y_2)
\int
\mathrm{d}^{2} z_1
\overline{\Psi_{y_1}(z_1)}
[\hat{p}_1]^{2s-1} [z_{10}]^{2s+x-1}
[\hat{p}_1]^{-s-y_2}
\Psi_{x_{1}}(z_1) \\
&\qquad =
\pi \mathbf{B}(s-y_2,s+y_2+x) \lambda_2^\ast(y_2)
\bigl\langle \Psi_{y_1}|[\hat{p}_1]^{2s-1} [z_{10}]^{2s+x-1}
[\hat{p}_1]^{-s-y_2}
\Psi_{x_{1}}\bigr\rangle .
\end{align*}
In the next step, we transform the operator in the middle to
the product of $Q$-operators
\begin{align*}
&
[\hat{p}_1]^{2s-1} [z_{10}]^{2s+x-1}
[\hat{p}_1]^{-s-y_2} \\
&
\qquad{}= {\blue[\hat{p}_1]^{2s-1} [z_{10}]^{2s-1}}
[z_{10}]^{s+x+y_2}
{\blue [z_{10}]^{1-2s} [\hat{p}_1]^{1-2s}}
{\red[\hat{p}_1]^{2s-1} [z_{10}]^{-y_2-1+s}
[\hat{p}_1]^{-s-y_2}} \\
&
\qquad{}= {\blue \mathcal{R}^{\dagger}_{10}(1-s)}
[z_{10}]^{s+x+y_2}
{\blue \mathcal{R}_{10}(1-s)} {\red \mathcal{R}^{\dagger}_{10}(y_2)} =
Q^{\dagger}_1(1-s)
[z_{10}]^{s+x+y_2}
Q_1(1-s) Q^{\dagger}_1(y_2) ,
\end{align*}
where we used the following expressions for $R$-operators:
\begin{align*}
\mathcal{R}_{10}(x) =
[\hat{p}_1]^{x+s-1} [z_{10}]^{x-s} [\hat{p}_1]^{1-2s},\qquad
\mathcal{R}_{10}(1-s) = [z_{10}]^{1-2s} [\hat{p}_1]^{1-2s} .
\end{align*}
The functions $\Psi_{y_1}$ and $\Psi_{x_1}$
are eigenfunctions of $Q$-operator and $Q^{\dagger}(y_2) = Q^{-1}(y_2+1)$ so that
\begin{align*}
&
\bigl\langle \Psi_{y_1}|Q^{\dagger}_1(1-s)
[z_{10}]^{s+x+y_2}
Q_1(1-s) Q^{\dagger}(y_2)
\Psi_{x_{1}}\bigr\rangle \\
& \qquad =
\frac{q^*(1-s;y_1) q(1-s;x_1)}
{q(y_2+1;x_1)}
\langle \Psi_{y_1}|[z_{10}]^{s+x+y_2}
\Psi_{x_{1}}\rangle .
\end{align*}
Collecting everything together, we obtain the relation~\eqref{Jn1}
in the simplest case
\begin{align*}
J(\bm{y}_2,x_{1};x) = \pi \mathbf{B}(s-y_2,s+y_2+x)
\frac{\lambda_2^\ast(y_2) q^*(1-s;y_1) q(1-s;x_1)}
{q(y_2+1;x_1)} I(y_1,x_1;s+x+y_2) .
\end{align*}
To avoid cumbersome formulas in general case, we consider the next nontrivial
example and hope that generalization will be evident.
We have
\begin{align*}
\Psi_{\bm{y}_3}(\bm{z}_3) = \lambda_3(y_3)
[z_{30}]^{-s-y_3} \mathcal{R}_{12}(y_3) \mathcal{R}_{23}(y_3)
\Psi_{\bm{y}_2}(\bm{z}_2)
\end{align*}
so that
\begin{align*}
J(\bm{y}_3,\bm{x}_2;x) & =
\langle \Psi_{\bm{y}_3}|[z_{30}]^{x} \Psi_{\bm{x}_2} \rangle =
\langle \lambda_3(y_3) [z_{30}]^{-s-y_3} \mathcal{R}_{12}(y_3)
\mathcal{R}_{23}(y_3) \Psi_{\bm{y}_2}|
[z_{30}]^{x} \Psi_{\bm{x}_2} \rangle \\
& =
\lambda_3^\ast(y_3)
\bigl\langle \Psi_{\bm{y}_2}|
\mathcal{R}^{\dagger}_{23}(y_3)
[z_{30}]^{s-1+y_3+x} \mathcal{R}^{\dagger}_{12}(y_3) \Psi_{\bm{x}_2} \bigr\rangle \\
& =
\lambda_3^\ast(y_3)
\bigl\langle \Psi_{\bm{y}_2}|
[\hat{p}_2]^{2s-1} [z_{23}]^{s-y_3-1} [z_{30}]^{s-1+y_3+x}
[\hat{p}_2]^{-s-y_3}
\mathcal{R}^{\dagger}_{12}(y_3) \Psi_{\bm{x}_2} \bigr\rangle ,
\end{align*}
or explicitly
\begin{gather*}
J(\bm{y}_3,\bm{x}_2;x) \\
\quad =
\lambda_3^\ast(y_3)
\int
\mathrm{d}^{2}\bm{z}_2
\mathrm{d}^{2} z_3
\overline{\Psi_{\bm{y}_2}(\bm{z}_2)}
[\hat{p}_2]^{2s-1} [z_{23}]^{s-y_3-1} [z_{30}]^{s-1+y_3+x}
[\hat{p}_2]^{-s-y_3}
\mathcal{R}^{\dagger}_{12}(y_3) \Psi_{\bm{x}_2}(\bm{z}_2).
\end{gather*}
The integral over $z_3$ can be calculated with the help of the
chain rule~\eqref{Chain}
\begin{align*}
\int \mathrm{d}^2 z_3 [z_{23}]^{s-y_3-1} [z_{30}]^{s+y_3+x-1} =
\pi \mathbf{B}(s-y_3,s+y_3+x)
[z_{20}]^{2s+x-1},
\end{align*}
and everything is reduced to the integral
\begin{align*}
&
\int
\mathrm{d}^{2}\bm{z}_2
\overline{\Psi_{\bm{y}_2}(\bm{z}_2)}
[\hat{p}_2]^{2s-1} [z_{20}]^{2s+x-1}
[\hat{p}_2]^{-s-y_3}
\mathcal{R}^{\dagger}_{12}(y_3) \Psi_{\bm{x}_2}(\bm{z}_2) \\
& \qquad =
\bigl\langle \Psi_{\bm{y}_2}|
[\hat{p}_2]^{2s-1} [z_{20}]^{2s+x-1}
[\hat{p}_2]^{-s-y_3}
\mathcal{R}^{\dagger}_{12}(y_3) \Psi_{\bm{x}_2}
\bigr\rangle .
\end{align*}
Next, we transform the operator in the middle to
the product of $\mathcal{R}$-operators or, equivalently, $Q$-operators
\begin{align*}
&
[\hat{p}_2]^{2s-1} [z_{20}]^{2s+x-1}
[\hat{p}_2]^{-s-y_3} \mathcal{R}^{\dagger}_{12}(y_3) \\
&\qquad{} =
{\blue [\hat{p}_2]^{2s-1} [z_{20}]^{2s-1}}
[z_{20}]^{s+x+y_3}
{\blue [z_{20}]^{1-2s} [\hat{p}_2]^{1-2s}}
{\red [\hat{p}_2]^{2s-1} [z_{20}]^{-y_3-1+s}
[\hat{p}_2]^{-s-y_3}} \mathcal{R}^{\dagger}_{12}(y_3) \\
&\qquad{} =
\mathcal{R}^{\dagger}_{20}(1-s)
[z_{20}]^{s+x+y_3}
\mathcal{R}_{20}(1-s)
\mathcal{R}^{\dagger}_{20}(y_3)
\mathcal{R}^{\dagger}_{12}(y_3) \\
&\qquad{} =
\mathcal{R}^{\dagger}_{20}(1-s) \mathcal{R}^{\dagger}_{12}(1-s)
[z_{20}]^{s+x+y_3}
\mathcal{R}_{12}(1-s) \mathcal{R}_{20}(1-s)
\mathcal{R}^{\dagger}_{20}(y_3)
\mathcal{R}^{\dagger}_{12}(y_3) \\
&\qquad{} =
Q^{\dagger}_2(1-s)
[z_{20}]^{s+x+y_3}
Q_2(1-s) Q^{\dagger}_2(y_3) .
\end{align*}
Note that we have inserted
\smash{$\mathcal{R}^{\dagger}_{12}(1-s) \mathcal{R}_{12}(1-s) = \II$}
to obtain the complete expressions for \smash{$Q^{\dagger}_2(1-s)$}
and $Q_2(1-s)$ and the similar insertion is needed in general case.
Next, we have
\begin{align*}
&
\bigl\langle \Psi_{\bm{y}_2}|
Q^{\dagger}_2(1-s)
[z_{20}]^{s+x+y_3}
Q_2(1-s) Q^{\dagger}_2(y_3)
\Psi_{\bm{x}_2}\bigr\rangle \\
& \qquad =
\frac{q^*(1-s;\bm{y}_2) q(1-s;\bm{x}_2)}
{q(y_3+1;\bm{x}_2)}
\langle \Psi_{\bm{y}_2}|[z_{10}]^{s+x+y_3}
\Psi_{\bm{x}_2}\rangle
\end{align*}
and collecting everything together we obtain the relation~\eqref{Jn1}
in the case $n=3$
\begin{align*}
J(\bm{y}_3,\bm{x}_{2};x) =
\pi \mathbf{B}(s-y_3,s+y_3+x)
\frac{\lambda_3^\ast(y_3) q^{*}(1-s;\bm{y}_{2}) q(1-s;\bm{x}_{2})}
{q(y_{3}+1;\bm{x}_{2})}
 I(\bm{y}_{2},\bm{x}_{2};s+x+y_{3}).
\end{align*}
Now we are going to the derivation of relation~\eqref{In1}. We have
\begin{equation*}
I(\bm{y}_n,\bm{x}_{n};x) =
\langle\Psi_{\bm{y}_n}| [z_{n0}]^{x} \Psi_{\bm{x}_{n}}\rangle =
\langle\Psi_{\bm{y}_n}|
\lambda_n(x_n) [z_{n0}]^{x-s-x_n} \mathcal{R}_{12}(x_n)\cdots\mathcal{R}_{n-1 n}(x_n)
\Psi_{\bm{x}_{n-1}}\rangle .
\end{equation*}
The product of $\mathcal{R}$-operators is almost coincides with the
expression for $Q$-operator but the operator $\mathcal{R}_{n0}(x_n)$ is missing.
To fill this gap, we use the trick based on the general formula
\begin{align}\label{Rn0}
\mathcal{R}_{n0}(x_n) [z_{n0}]^{a} =
[z_{n0}]^{1-2s} [\hat{p}_n]^{x_n-s} [z_{n0}]^{x_n+s-1+a} =
\frac{[\imath]^{1+s-x_n} \mathbf{\Gamma}(x_n+s+a)}
{\mathbf{\Gamma}(2s+a)}
[z_{n0}]^{a} ,
\end{align}
which is derived with the help of the chain rule~\eqref{Chain1}.
Choosing $a = x-s-x_n$, we obtain
\begin{align*}
I(\bm{y}_n,\bm{x}_{n};x) ={}& \langle\Psi_{\bm{y}_n}| \lambda_n(x_n)
[z_{n0}]^{x-s-x_n} \mathcal{R}_{12}(x_n)\cdots\mathcal{R}_{n-1 n}(x_n)
\Psi_{\bm{x}_{n-1}}\rangle \\
={}&
\frac{[\imath]^{x_n-s-1} \mathbf{\Gamma}(s+x-x_n) \lambda_n(x_n)}
{\mathbf{\Gamma}(x)} \\
& {\times}\, \langle\Psi_{\bm{y}_n}|
\mathcal{R}_{12}(x_n)\cdots\mathcal{R}_{n-1 n}(x_n)
\mathcal{R}_{n0}(x_n) [z_{n0}]^{x-s-x_n} \Psi_{\bm{x}_{n-1}}\rangle .
\end{align*}
The appeared product of $\mathcal{R}$-operators is the $Q$-operator so that
\begin{align*}
&
\langle\Psi_{\bm{y}_n}|
\mathcal{R}_{12}(x_n)\cdots\mathcal{R}_{n-1 n}(x_n)
\mathcal{R}_{n0}(x_n) [z_{n0}]^{x-s-x_n} \Psi_{\bm{x}_{n-1}}\rangle \\
& \qquad =
\langle\Psi_{\bm{y}_n}| Q_n(x_n) [z_{n0}]^{x-s-x_n} \Psi_{\bm{x}_{n-1}}\rangle =
\bigl\langle Q^{\dagger}_n(x_n)\Psi_{\bm{y}_n}|
 [z_{n0}]^{x-s-x_n} \Psi_{\bm{x}_{n-1}}\bigr\rangle \\
& \qquad =
\langle Q^{-1}_n(x_n+1)\Psi_{\bm{y}_n}|
 [z_{n0}]^{x-s-x_n} \Psi_{\bm{x}_{n-1}}\rangle =
\frac{1}{q^*(x_n+1;\bm{y}_n)} \langle \Psi_{\bm{y}_n}|
 [z_{n0}]^{x-s-x_n} \Psi_{\bm{x}_{n-1}}\rangle .
\end{align*}
Collecting everything together, we obtain relation~\eqref{In1}
\begin{align*}
I(\bm{y}_n,\bm{x}_{n};x) =
\frac{[\imath]^{x_n-s-1} \mathbf{\Gamma}(s+x-x_n) \lambda_n(x_n)}
{\mathbf{\Gamma}(x) q^*(x_n+1;\bm{y}_n)}
\langle \Psi_{\bm{y}_n}|
 [z_{n0}]^{x-s-x_n} \Psi_{\bm{x}_{n-1}}\rangle .
\end{align*}

\subsection{The overlap integral}

The eigenfunction $\Psi_{\bm{x}_{n}}$ depends on the external point
$z_0$ and we show this dependence using more detailed
notation \smash{$\Psi^{0}_{\bm{x}_{n}}$}.
In this subsection, we are going to calculate the following integral:
\begin{align}\label{Over}
I_{0 0^{\prime}}(\bm{y}_n,\bm{x}_{n}) =
\int \mathrm{d}^{2}\bm{z}_n
\overline{\Psi^{0^{\prime}}_{\bm{y}_n}(\bm{z}_n)}
\Psi^{0}_{\bm{x}_{n}}(\bm{z}_n) =
\bigl\langle\Psi^{0^{\prime}}_{\bm{y}_n} | \Psi^{0}_{\bm{x}_{n}}\bigr\rangle .
\end{align}
The derivation again is based on the fact that integral~\eqref{Over} contains $Q$-operator in a hidden form. There exists the following recurrence formula:
\begin{align}
\nonumber
I_{0 0^{\prime}}(\bm{y}_n,\bm{x}_{n}) ={}&
\frac{\pi [\imath]^{x_n-s-1} [-1]^{y_n-x_n} [z_{0 0^{\prime}}]^{y_n-x_n}\lambda_n^\ast(y_n) \lambda_n(x_n)}
{\mathbf{\Gamma}(1-2s,s+x_n) \mathbf{\Gamma}^{*}(s-y_n,s+y_n,2-2s)} \\
& \label{I0}
{\times}\, \frac{q^*(1-s;\bm{y}_{n-1}) q(1-s;\bm{x}_{n-1})}
{q^{*}(x_n+1;\bm{y}_n) q(y_n+1;\bm{x}_{n-1})}
I_{0 0^{\prime}}(\bm{y}_{n-1},\bm{x}_{n-1}) ,
\end{align}
where $q(u ,\bm x_{n})$ is the eigenvalue~\eqref{qn} of $Q$-operator.
This recurrence relation can be solved using the initial condition
\begin{align*}
I_{0 0^{\prime}}(y_1,x_{1}) & =
\int \mathrm{d}^{2} z
\overline{\Psi^{0^{\prime}}_{y_1}(z)}
\Psi^{0}_{x_{1}}(z) =
\int \mathrm{d}^{2} z
[z-z_{0^{\prime}}]^{s-1+y_1} [z-z_0]^{-s-x_1} \\
& =
\pi \mathbf{B}(s+y_1,1-s-x_1)
[z_{0 0^{\prime}}]^{y_1-x_1} [-1]^{-s-x_1}
\end{align*}
and the general solution has the following form:
\begin{align}
\label{In0}
I_{0 0^{\prime}}(\bm{y}_n,\bm{x}_{n}) = \pi^{n^2} [-1]^{n(n+1)s-n\underline{\bm{y}_n}-\underline{\bm{x}_n}}
[z_{00'}]^{\underline{\bm{y}_n}-\underline{\bm{x}_n}}
\frac{\mathbf{\Gamma}^{n}(s ,-\bm{y}_n) \mathbf{\Gamma}(\bm{x}_n,\bm{y}_n)}
{\mathbf{\Gamma}^{n}(s ,-\bm{x}_n)} .
\end{align}
This integral was calculated in the paper \cite{DMV2} using
diagram technique.
Here we present the recurrent calculation which is similar to the
calculation of the previous integrals and again one uses the presence
of $Q$-operator in a hidden form.

Let us derive the recurrence relation. We have
\begin{align*}
I_{0 0^{\prime}}(\bm{y}_n,\bm{x}_{n}) & =
\bigl\langle\Psi^{0^{\prime}}_{\bm{y}_n}|
\lambda_n(x_n) [z_{n0}]^{-s-x_n} \mathcal{R}_{12}(x_n)\cdots\mathcal{R}_{n-1 n}(x_n)
\Psi^{0}_{\bm{x}_{n-1}}\bigr\rangle \\
& =
\lambda_n(x_n) \bigl\langle\Psi^{0^{\prime}}_{\bm{y}_n}|
{\blue \mathcal{R}_{12}(x_n)\cdots\mathcal{R}_{n-1 n}(x_n)
\mathcal{R}_{n 0^{\prime}}(x_n)}
\mathcal{R}^{-1}_{n 0^{\prime}}(x_n) [z_{n0}]^{-s-x_n}
\Psi^{0}_{\bm{x}_{n-1}}\bigr\rangle \\
& =
\lambda_n(x_n) \bigl\langle\Psi^{0^{\prime}}_{\bm{y}_n}|
Q^{0^{\prime}}_{n}(x_n)
\mathcal{R}^{-1}_{n 0^{\prime}}(x_n) [z_{n0}]^{-s-x_n}
\Psi^{0}_{\bm{x}_{n-1}}\bigr\rangle \\
& =
\frac{\lambda_n(x_n) }{q^{*}(x_n+1;\bm{y}_n)}
\bigl\langle\Psi^{0^{\prime}}_{\bm{y}_n}|
\mathcal{R}^{-1}_{n 0^{\prime}}(x_n) [z_{n0}]^{-s-x_n}
\Psi^{0}_{\bm{x}_{n-1}}\bigr\rangle .
\end{align*}
The star-triangle relation~\eqref{Star1} allows to
calculate the result of the action of $\mathcal{R}$-operator
in explicit form
\begin{align*}
\mathcal{R}^{-1}_{n 0^{\prime}}(x_n) [z_{n0}]^{-s-x_n} & =
[z_{n0^{\prime}}]^{1-s-x_n} [\hat{p}_n]^{s-x_n}
[z_{n0^{\prime}}]^{2s-1} [z_{n0}]^{-s-x_n} \\
& =
\frac{[\imath]^{x_n-s-1} [-1]^{s+x_n}}
{\mathbf{\Gamma}(1-2s,s+x_n)} [z_{n0}]^{-2s}
[z_{0 0^{\prime}}]^{s-x_n}
\end{align*}
so that one obtains
\begin{align*}
I_{0 0^{\prime}}(\bm{y}_n,\bm{x}_{n}) =
\frac{[\imath]^{x_n-s-1} [-1]^{s+x_n} [z_{0 0^{\prime}}]^{s-x_n} \lambda_n(x_n)}
{\mathbf{\Gamma}(1-2s,s+x_n) q^{*}(x_n+1;\bm{y}_n)}
\bigl\langle\Psi^{0^{\prime}}_{\bm{y}_n}| [z_{n0}]^{-2s}
\Psi^{0}_{\bm{x}_{n-1}}\bigr\rangle .
\end{align*}
Next step is very similar to the previous ones
\begin{align*}
&
\bigl\langle\Psi^{0^{\prime}}_{\bm{y}_n}| [z_{n0}]^{-2s}
\Psi^{0}_{\bm{x}_{n-1}}\bigr\rangle \\
& \qquad =
\lambda_n^\ast(y_n) \bigl\langle Q^{0}_{n-1}(y_n)
\mathcal{R}^{-1}_{n-1 0}(y_n)
\mathcal{R}_{n-1 n}(y_n)
[z_{n 0^{\prime}}]^{-s-y_n} [z_{n0}]^{2s-2}
\Psi^{0^{\prime}}_{\bm{y}_{n-1}}| \Psi^{0}_{\bm{x}_{n-1}}\bigr\rangle \\
& \qquad =
\frac{\lambda_n^\ast(y_n)}{q(y_n+1;\bm{x}_{n-1})} \bigl\langle \mathcal{R}^{-1}_{n-1 0}(y_n)
\mathcal{R}_{n-1 n}(y_n)
[z_{n 0^{\prime}}]^{-s-y_n} [z_{n0}]^{2s-2}
\Psi^{0^{\prime}}_{\bm{y}_{n-1}}| \Psi^{0}_{\bm{x}_{n-1}}\bigr\rangle .
\end{align*}
The operator in the middle has the following explicit form:
\begin{gather*}
\mathcal{R}_{n-1 n}(y_n)
[z_{n 0^{\prime}}]^{-s-y_n} [z_{n0}]^{2s-2} =
[\hat{p}_{n-1}]^{y_n+s-1} [z_{n-1 n}]^{y_n-s}
[z_{n 0^{\prime}}]^{-s-y_n} [z_{n0}]^{2s-2}
[\hat{p}_{n-1}]^{1-2s}.
\end{gather*}
Note that the whole dependence on $z_n$ is localized now
in the product of three factors so that the
corresponding integral can be calculated using
star-triangle relation in the form~\eqref{Star}
\begin{align*}
&
\int \mathrm{d}^2 z_n [z_{n-1 n}]^{y_n-s}
[z_{n 0^{\prime}}]^{-s-y_n} [z_{n0}]^{2s-2} \\
& \qquad =
\frac{\pi [-1]^{s+y_n}}{\mathbf{\Gamma}(s-y_n,s+y_n,2-2s)}
[z_{n-1 0^{\prime}}]^{1-2s} [z_{n-1 0}]^{s+y_n-1} [z_{0 0^{\prime}}]^{s-y_n-1} .
\end{align*}
Finally, we obtain the following expression for the initial integral:
\begin{align*}
&
I_{0 0^{\prime}}(\bm{y}_n,\bm{x}_{n}) \\
& \qquad =
\frac{[\imath]^{x_n-s-1} [-1]^{s+x_n} [z_{0 0^{\prime}}]^{s-x_n}
[z_{0 0^{\prime}}]^{y_n-s} \lambda_n^\ast(y_n) \lambda_n(x_n)}
{\mathbf{\Gamma}(1-2s,s+x_n) q^{*}(x_n+1;\bm{y}_n) q(y_n+1;\bm{x}_{n-1})} \\
& \qquad\quad \times
\frac{\pi [-1]^{s+y_n}}{\mathbf{\Gamma}^{*}(s-y_n,s+y_n,2-2s)} \\
& \qquad\quad \times
\bigl\langle \mathcal{R}^{-1}_{n-1 0}(y_n)
[\hat{p}_{n-1}]^{y_n+s-1}
[z_{n-1 0^{\prime}}]^{1-2s} [z_{n-1 0}]^{s+y_n-1}
 [\hat{p}_{n-1}]^{1-2s} \Psi^{0^{\prime}}_{\bm{y}_{n-1}}|
\Psi^{0}_{\bm{x}_{n-1}}\bigr\rangle .
\end{align*}
It appears that the remaining product of $\mathcal{R}$-operators
can be reduced to the product of $Q$-operators
\begin{align*}
&
\mathcal{R}^{-1}_{n-1 0}(y_n)
[\hat{p}_{n-1}]^{y_n+s-1}
[z_{n-1 0^{\prime}}]^{1-2s} [z_{n-1 0}]^{s+y_n-1}
 [\hat{p}_{n-1}]^{1-2s} \\
&\qquad{} =
[\hat{p}_{n-1}]^{2s-1}
[z_{n-1 0}]^{s-y_n}
[\hat{p}_{n-1}]^{1-y_n-s}
[\hat{p}_{n-1}]^{y_n+s-1}
[z_{n-1 0^{\prime}}]^{1-2s} [z_{n-1 0}]^{s+y_n-1}
 [\hat{p}_{n-1}]^{1-2s} \\
&\qquad{} =
[\hat{p}_{n-1}]^{2s-1}
[z_{n-1 0}]^{2s-1}
[z_{n-1 0^{\prime}}]^{1-2s}
 [\hat{p}_{n-1}]^{1-2s} =
\mathcal{R}^{-1}_{n-1 0}(1-s) \mathcal{R}_{n-1 0^{\prime}}(1-s) \\
&\qquad{} =
\mathcal{R}^{-1}_{n-1 0}(1-s)\cdots
\mathcal{R}^{-1}_{12}(1-s)
\mathcal{R}_{12}(1-s)\cdots\mathcal{R}_{n-1 0^{\prime}}(1-s) \\
&\qquad{} =
\bigl(Q^{0}_{n-1}(1-s)\bigr)^{\dagger} Q^{0^{\prime}}_{n-1}(1-s) .
\end{align*}
It allows us to reduce everything to the needed form
\begin{align*}
&
\bigl\langle \mathcal{R}^{-1}_{n-1 0}(y_n)
[\hat{p}_{n-1}]^{y_n+s-1}
[z_{n-1 0^{\prime}}]^{1-2s} [z_{n-1 0}]^{s+y_n-1}
 [\hat{p}_{n-1}]^{1-2s} \Psi^{0^{\prime}}_{\bm{y}_{n-1}}|
\Psi^{0}_{\bm{x}_{n-1}}\bigr\rangle \\
& \qquad =
\bigl\langle
\bigl(Q^{0}_{n-1}(1-s)\bigr)^{\dagger}
Q^{0^{\prime}}_{n-1}(1-s)
 \Psi^{0^{\prime}}_{\bm{y}_{n-1}}|
\Psi^{0}_{\bm{x}_{n-1}}\bigr\rangle \\
& \qquad =
\bigl\langle
Q^{0^{\prime}}_{n-1}(1-s)
 \Psi^{0^{\prime}}_{\bm{y}_{n-1}}|
Q^{0}_{n-1}(1-s) \Psi^{0}_{\bm{x}_{n-1}}\bigr\rangle \\
& \qquad =
q^*(1-s;\bm{y}_{n-1}) q(1-s;\bm{x}_{n-1})
\bigl\langle
\Psi^{0^{\prime}}_{\bm{y}_{n-1}}|
\Psi^{0}_{\bm{x}_{n-1}}\bigr\rangle
\end{align*}
and obtain the recurrence relation
\begin{align*}
I_{0 0^{\prime}}(\bm{y}_n,\bm{x}_{n}) ={}&
\frac{[\imath]^{x_n-s-1} [-1]^{s-x_n} [z_{0 0^{\prime}}]^{s-x_n}
[z_{0 0^{\prime}}]^{y_n-s} \lambda_n^\ast(y_n) \lambda_n(x_n)}
{\mathbf{\Gamma}(1-2s,s+x_n) q^{*}(x_n+1;\bm{y}_n) q(y_n+1;\bm{x}_{n-1})} \\
& {\times}\,
\frac{\pi [-1]^{s-y_n} q^*(1-s;\bm{y}_{n-1})
q(1-s;\bm{x}_{n-1})}{\mathbf{\Gamma}^{*}(s-y_n,s+y_n,2-2s)}
I_{0 0^{\prime}}(\bm{y}_{n-1},\bm{x}_{n-1}) .
\end{align*}

\section{Orthogonality and completeness}
\label{sect:Ort}

The functions $\Psi_{\bm{x}_n}(\bm{z}_n)$ form a complete orthogonal set in the Hilbert space~\eqref{HN} which is the space $\mathrm{L}^2\left(\mathbb C^{n}\right)$ of functions of
variables $\bm{z}_n$ with the scalar product
\begin{equation*}
	\langle\Phi|\Psi\rangle =
	\int \mathrm{d}^{2}\bm{z}_n \overline{\Phi(\bm{z}_n)} \Psi(\bm{z}_n) .
\end{equation*}
The orthogonality relation has the following form:
\begin{equation} \label{orth}
\langle\Psi_{\bm{y}_n}|\Psi_{\bm{x}_n}\rangle = \int \mathrm{d}^{2}\bm{z}_n \overline{\Psi_{\bm{y}_n}(\bm{z}_n)} \Psi_{\bm{x}_n}(\bm{z}_n) =
	\mu^{-1}(\bm{x}_n) \delta^{(2)}\bigl(\bm{x}_n,\bm{y}_n \bigr) ,
\end{equation}
where \smash{$\delta^{(2)}\bigl(\bm{x}_n,\bm{y}_n \bigr)$} is the kernel of the
identity operator on the space of symmetric functions
\begin{align}\label{deltasym}
\delta^{(2)}\bigl(\bm{x}_n,\bm{y}_n \bigr) =
\frac{1}{n!}\sum\limits_{\tau\in\mathfrak{S}_n}
	\delta^{(2)}\bigl(x_1-y_{\tau(1)}\bigr) \cdots \delta^{(2)}\bigl(x_n-y_{\tau(n)}\bigr)
\end{align}
and $\mathfrak{S}_n$ is the symmetric group of degree $n$.
The parameters $x_i$, $\bar{x}_i$, $y_j$, $\bar{y}_j$ have the form
\begin{equation} \label{xjyk}
	x_i = \tfrac{n_i}{2}+\imath\nu_i, \qquad
	\bar{x}_i = -\tfrac{n_i}{2}+\imath\nu_i ,\qquad
	y_j=\tfrac{m_j}{2}+\imath\eta_j, \qquad
	\bar{y}_j=-\tfrac{m_j}{2}+\imath\eta_j ,
\end{equation}
where $n_i, m_j \in \mathbb{Z}+\sigma$, the quantity $\sigma$ is defined in~\eqref{sigma}, and numbers $\nu_i$, $\eta_j$ are real.
The~delta function $\delta^{(2)}$ is defined in~\eqref{delta_sum}, the corresponding factors in~\eqref{orth} read
\begin{equation*}
	\delta^{(2)}(x_i-y_j) = \delta_{n_i,m_j} \delta(\nu_i-\eta_j) .
\end{equation*}
The so-called Sklyanin measure $\mu(\bm{x}_n)$ is defined by formula
\begin{equation} \label{mu}
\mu(\bm{x}_n) =
\frac{1}{\pi^{n^2}} \frac{1}{(2\pi)^n n!}
\prod\limits_{1\leq i<j\leq n}|x_i-x_j|^2 ,
\end{equation}
where $|x_i-x_j|$ is the absolute value of $x_i-x_j$.

The completeness relation has the form
\begin{equation*}
	\int \mathcal{D}\bm{x}_n \mu(\bm{x}_n)
	\Psi_{\bm{x}_n}(\bm{z}_n) \overline{\Psi_{\bm{x}_n}(\bm{w}_n)} =
	\delta^2(z_1-w_1) \cdots \delta^2(z_n-w_n) ,
\end{equation*}
or, using Dirac notations
\begin{equation*}
\int \mathcal{D}\bm{x}_n \mu(\bm{x}_n)
|\Psi_{\bm{x}_n}\rangle \langle\Psi_{\bm{x}_n}| = \II ,
\end{equation*}
where $\II$ is the identity operator in the initial coordinate space
which can be formally represented as integral operator with the
integral kernel $\delta^2(z_1-w_1) \cdots \delta^2(z_n-w_n)$.

The integral over variables $\bm{x}_n$ is defined in a natural way
as integral over continues variables and the sum over discrete variables
\begin{equation}\label{sumint}
	\int \mathcal{D}\bm{x}_n = \prod\limits_{k=1}^{n}\int\mathcal{D}x_k,\qquad 	\int\mathcal{D}x_k = \sum\limits_{n_k\in\mathbb{Z}+\sigma}\int\limits_{\mathbb{R}}\mathrm{d}\nu_k
\end{equation}
and $\delta^2(z)$ is the two-dimensional delta function
\begin{equation*}
	\delta^2(z) = \delta(\Re z) \delta(\Im z) .
\end{equation*}
The completeness of the set of eigenfunctions $\Psi_{\bm{x}_n}(\bm{z}_n)$ is proved quite recently \cite{M}. We will consider the calculation of the scalar product~\eqref{orth} and for a detailed analysis of the completeness issues we refer the reader to~\cite{M}.

There exist several ways to calculate the scalar product~\eqref{orth}.
The first is the diagrammatic approach~\cite{DM2} which allows to calculate
the scalar product~\eqref{orth} under assumption $x_i \neq y_j$, ${i\neq n+1-j}$.
By these conditions, only one contribution in the sum over permutations%
survives
\begin{equation*}
	\langle \Psi_{\bm{y}_n} | \Psi_{\bm{x}_n} \rangle =
	\mu^{-1}(\bm{x}_n) \frac{1}{n!} \delta^{(2)}(x_1-y_n) \cdots \delta^{(2)}(x_n-y_1),
\end{equation*}
and then the complete formula is restored using the symmetry under permutations of $x_1,\dots,x_n$ and of $y_1, \dots, y_n$.

Second possible way is to use the exact expression~\eqref{In0} for the overlap integral~\eqref{Over} and then analyse the obtained delta-sequence when $z_{0'} \to z_{0}$.

In the rest of this section, we present the third possible way which
is based on the use of the explicit formula~\eqref{Jn2} for the basic
integral~\eqref{Jn}.

The main ideas are very the same for all approaches to the calculation of the scalar product. First of all, we have to introduce some effective regularization allowing to calculate regularized expression for the scalar product in a closed form. Then we have to prove that one obtains the delta-sequence in the limit when regularization is removed.
We will use regularization of the type~\eqref{RedQL}, which was applied to reduce the $Q$-operator to a raising operator. We define it as follows:
\begin{align}
\nonumber
	\Psi_{\bm{x}_n}(\bm{z}_n) &= \Lambda_n(x_n) \Psi_{\bm{x}_{n-1}}(\bm{z}_{n-1}) \\
	&\label{Psi_expr_Q}
	=\lim\limits_{\varepsilon\to 0}
	\frac{\lambda_n(x_n+\varepsilon)}{q(x_n+\varepsilon, x_n-(n-1)\varepsilon)}
	Q_n(x_n+\varepsilon) [z_{n0}]^{(n-1)\varepsilon-s-x_n} \Psi_{\bm{x}_{n-1}+\varepsilon}(\bm{z}_{n-1}) ,
\end{align}
where
\begin{equation*}
\bm{x}_{n-1}+\varepsilon=(x_1+\varepsilon, \bar{x}_1+\varepsilon, \dots, x_{n-1}+\varepsilon, \bar{x}_{n-1}+\varepsilon) .
\end{equation*}
The formula~\eqref{Psi_expr_Q} follows from~\eqref{QL1}. Note that in Section~\ref{sect:SymmNorm} we changed the normalization of raising operator $\Lambda_n(x)$ multiplying it by $\lambda_n(x)$ so that the relation~\eqref{QL1} takes the form
\begin{align*}
	Q_n(x) [z_{n 0}]^{-s-\alpha} \Psi(\bm z_{n-1}) =
	q(x,\alpha) \lambda_n^{-1}(x) [z_{n0}]^{x-\alpha}
	\Lambda_n( x) \Psi(\bm z_{n-1}) ,
\end{align*}
or, equivalently,
\begin{align*}
	\Lambda_n(x) \Psi(\bm z_{n-1}) = \frac{\lambda_n(x)}{q(x,\alpha)}
	[z_{n0}]^{\alpha-x} Q_n(x) [z_{n 0}]^{-s-\alpha} \Psi(\bm z_{n-1}) .
\end{align*}
Now we choose
\begin{align*}
&(x,\bar{x})=(x_n+\varepsilon,\bar{x}_n+\varepsilon) ,\qquad \Psi(\bm z_{n-1})=\Psi_{\bm{x}_{n-1}+\varepsilon}(\bm{z}_{n-1}) , \\
&(\alpha,\bar{\alpha})=(x-(n-1)\varepsilon, \bar{x}-(n-1)\varepsilon) ,
\end{align*}
where $\varepsilon \geq 0$ and obtain~\eqref{Psi_expr_Q} in the limit $\varepsilon\to 0$.

Substitution of expression~\eqref{Psi_expr_Q} for $\Psi_{\bm{x}_n}$ into
the scalar product $\langle \Psi_{\bm{y}_n} | \Psi_{\bm{x}_n} \rangle$ gives
\begin{align*}
	\langle \Psi_{\bm{y}_n} | \Psi_{\bm{x}_n} \rangle =
	\lim\limits_{\varepsilon\to 0_+}
	\frac{\lambda_n(x_n+\varepsilon)}{q(x_n+\varepsilon, x_n-(n-1)\varepsilon)}
	 \bigl\langle \Psi_{\bm{y}_n} |
	Q_n(x_n+\varepsilon) [z_{n0}]^{(n-1)\varepsilon-s-x_n}
	\Psi_{\bm{x}_{n-1}+\varepsilon} \bigr\rangle .
\end{align*}
Next transformation
\begin{align*}
&\bigl\langle \Psi_{\bm{y}_n} |
Q_n(x_n+\varepsilon) [z_{n0}]^{(n-1)\varepsilon-s-x_n}
\Psi_{\bm{x}_{n-1}+\varepsilon}\bigr\rangle \\
& \qquad= \bigl\langle Q_n^\dagger(x_n+\varepsilon) \Psi_{\bm{y}_n} |
[z_{n0}]^{(n-1)\varepsilon-s-x_n}
\Psi_{\bm{x}_{n-1}+\varepsilon}\bigr\rangle \\
& \qquad =
\frac{1}{q^\ast(1+x_n-\varepsilon; \bm{y}_n)} \bigl\langle \Psi_{\bm{y}_n} |
[z_{n0}]^{(n-1)\varepsilon-s-x_n}
\Psi_{\bm{x}_{n-1}+\varepsilon}\bigr\rangle
\end{align*}
is based on the properties of the $Q$-operator
\begin{align*}
Q_n^\dagger(x_n+\varepsilon) = Q_n^{-1}(1-\bar{x}_n^\ast-\varepsilon) =
Q_n^{-1}(1+x_n-\varepsilon)
\end{align*}
and the fact that $\Psi_{\bm{y}_n}$ is eigenfunction of the
$Q$-operator with eigenvalue $q(u; \bm{y}_n)$.
After all, we reduced the regularized scalar product to the expression
\begin{equation} \label{PsiyPsix2}
	\langle \Psi_{\bm{y}_n} | \Psi_{\bm{x}_n} \rangle =
	\lim\limits_{\varepsilon\to 0_+}
	\frac{\lambda_n(x_n+\varepsilon) J(\bm{y}_n, \bm{x}_{n-1}+\varepsilon; (n-1)\varepsilon-s-x_n)}
	{q(x_n+\varepsilon, x_n-(n-1)\varepsilon)
		q^\ast(1+x_n-\varepsilon,\bm{y}_n)},
\end{equation}
which contains the integral $J(\bm{y}_n,\bm{x}_{n-1}; x)$ calculated in Section~\ref{sect:Basic}
\begin{align*}
J(\bm{y}_n,\bm{x}_{n-1};x) ={}&
\langle\Psi_{\bm{y}_n}|[z_{n0}]^{x} \Psi_{\bm{x}_{n-1}}\rangle =
2\pi^{2+n(n-1)}[-1]^{(n-1)(ns+\underline{\bm{y}_n})} \\
& {\times}\,
\frac{\mathbf{\Gamma}^{n-1}(s ,-\bm{y}_n) \mathbf{\Gamma}(\bm{x}_{n-1},\bm{y}_n)
\mathbf{\Gamma}(s ,\bm{y}_n)}
{\mathbf{\Gamma}^{n-1}(s ,-\bm{x}_{n-1}) \mathbf{\Gamma}(2s+x)}
 \delta^{(2)}\bigl(x+s-\underline{\bm{x}_{n-1}}+\underline{\bm{y}_n}\bigr) .
\end{align*}
The explicit expression for the building block containing
$\lambda_n$ and $q$-factors in~\eqref{PsiyPsix2} reads
\begin{align*}
&\frac{\lambda_n(x_n+\varepsilon)}
{q(x_n+\varepsilon, x_n-(n-1)\varepsilon) q^\ast(1+x_n-\varepsilon,\bm{y}_n)} \\
& \qquad =
\frac{\Gamma(1-n\varepsilon)}{\Gamma(1+n\varepsilon)} \frac{\pi^{n-1} n\varepsilon
\mathbf{\Gamma}(s-x_n+(n-1)\varepsilon)}{\mathbf{\Gamma}^{n-1}(s+x_n+\varepsilon)}
\frac{\mathbf{\Gamma}(x_n+\varepsilon ,\bm{y}_n)}
{\mathbf{\Gamma}(s ,\bm{y}_n)} .
\end{align*}
Collecting everything together, we rewrite~\eqref{PsiyPsix2} in a following way:
\begin{align*}
\langle\Psi_{\bm{y}_n}|\Psi_{\bm{x}_n}\rangle =
2\pi \pi^{n^2} [-1]^{(n-1)(ns+\underline{\bm{y}_n})}
\frac{\mathbf{\Gamma}^{n-1}(s ,-\bm{y}_n)}
{\mathbf{\Gamma}^{n-1}(s ,-\bm{x}_n)}
\delta^{(2)}\bigl(\underline{\bm{x}_n}-\underline{\bm{y}_n}\bigr)
\lim_{\varepsilon\to 0_+}
n \varepsilon \mathbf{\Gamma}(\bm{x}_n+\varepsilon, \bm{y}_n) .
\end{align*}
Due to the presence of $\varepsilon$ in the front of $\mathbf{\Gamma}$-function, the nontrivial contribution
in the limit $\varepsilon \to 0$ is defined by the singular
part of the function $\mathbf{\Gamma}(\bm{x}_n+\varepsilon, \bm{y}_n)$.
The singular part arises from the $\Gamma$-functions in the numerator
and can be extracted explicitly
\begin{align*}
\mathbf{\Gamma}(\bm{x}_n+\varepsilon, \bm{y}_n) \to
\prod\limits_{i,j=1}^n
\frac{\Gamma(1+x_i-y_j)}{\Gamma(1 -\bar{x}_i+\bar{y}_j)}
\prod\limits_{i,j=1}^n
\frac{1}{(x_i-y_j+\varepsilon)} .
\end{align*}
The Cauchy determinant identity ($s(\tau)$ is the sign of the permutation $\tau$)
\begin{align*}
\frac{\prod\limits_{k<j}^n (x_{k}-x_{j}) (y_{j}-y_{k})}{\prod\limits_{k,j=1}^{n}(x_k-y_j+\varepsilon)}
= \det\biggl(\frac{1}{x_k-y_{j}+\varepsilon}\biggr) =
\sum_{\tau\in \mathfrak{S}_{n}} (-1)^{s(\tau)}
\prod_{k=1}^{n}\frac{1}{x_k-y_{\tau(k)}+\varepsilon} .
\end{align*}
allows us to reduce the nontrivial part of the considered limit to the form
\begin{align*}
&
\delta^{(2)}\bigl(\underline{\bm{x}_n}-\underline{\bm{y}_n}\bigr)
\lim_{\varepsilon\to 0_+}
n \varepsilon \mathbf{\Gamma}(\bm{x}_n+\varepsilon, \bm{y}_n) \\
& \qquad =
\prod\limits_{i,j=1}^n
\frac{\Gamma(1+x_i-y_j)}{\Gamma(1 -\bar{x}_i+\bar{y}_j)}
\prod\limits_{k<p}^n \frac{1}
{(x_{k}-x_{p}) (y_{p}-y_{k})} \\
& \qquad =
\delta^{(2)}\bigl(\underline{\bm{x}_n}-\underline{\bm{y}_n}\bigr)
\sum_{\tau\in \mathfrak{S}_{n}} (-1)^{s(\tau)}
\lim_{\varepsilon\to 0_+}
\frac{n \varepsilon}{\prod_{k=1}^{n}(x_k-y_{\tau(k)}+\varepsilon)} .
\end{align*}
Due to presence of \smash{$\delta^{(2)}\bigl(\underline{\bm{x}_n}-\underline{\bm{y}_n}\bigr)$}, we have to
calculate everything by condition $\underline{\bm{x}_n}=\underline{\bm{y}_n}$.
Let us consider the term in the sum which corresponds to the trivial permutation $\tau(k) = k$ and transform it using equivalent condition
$x_n-y_n = \underline{\bm{y}_{n-1}}-\underline{\bm{x}_{n-1}}$ to the form
\begin{align*}
\lim_{\varepsilon\to 0_+}
\frac{n \varepsilon}{\prod_{k=1}^{n}(x_k-y_{k}+\varepsilon)} & =
\lim_{\varepsilon\to 0_+}
\frac{n \varepsilon}{(x_n-y_{n}+\varepsilon)
\prod_{k=1}^{n-1}(x_k-y_k+\varepsilon)} \\
& =
\lim_{\varepsilon\to 0_+}
\frac{n \varepsilon}
{\bigl(\underline{\bm{y}_{n-1}}-\underline{\bm{x}_{n-1}}+\varepsilon\bigr)
\prod_{k=1}^{n-1}(x_k-y_k+\varepsilon)} .
\end{align*}
Now it is possible to apply the formula~\eqref{deltan}
\begin{align*}
\lim\limits_{\varepsilon\to 0_+}
\frac{n \varepsilon}
{\bigl(\underline{\bm{y}_{n-1}}-\underline{\bm{x}_{n-1}}+\varepsilon\bigr)
\prod_{k=1}^{n-1}(x_k-y_k+\varepsilon)} =
(2\pi)^{n-1} \delta^{(2)}(x_1-y_1) \cdots \delta^{(2)}(x_{n-1}-y_{n-1})
\end{align*}
so that after all we obtain
\begin{align*}
\delta^{(2)}\bigl(\underline{\bm{x}_n}-\underline{\bm{y}_n}\bigr)
\lim\limits_{\varepsilon\to 0_+}
\frac{n \varepsilon}
{\prod_{k=1}^{n}(x_k-y_k+\varepsilon)} =
(2\pi)^{n-1} \prod_{k=1}^{n}
\delta^{(2)}(x_k-y_k) .
\end{align*}
The term in the sum which corresponds the nontrivial
permutation $\tau$ is calculated in the same way with evident result
\begin{align*}
\delta^{(2)}\bigl(\underline{\bm{x}_n}-\underline{\bm{y}_n}\bigr)
\lim\limits_{\varepsilon\to 0_+}
\frac{n \varepsilon}
{\prod_{k=1}^{n}(x_k-y_{\tau(k)}+\varepsilon)} =
(2\pi)^{n-1} \prod_{k=1}^{n}
\delta^{(2)}(x_k-y_{\tau(k)})
\end{align*}
so that it is possible to calculate the whole sum.
Note that sign factor $(-1)^{s(\tau)}$ disappears in the whole sum
due to presence of Vandermonde determinant
\smash{$\prod_{k<p}(y_{p}-y_{k})$} and after all we obtain the
symmetric expression
\begin{align*}
&
\delta^{(2)}\bigl(\underline{\bm{x}_n}-\underline{\bm{y}_n}\bigr)
\lim_{\varepsilon\to 0_+}
n \varepsilon \mathbf{\Gamma}(\bm{x}_n+\varepsilon, \bm{y}_n) \\
& \qquad =
(2\pi)^{n-1} \prod\limits_{i,j=1}^n
\frac{\Gamma(1+x_i-x_j)}{\Gamma(1 -\bar{x}_i+\bar{x}_j)}
\prod\limits_{k<p}^n \frac{1}
{(x_{k}-x_{p}) (x_{p}-x_{k})}
\sum_{\tau\in \mathfrak{S}_{n}} \prod_{k=1}^{n}
\delta^{(2)}(x_k-y_{\tau(k)}) .
\end{align*}
It remains to perform some simplifications using
reflection formula for $\Gamma$-functions
\begin{align*}
&
\prod\limits_{i,j=1}^n
\frac{\Gamma(1+x_i-x_j)}{\Gamma(1 -\bar{x}_i+\bar{x}_j)}
\prod\limits_{k<p}^n \frac{1}
{(x_{k}-x_{p}) (x_{p}-x_{k})} =
\prod\limits_{i<j}^n
\frac{\Gamma(1+x_{ij})\Gamma(1-x_{ij})}
{\Gamma(1+\bar{x}_{ij})\Gamma(1 -\bar{x}_{ij})} \frac{1}
{x_{ij} x_{ji}} \\
& \qquad=
\prod\limits_{i<j}^n
\frac{\Gamma(x_{ij})\Gamma(1-x_{ij})}
{\Gamma(\bar{x}_{ij})\Gamma(1 -\bar{x}_{ij})} \frac{1}
{\bar{x}_{ij} x_{ji}} = \prod\limits_{i<j}^n
\frac{\sin(\pi \bar{x}_{ij})}
{\sin(\pi x_{ij})} \frac{1}
{\bar{x}_{ij} x_{ji}} = \frac{1}
{\prod\limits_{i<j}^n |x_i-x_j|^2}
\prod\limits_{i<j}^n (-1)^{k_i-k_j} ,
\end{align*}
where $x_{ij} = x_i-x_j$ and
$x_i = \tfrac{k_i}{2}+\imath\nu_i$, $\bar{x}_i = -\tfrac{k_i}{2}+\imath\nu_i$
with $k_i \in \mathbb{Z}+\sigma$ and $\sigma \in \bigl\{0, \tfrac{1}{2}\bigr\}$.
It is possible to check that
\begin{align*}
\prod\limits_{i<j}^n (-1)^{k_i-k_j} = [-1]^{(n-1)(ns+\underline{\bm{x}_n})}
\end{align*}
so that after combining everything together we obtain
\begin{align*}
\delta^{(2)}\bigl(\underline{\bm{x}_n}-\underline{\bm{y}_n}\bigr)
\lim_{\varepsilon\to 0_+}
n \varepsilon \mathbf{\Gamma}(\bm{x}_n+\varepsilon, \bm{y}_n) =
[-1]^{(n-1)(ns+\underline{\bm{x}_n})}
\frac{(2\pi)^{n-1} n! }{\prod\limits_{i<j}^n |x_i-x_j|^2}
\delta^{(2)}\bigl(\bm{x}_n,\bm{y}_n \bigr)
\end{align*}
and as consequence
\begin{align*}
\langle\Psi_{\bm{y}_n}|\Psi_{\bm{x}_n}\rangle =
\frac{\pi^{n^2} (2\pi)^{n} n!}{\prod\limits_{i<j}^n |x_i-x_j|^2}
\delta^{(2)}\bigl(\bm{x}_n,\bm{y}_n \bigr) .
\end{align*}

\section{Mellin--Barnes representation}
\label{sect:MB}

We have the recurrence formula for eigenfunctions
\begin{align*}
\Psi_{\bm{x}_n}(\bm{z}_n) =
\Lambda_n( x_n) \Psi_{\bm{x}_{n-1}} = \int \mathrm{d}^{2} \bm w_{k-1}
\Lambda(\bm z_k,\bm w_{k-1};x_n)
\Psi_{\bm{x}_{n-1}}(\bm w_{k-1}) ,
\end{align*}
where the integral kernel of the raising operator has the form
\begin{align*}
\Lambda(\bm z_n,\bm w_{n-1};x) =
c^{n-1}(x-s) [z_{n0}]^{-s-x}
\prod\limits_{k=1}^{n-1}
[z_{k k+1}]^{1-2s}[z_k-w_k]^{s-x-1}[w_k -z_{k+1}]^{s+x-1} .
\end{align*}
In this expression, the spectral parameters $\bm{x}_{n-1}$ in the function
$\Psi_{\bm{x}_{n-1}}(\bm w_{k-1})$ are fixed and the integration is performed with respect to variables $\bm{w}_{n-1}$.

There exists an alternative recurrence relation expressing
$\Psi_{\bm{x}_n}$ in terms of $\Psi_{\bm{x}_{n-1}}$.
In analogy with similar representation for eigenfunctions for
the quantum Toda chain \cite{KL1, KL2,KK}, we call it the Mellin--Barnes representation.
It has the form \cite{Val20}
\begin{equation} \label{PsiMB}
\Psi_{\bm{y}_n}(\bm{z}_n) = \hat{\Lambda}_n(z_n) \Psi_{\bm{x}_{n-1}} =
\lim\limits_{\varepsilon\to 0_+}
\int \mathcal{D}\bm{x}_{n-1}
\mu(\bm{x}_{n-1})
\hat{\Lambda}_\varepsilon(\bm{y}_{n},\bm{x}_{n-1}; z_n)
\Psi_{\bm{x}_{n-1}}(\bm{z}_{n-1}) .
\end{equation}
Here, in contrast to the first recurrence relation, the arguments $\bm{z}_{n-1}$ in $\Psi_{\bm{x}_{n-1}}(\bm z_{k-1})$ are fixed and the integration is performed with respect to the spectral variables $\bm{x}_{n-1}$.
In analogy with raising operator $\Lambda_n( x_n)$ acting on coordinate variables $\bm{z}_{n-1}$, we introduced the dual raising operator $\hat{\Lambda}_n(z_n)$ acting on
spectral variables $\bm{x}_{n-1}$.
It is an integral operator with the following integral kernel:
\begin{align}
\nonumber
\hat{\Lambda}_\varepsilon(\bm{y}_{n},\bm{x}_{n-1};z_n) ={}&
(\pi [-1]^{s})^{n(n-1)} [-1]^{n\underline{\bm{x}_{n-1}}}
[z_{n0}]^{\underline{\bm{x}_{n-1}}-\underline{\bm{y}_n}-s} \\
&\label{dual}
{\times}\,
\mathbf{\Gamma}(\underline{\bm{x}_{n-1}}-\underline{\bm{y}_n}+s) 	\frac{\mathbf{\Gamma}^{n-1}(s ,-\bm{x}_{n-1})
\mathbf{\Gamma}(\bm{y}_n+\varepsilon,\bm{x}_{n-1})}
{\mathbf{\Gamma}^{n-1}(s ,-\bm{y}_n) \mathbf{\Gamma}(s ,\bm{y}_n)} .
\end{align}
We remind that the parameters $x_i$, $\bar{x}_i$ and $y_j$, $\bar{y}_j$ have the form
\eqref{xjyk} and the integral over variables $\bm{x}_n$ is defined
as integral over continues variables and the sum over discrete variables~\eqref{sumint}.
Applying the dual raising operator iteratively, we derive
the dual analog of the formula~\eqref{reprPsi}
\begin{align*}
\Psi_{\bm{x}_n}(\bm z_n) =
\hat{\Lambda}_n( z_n)
\hat{\Lambda}_{n-1}( z_{n-1})\cdots
\hat{\Lambda}_{2}(z_{2}) [z_{10}]^{-s-x_1} .
\end{align*}

The origin and the whole sense of the formula~\eqref{PsiMB} is very simple.
Eigenfunctions $\Psi_{\bm{y}_n}$ form a complete and orthogonal set in the Hilbert space $\mathrm{L}^2(\mathbb{C}^n)$ of functions of $n$ variables $\bm{z}_n$.
It is possible to construct another complete set using functions $\Psi_{\bm{x}_{n-1}}(\bm{z}_{n-1}) [z_{n0}]^{-s-x_n}$ where functions $\Psi_{\bm{x}_{n-1}}(\bm{z}_{n-1})$ form a complete and orthogonal set in the Hilbert space $\mathrm{L}^2\bigl(\mathbb{C}^{n-1}\bigr)$ of functions of $n-1$ variables $\bm{z}_{n-1}$ and
functions $[z_{n0}]^{-s-x_n}$ form a complete and orthogonal set in the Hilbert space $\mathrm{L}^2(\mathbb{C})$ of functions of remaining variable $z_n$.

The function $\Psi_{\bm{y}_n}(\bm{z}_n)$ can be expanded over the second complete set $\Psi_{\bm{x}_{n-1}}(\bm{z}_{n-1}) [z_{n0}]^{-s-x_n}$
\begin{align} \label{Psi_n_Psi_n-1_Psi_1}
\Psi_{\bm{y}_n}(\bm{z}_n) =
\int \mathcal{D}\bm{x}_{n-1} \mathcal{D} x_n \mu(\bm{x}_{n-1})
\Psi_{\bm{x}_{n-1}}(\bm{z}_{n-1}) [z_{n0}]^{-s-x_n} \langle[z_{n0}]^{-s-x_n} \Psi_{\bm{x}_{n-1}}|\Psi_{\bm{y}_n}\rangle .
\end{align}
This formula is obtained in a usual way by inserting of the resolution of unity
\begin{align*}
\II = \int \mathcal{D}\bm{x}_{n-1} \mathcal{D} x_n \mu(\bm{x}_{n-1})
|\Psi_{\bm{x}_{n-1}} [z_{n0}]^{-s-x_n}\rangle \langle[z_{n0}]^{-s-x_n} \Psi_{\bm{x}_{n-1}}| .
\end{align*}
In fact, the integral kernel $\langle[z_{n0}]^{-s-x_n} \Psi_{\bm{x}_{n-1}}|\Psi_{\bm{y}_n}\rangle$ was calculated in Section~\ref{sect:Basic}, since it is the scalar product of the type~\eqref{Jn}
\begin{equation*}
	\langle[z_{n0}]^{-s-x_n} \Psi_{\bm{x}_{n-1}}|\Psi_{\bm{y}_n}\rangle = J^\ast(\bm{y}_n,\bm{x}_{n-1}; -s-x_n) ,
\end{equation*}
where $J^\ast$ is the complex conjugation of $J$. This scalar product was first computed in \cite{Val20} with the help of diagram technique given in Appendix~\ref{App-DiagTech}. In the present paper, we developed a~method of its calculation based on recurrence relations. It is shown that $J(\bm{y}_n,\bm{x}_{n-1}; x)$ can be expressed in terms of eigenvalues of $Q$-operators. This allows us to see the structure of the resulting expression~\eqref{Jn2} for $J$, which is written in terms of $\mathbf{\Gamma}$-functions.
For $x=-s-x_n$, the function $J(\bm{y}_n,\bm{x}_{n-1}; x)$ reads
\begin{align}
\nonumber
J(\bm{y}_n,\bm{x}_{n-1}; -s-x_n) ={}&
2\pi^{2+n(n-1)}[-1]^{(n-1)(ns+\underline{\bm{y}_n})} \\
& \label{Pn_formula2}
{\times}\,
\frac{\mathbf{\Gamma}^{n-1}(s ,-\bm{y}_n)
\mathbf{\Gamma}(\bm{x}_{n-1},\bm{y}_n)
\mathbf{\Gamma}(s ,\bm{y}_n)}
{\mathbf{\Gamma}^{n-1}(s ,-\bm{x}_{n-1}) \mathbf{\Gamma}(s-x_n)}
\delta^{(2)}\bigl(x_n+\underline{\bm{x}_{n-1}}-\underline{\bm{y}_n}\bigr) .
\end{align}
Substituting~\eqref{Pn_formula2} into~\eqref{Psi_n_Psi_n-1_Psi_1} and integrating the delta function with respect to $x_n$, one obtains~\eqref{PsiMB}.

We should note that using relations~\eqref{gamma-diff} and
\eqref{gamma-refl} it is possible to transform the
product of the measure $\mu(\bm{x}_{n})$ and the corresponding
sign factor from~\eqref{PsiMB} to the following form:
\begin{align*}
[-1]^{sn(n-1)} [-1]^{(n-1)\underline{\bm{x}_{n}}}
\mu(\bm{x}_{n}) = \frac{1}{\pi^{n^2}}
\frac{1}{(2\pi)^n n!}
\prod\limits_{1\leq i<j\leq n}
\frac{1}{\mathbf{\Gamma}(x_i-x_j)\mathbf{\Gamma}(x_j-x_i)},
\end{align*}
which coincides up to the factor $\pi^{-n^2}$ with the measure in the following generalization \cite{DM,DM3,DMV2,DMV1,MN,N,SS0,SS1,SS2} of the A-type Gustafson integral \cite{G2,G1,G3}
\begin{align*}
\frac{1}{(2\pi)^n n!}
\lim\limits_{\varepsilon\to 0_+}
\int \mathcal{D}\bm{x}_{n}
\frac{\mathbf{\Gamma}(\bm \alpha_{n+1}+\varepsilon,\bm x_{n})
\mathbf{\Gamma}(\bm x_{n}+\varepsilon,\bm \beta_{n+1})}
{\prod\limits_{i<j}
\mathbf{\Gamma}(x_{ij})\mathbf{\Gamma}(x_{ji})} =
\frac{\mathbf{\Gamma}(\bm \alpha_{n+1},\bm \beta_{n+1})}
{\mathbf{\Gamma}\bigl(\underline{\bm \alpha_{n+1}}-\underline{\bm \beta_{n+1}}\bigr)} .
\end{align*}
The $\varepsilon$-prescription provides the needed separation of the series
of poles due to the $\mathbf{\Gamma}$-functions in numerators.
It allows us to rewrite the formula~\eqref{PsiMB} in equivalent form
\begin{align*}
\Psi_{\bm{y}_{n}}(\bm{z}_{n}) ={}& \hat{\Lambda}_n(z_n) \Psi_{\bm{x}_{n-1}} \\
={}&
\frac{\pi^{n-1}}{\mathbf{\Gamma}^{n-1}(s ,-\bm{y}_{n})
\mathbf{\Gamma}(s ,\bm{y}_{n})} \\
&{\times}\,
\lim\limits_{\varepsilon\to 0_+} \frac{1}{(2\pi)^{n-1} (n-1)!}
\int \mathcal{D}\bm{x}_{n-1}
\frac{\mathbf{\Gamma}(\bm y_{n}+\varepsilon,\bm x_{n-1})
\mathbf{\Gamma}^{n-1}(s ,-\bm{x}_{n-1})}
{\prod\limits_{i<j}\mathbf{\Gamma}(x_{ij})\mathbf{\Gamma}(x_{ji})} \\
&{\times}\,
\mathbf{\Gamma}\bigl(\underline{\bm{x}_{n-1}}-\underline{\bm{y}_{n}}+s\bigr)
[z_{n 0}]^{\underline{\bm{x}_{n-1}}-\underline{\bm{y}_{n}}-s}
\Psi_{\bm{x}_{n-1}}(\bm{z}_{n-1}) .
\end{align*}
Two iterative expressions for eigenfunctions look very different
so that it is instructive to give alternative and more direct proof of its equivalence.
We will present the inductive proof based on the application
of the following reduced Gustafson integral \cite{M}
\begin{align}
& \nonumber
\frac{1}{(2\pi)^n n!}
\lim\limits_{\varepsilon\to 0_+}
\int \mathcal{D}\bm{x}_{n}
\frac{\mathbf{\Gamma}(\bm \alpha_{n}+\varepsilon,\bm x_{n})
\mathbf{\Gamma}(\bm x_{n}+\varepsilon,\bm \beta_{n})}
{\prod\limits_{i<j}
\mathbf{\Gamma}(x_{ij})\mathbf{\Gamma}(x_{ji})}
\biggl[\frac{z_1}{z_2}\biggr]^{\underline{\bm x_{n}}} \\
\label{gus1}
& \qquad =
[z_1]^{\underline{\bm \alpha_{n}}} [z_2]^{-\underline{\bm \beta_{n}}}
[z_1+z_2]^{\underline{\bm \beta_{n}}-\underline{\bm \alpha_{n}}}
\mathbf{\Gamma}(\bm \alpha_{n},\bm \beta_{n}) .
\end{align}
We should note that the $\varepsilon$-prescription provides
the needed separation of the series of poles due to
the $\mathbf{\Gamma}$-functions in numerators.
The main steps of the transformation from one representation of
eigenfunctions to another are very similar to the ones used
by Kozlowski~\cite{KK} for the analogous transformation in the
case of the Toda chain. In fact, it will be illustration of the hidden
role of $Q$-operator again.
We start from the explicit formula for eigenfunction in Mellin--Barnes representation
\begin{align*}
\Psi_{\bm{y}_{n+1}}(\bm{z}_{n+1}) ={}&
\frac{\pi^n}{\mathbf{\Gamma}^{n}(s ,-\bm{y}_{n+1})
\mathbf{\Gamma}(s ,\bm{y}_{n+1})} \\
& {\times}\,
\lim\limits_{\varepsilon\to 0_+}
\frac{1}{(2\pi)^n n!} \int \mathcal{D}\bm{x}_{n}
\frac{\mathbf{\Gamma}(\bm y_{n+1}+\varepsilon,\bm x_{n})
\mathbf{\Gamma}^{n}(s ,-\bm{x}_{n})}
{\prod\limits_{i<j}\mathbf{\Gamma}(x_{ij})\mathbf{\Gamma}(x_{ji})} \\
& {\times}\,
\mathbf{\Gamma}\bigl(\underline{\bm{x}_{n}}-\underline{\bm{y}_{n+1}}+s\bigr)
[z_{n+1 0}]^{\underline{\bm{x}_{n}}-\underline{\bm{y}_{n+1}}-s}
\Psi_{\bm{x}_{n}}(\bm{z}_{n}) ,
\end{align*}
where $z_{n+1 0}= z_{n+1}-z_0$ and using the formula~\eqref{QPsi}
\begin{align*}
Q_n(u) \Psi_{\bm x_n}(\bm z_n)
= [\imath]^{(s-u)n} \frac{\mathbf{\Gamma}(u ,\bm{x}_n)}
{\mathbf{\Gamma}(s ,\bm{x}_n)} \Psi_{\bm x_n}(\bm z_n)
\end{align*}
transform it to the form
\begin{align*}
\Psi_{\bm{y}_{n+1}}(\bm{z}_{n+1}) ={}&
\frac{\pi^n [\imath]^{(y_{n+1}-s)n}}
{\mathbf{\Gamma}^{n}(s+y_{n+1})
\mathbf{\Gamma}^{n}(s ,-\bm{y}_{n})
\mathbf{\Gamma}(s ,\bm{y}_{n+1})} \\
& {\times}\, Q_n(y_{n+1})
\lim\limits_{\varepsilon\to 0_+}
\frac{1}{(2\pi)^n n!} \int \mathcal{D}\bm{x}_{n}
\frac{\mathbf{\Gamma}(\bm y_{n}+\varepsilon,\bm x_{n})
\mathbf{\Gamma}^{n}(s ,-\bm{x}_{n})\mathbf{\Gamma}(s ,\bm{x}_n)}
{\prod\limits_{i<j}\mathbf{\Gamma}(x_{ij})\mathbf{\Gamma}(x_{ji})} \\
& {\times}\,
\mathbf{\Gamma}\bigl(\underline{\bm{x}_{n}}-\underline{\bm{y}_{n+1}}+s\bigr)
[z_{n+1 0}]^{\underline{\bm{x}_{n}}-\underline{\bm{y}_{n+1}}-s}
\Psi_{\bm{x}_{n}}(\bm{z}_{n}) .
\end{align*}
Note that the formula~\eqref{QPsi} is derived in initial
coordinate representation so that at this step we have
used induction and suppose that for $\Psi_{\bm{x}_{n}}(\bm{z}_{n})$
both representations coincide.
Our goal is the transformation of the previous representation to the form
\begin{align*}
\Lambda_{n+1}(y_{n+1}) \Psi_{\bm{y}_{n}}
=
\lambda_{n+1}(y_{n+1}) \mathcal{R}_{12}(y_{n+1})\cdots
\mathcal{R}_{n-1 n}(y_{n+1})
\mathcal{R}_{n n+1}(y_{n+1}) [z_{n+1 0}]^{-y_{n+1}-s} \Psi_{\bm{y}_{n}} ,
\end{align*}
where
\begin{equation*}
\lambda_{n+1}(y_{n+1}) =
\biggl(\frac{\pi [\imath]^{y_{n+1}-s}}{\mathbf{\Gamma}(s+y_{n+1})}\biggr)^{n}.
\end{equation*}
The needed normalization factor $\lambda_{n+1}(y_{n+1})$ already appeared in a very natural way and the essential part of $\mathcal{R}$-operators entering
the raising operator is present inside of extracted $Q$-operator
\begin{align*}
Q_{n}(y_{n+1}) =
\mathcal{R}_{12}(y_{n+1})\cdots
\mathcal{R}_{n-1 n}(y_{n+1})
\mathcal{R}_{n 0}(y_{n+1}).
\end{align*}
In the next step, we will use the Mellin--Barnes representation for the function $\Psi_{\bm{x}_{n}}(\bm{z}_{n})$ in the form
\begin{align*}
\Psi_{\bm{x}_{n}}(\bm{z}_{n}) ={}&
\frac{\pi^{n-1}}{\mathbf{\Gamma}^{n-1}(s ,-\bm{x}_{n})
\mathbf{\Gamma}(s ,\bm{x}_{n})} \\
& {\times}\,
\lim\limits_{\varepsilon\to 0_+}
\frac{1}{(2\pi)^{n-1} (n-1)!} \int \mathcal{D}\bm{u}_{n-1}
\frac{\mathbf{\Gamma}(\bm x_{n}+\varepsilon,\bm u_{n-1})
\mathbf{\Gamma}^{n-1}(s ,-\bm{u}_{n-1})}
{\prod\limits_{i<j}\mathbf{\Gamma}(u_{ij})\mathbf{\Gamma}(u_{ji})} \\
& {\times}\,
\mathbf{\Gamma}\bigl(\underline{\bm{u}_{n-1}}-\underline{\bm{x}_{n}}+s\bigr)
[z_{n 0}]^{\underline{\bm{u}_{n-1}}-\underline{\bm{x}_{n}}-s}
\Psi_{\bm{u}_{n-1}}(\bm{z}_{n-1}) .
\end{align*}
Substitution of this expression for $\Psi_{\bm{x}_{n}}(\bm{z}_{n})$ in the previous
representation for the function $\Psi_{\bm{y}_{n+1}}(\bm{z}_{n+1})$ gives
\begin{gather*}
\Psi_{\bm{y}_{n+1}}(\bm{z}_{n+1}) =
\frac{\pi^{n-1} \lambda_{n+1}(y_{n+1})}{\mathbf{\Gamma}^{n}(s ,-\bm{y}_{n})
\mathbf{\Gamma}(s ,\bm{y}_{n+1})}
\mathcal{R}_{12}(y_{n+1})\cdots
\mathcal{R}_{n-1 n}(y_{n+1})\mathcal{R}_{n 0}(y_{n+1})\\
\quad \times \frac{1}{(2\pi)^{n-1} (n-1)!} \int \mathcal{D}\bm{u}_{n-1}
\frac{\mathbf{\Gamma}^{n-1}(s ,-\bm{u}_{n-1})}
{\prod\limits_{i<j}\mathbf{\Gamma}(u_{ij})\mathbf{\Gamma}(u_{ji})}
[z_{n 0}]^{\underline{\bm{u}_{n-1}}-s}
\Psi_{\bm{u}_{n-1}}(\bm{z}_{n-1})
\\
\quad\times \frac{1}{(2\pi)^n n!} \int \mathcal{D}\bm{x}_{n}
\frac{\mathbf{\Gamma}(\bm y_{n},\bm x_{n})
\mathbf{\Gamma}(\bm x_{n},\bm u_{n-1})
\mathbf{\Gamma}(s ,-\bm{x}_{n})}
{\prod\limits_{i<j}\mathbf{\Gamma}(x_{ij})\mathbf{\Gamma}(x_{ji})}
\mathbf{\Gamma}\bigl(\underline{\bm{x}_{n}}-\underline{\bm{y}_{n+1}}+s\bigr)
\mathbf{\Gamma}\bigl(\underline{\bm{u}_{n-1}}-\underline{\bm{x}_{n}}+s\bigr)\\
\quad\hphantom{\times \frac{1}{(2\pi)^n n!} \int}{}\times [z_{n+1 0}]^{\underline{\bm{x}_{n}}-\underline{\bm{y}_{n+1}}-s}
[z_{n 0}]^{-\underline{\bm{x}_{n}}} ,
\end{gather*}
where we interchanged orders of integrals and used the
representation~\eqref{Q} for $Q$-operator as the product
of $\mathcal{R}$-operators.
Note that we assume the necessary $\varepsilon$-prescription everywhere, but omit it for the sake of simplicity.
The integral over $\bm{x}_{n}$ is very similar to the reduced Gustafson
integral~\eqref{gus1} and the transformation to the needed form
\eqref{gus1} can be performed with the help of the
special case of the general formula~\eqref{Rn0}
\begin{align*}
&
\mathbf{\Gamma}\bigl(\underline{\bm{x}_{n}}-\underline{\bm{y}_{n+1}}+s\bigr)
\mathbf{\Gamma}\bigl(\underline{\bm{u}_{n-1}}-\underline{\bm{x}_{n}}+s\bigr)
[z_{n+1 0}]^{\underline{\bm{x}_{n}}-\underline{\bm{y}_{n+1}}-s} \\
& \qquad =
[-1]^{\underline{\bm{x}_{n}}-\underline{\bm{u}_{n-1}}-s}
[\imath]^{\underline{\bm{u}_{n-1}}-\underline{\bm{y}_{n+1}}+2s}
\mathcal{R}^{-1}_{n+1 0}
\bigl(\underline{\bm{y}_{n+1}}-\underline{\bm{u}_{n-1}}+1-s\bigr)
[z_{n+1 0}]^{\underline{\bm{x}_{n}}-\underline{\bm{y}_{n+1}}-s} .
\end{align*}
It allows us to calculate the integral over $\bm{x}_{n}$
in a closed form
\begin{align*}
&
\frac{1}{(2\pi)^n n!} \int \mathcal{D}\bm{x}_{n}
\frac{\mathbf{\Gamma}(\bm y_{n},\bm x_{n})
\mathbf{\Gamma}(\bm x_{n},\bm u_{n-1})
\mathbf{\Gamma}(\bm{x}_{n} ,-s)}
{\prod\limits_{i<j}\mathbf{\Gamma}(x_{ij})\mathbf{\Gamma}(x_{ji})}
[z_0 -z_{n+1}]^{\underline{\bm{x}_{n}}}
[z_n-z_{0}]^{-\underline{\bm{x}_{n}}} \\
& \qquad =
\mathbf{\Gamma}(\bm y_{n},\bm u_{n-1})
\mathbf{\Gamma}(\bm{y}_{n} ,-s)
[z_0 - z_{n+1}]^{\underline{\bm{y}_{n}}}
[z_{n}-z_0]^{s-\underline{\bm{u}_{n-1}}}
[z_n-z_{n+1}]^{\underline{\bm{u}_{n-1}}-\underline{\bm{y}_{n}}-s}
\end{align*}
so that after combining everything together we obtain
\begin{align*}
	&
\Psi_{\bm{y}_{n+1}}(\bm{z}_{n+1}) =
\frac{\pi^{n-1} \lambda_{n+1}(y_{n+1})}
{\mathbf{\Gamma}^{n-1}(s ,-\bm{y}_{n})
\mathbf{\Gamma}(s ,\bm{y}_{n+1})}
\mathcal{R}_{12}(y_{n+1})\cdots
\mathcal{R}_{n-1 n}(y_{n+1})\\
& \quad\times
\frac{1}
{(2\pi)^{n-1} (n-1)!} \int \mathcal{D}\bm{u}_{n-1}
\frac{\mathbf{\Gamma}(\bm y_{n},\bm u_{n-1}) \mathbf{\Gamma}^{n-1}(s ,-\bm{u}_{n-1})}
{\prod\limits_{i<j}\mathbf{\Gamma}(u_{ij})\mathbf{\Gamma}(u_{ji})}
\Psi_{\bm{u}_{n-1}}(\bm{z}_{n-1}) \\
& \qquad \times
[-1]^{\underline{\bm{y}_{n}}-\underline{\bm{u}_{n-1}}-s}
[\imath]^{\underline{\bm{u}_{n-1}}-\underline{\bm{y}_{n+1}}+2s}
\\
& \qquad \times
\mathcal{R}^{-1}_{n+1 0}
\bigl(\underline{\bm{y}_{n+1}}-\underline{\bm{u}_{n-1}}+1-s\bigr)
[z_{n+1}-z_0]^{\underline{\bm{y}_{n}}-\underline{\bm{y}_{n+1}}-s}
\mathcal{R}_{n 0}(y_{n+1})
[z_n-z_{n+1}]^{\underline{\bm{u}_{n-1}}-\underline{\bm{y}_{n}}-s}.
\end{align*}
In the next step, we use chain rule~\eqref{Chain1} to rewrite the
last term in the form
\begin{align*}
&
\mathcal{R}_{n 0}(y_{n+1})
[z_n-z_{n+1}]^{\underline{\bm{u}_{n-1}}-\underline{\bm{y}_{n}}-s} \\
& \qquad =
[\imath]^{2s}\frac{\mathbf{\Gamma}\bigl(\underline{\bm{u}_{n-1}}-
\underline{\bm{y}_{n}}+1-s\bigr)}
{\mathbf{\Gamma}\bigl(\underline{\bm{u}_{n-1}}-
\underline{\bm{y}_{n}}+s\bigr)}
[\hat{p}_n]^{y_{n+1}+s-1}
[z_{n0}]^{y_{n+1}-s}
[z_n-z_{n+1}]^{\underline{\bm{u}_{n-1}}-\underline{\bm{y}_{n}}+s-1} ,
\end{align*}
and now it is possible to apply the star-triangle relation~\eqref{Star1}
in order to perform the last essential transformation
\begin{align*}
&
\mathcal{R}^{-1}_{n+1 0}
\bigl(\underline{\bm{y}_{n+1}}-\underline{\bm{u}_{n-1}}+1-s\bigr)
[z_{n+1}-z_0]^{\underline{\bm{y}_{n}}-\underline{\bm{y}_{n+1}}-s}
[z_n-z_{n+1}]^{\underline{\bm{u}_{n-1}}-\underline{\bm{y}_{n}}+s-1} \\
& \qquad =
\frac{[\imath]^{\underline{\bm{y}_{n+1}}-
\underline{\bm{u}_{n-1}}-2s} \mathbf{\Gamma}(s-y_{n+1})}
{\mathbf{\Gamma}\bigl(1-s+\underline{\bm{y}_{n}}-
\underline{\bm{u}_{n-1}}\bigr)} \\
& \qquad\quad \times
[z_{n+1}-z_0]^{-y_{n+1}-s}
[z_n-z_{n+1}]^{y_{n+1}-s}
[z_n-z_0]^{\underline{\bm{u}_{n-1}}-
\underline{\bm{y}_{n+1}}+2s-1} .
\end{align*}
The last two transformations can be summarized as follows:
\begin{align*}
&
\mathcal{R}^{-1}_{n+1 0}
\bigl(\underline{\bm{y}_{n+1}}-\underline{\bm{u}_{n-1}}+1-s\bigr)
[z_{n+1 0}]^{\underline{\bm{y}_{n}}-\underline{\bm{y}_{n+1}}-s}
\mathcal{R}_{n 0}(y_{n+1})
[z_{n n+1}]^{\underline{\bm{u}_{n-1}}-\underline{\bm{y}_{n}}-s} \\
& \qquad =
[\imath]^{\underline{\bm{y}_{n+1}}-
\underline{\bm{u}_{n-1}}}
\frac{\mathbf{\Gamma}\bigl(1-s+\underline{\bm{u}_{n-1}}-
\underline{\bm{y}_{n}}\bigr) \mathbf{\Gamma}(s-y_{n+1})}
{\mathbf{\Gamma}\bigl(s+\underline{\bm{u}_{n-1}}-
\underline{\bm{y}_{n}}\bigr)\mathbf{\Gamma}\bigl(1-s+\underline{\bm{y}_{n}}-
\underline{\bm{u}_{n-1}}\bigr)} \\
& \qquad\quad \times
[z_{n+1 0}]^{-y_{n+1}-s}
[\hat{p}_n]^{y_{n+1}+s-1} [z_{n n+1}]^{y_{n+1}-s}
{\blue [\hat{p}_n]^{1-2s} [\hat{p}_n]^{2s-1}}
[z_{n 0}]^{\underline{\bm{u}_{n-1}}-
\underline{\bm{y}_{n}}+s-1} ,
\end{align*}
and it remains to extract the needed $\mathcal{R}$-operator
\[
\mathcal{R}_{n n+1}(y_{n+1}) = [\hat{p}_n]^{y_{n+1}+s-1}
[z_{n n+1}]^{y_{n+1}-s} [\hat{p}_n]^{1-2s}
\]
inserting $\II = [\hat{p}_n]^{1-2s} [\hat{p}_n]^{2s-1}$ and using formula
\begin{align*}
[\hat{p}_n]^{2s-1}
[z_{n 0}]^{\underline{\bm{u}_{n-1}}-
\underline{\bm{y}_{n}}+s-1} =
[\imath]^{-2s}
\frac{\mathbf{\Gamma}\bigl(\underline{\bm{u}_{n-1}}-
\underline{\bm{y}_{n}}+s\bigr)}
{\mathbf{\Gamma}\bigl(\underline{\bm{u}_{n-1}}-
\underline{\bm{y}_{n}}+1-s\bigr)}
[z_n-z_0]^{\underline{\bm{u}_{n-1}}-
\underline{\bm{y}_{n}}-s}
\end{align*}
so that finally one obtains
\begin{align*}
&
\mathcal{R}^{-1}_{n+1 0}
\bigl(\underline{\bm{y}_{n+1}}-\underline{\bm{u}_{n-1}}+1-s\bigr)
[z_{n+1 0}]^{\underline{\bm{y}_{n}}-\underline{\bm{y}_{n+1}}-s}
\mathcal{R}_{n 0}(y_{n+1})
[z_{n n+1}]^{\underline{\bm{u}_{n-1}}-\underline{\bm{y}_{n}}-s} \\
& \qquad =
 [\imath]^{\underline{\bm{y}_{n+1}}-\underline{\bm{u}_{n-1}}-2s}
[-1]^{\underline{\bm{y}_{n}}-\underline{\bm{u}_{n-1}}-s}
\mathbf{\Gamma}(s-y_{n+1})
\mathbf{\Gamma}\bigl(\underline{\bm{u}_{n-1}}-\underline{\bm{y}_{n}}+s\bigr) \\
& \qquad\quad \times
[z_{n+1 0}]^{-y_{n+1}-s} \mathcal{R}_{n n+1}(y_{n+1})
[z_{n 0}]^{\underline{\bm{u}_{n-1}}-
\underline{\bm{y}_{n}}-s} ,
\end{align*}
and this provides the necessary restructuring of the product of
$\mathcal{R}$-operators into the raising operator
\begin{align*}
&
\Psi_{\bm{y}_{n+1}}(\bm{z}_{n+1}) \\
& \qquad =
\frac{\pi^{n-1} \lambda_{n+1}(y_{n+1})}
{\mathbf{\Gamma}^{n-1}(s ,-\bm{y}_{n})
\mathbf{\Gamma}(s ,\bm{y}_{n})}
\mathcal{R}_{12}(y_{n+1})\cdots
\mathcal{R}_{n-1 n}(y_{n+1})
[z_{n+1 0}]^{-y_{n+1}-s} \mathcal{R}_{n n+1}(y_{n+1})\\
& \qquad\quad \times
\int \frac{\mathcal{D}\bm{u}_{n-1}}
{(2\pi)^{n-1} (n-1)!}
\frac{\mathbf{\Gamma}(\bm y_{n},\bm u_{n-1})
\mathbf{\Gamma}^{n-1}(s ,-\bm{u}_{n-1})}
{\prod\limits_{i<j}\mathbf{\Gamma}(u_{ij})\mathbf{\Gamma}(u_{ji})} \\
& \qquad\qquad\quad \times
\mathbf{\Gamma}\bigl(\underline{\bm{u}_{n-1}}-\underline{\bm{y}_{n}}+s\bigr)
\Psi_{\bm{u}_{n-1}}(\bm{z}_{n-1})
[z_{n 0}]^{\underline{\bm{u}_{n-1}}-
\underline{\bm{y}_{n}}-s} \\
& \qquad =
\lambda_{n+1}(y_{n+1})
[z_{n+1 0}]^{-y_{n+1}-s}
\mathcal{R}_{12}(y_{n+1})\cdots
\mathcal{R}_{n-1 n}(y_{n+1}) \mathcal{R}_{n n+1}(y_{n+1})
\Psi_{\bm{y}_{n-1}}(\bm{z}_{n-1}) \\
& \qquad =
\Lambda_{n+1}(y_{n+1}) \Psi_{\bm{y}_{n-1}}(\bm{z}_{n-1}) .
\end{align*}

\section{Conclusions}
\label{con}

Let us collect some formulas together.
We should note that the whole structure of our paper
is very similar to the structure of the series of
papers \cite{BDKK1,BDKK2,BDKK3,BDKK4} devoted to
the Ruijsenaars hyperbolic system.
It is very instructive to draw parallels between the two integrable systems.

The main local building block is the $\mathcal{R}$-operator~\eqref{R}.
It is defined as the solution of the relation~\eqref{defR}
\begin{align*}
	\mathcal{R}_{1 2}(x) L_1(u_1, u_2) L_2(u_1, u-x) =
	L_1(u_1, u-x) L_2(u_1, u_2) \mathcal{R}_{1 2}(x)
\end{align*}
and has the following form~\eqref{R1}:
\begin{align*}
	\mathcal{R}_{k j}(x) = [z_{kj}]^{1-2s} [\hat{p}_k]^{x-s} [z_{kj}]^{s+x-1} ,
\end{align*}
where the integral operator $[\hat{p}]^\alpha$ is defined in~\eqref{d}.
There are three commuting operators -- combination of elements
of the monodromy matrix $A(u,z_0) = A(u)+z_0 B(u)$, its antiholomorphic counterpart $\bar{A}(\bar{u},\bar{z}_0) = \bar{A}(\bar{u})+\bar{z}_0 \bar{B}(\bar{u})$ and the composite
operator~\eqref{Q}
\begin{align*}
	Q_n(x) =
	\mathcal{R}_{12}(x) \mathcal{R}_{23}(x)\cdots
	\mathcal{R}_{n-1 n}(x) \mathcal{R}_{n 0}(x) .
\end{align*}
Operator $Q_n(x)$ has all characteristic properties of the Baxter $Q$-operator
\begin{align*}
	& Q_n(x) A(u,z_0) = A(u,z_0) Q_n(x) , \\
	& Q_n(x) Q_n(y) = Q_n(y) Q_n(x) ,\\
	& A(u,z_0) Q_n(u,\bar{u}) = \imath^n Q_n(u+1,\bar{u}) ,
\end{align*}
the relations between $Q_n$ and $\bar{A}(\bar{u},\bar{z}_0)$ are similar to those between $Q_n$ and $A(u,z_0)$.
The common eigenfunctions of $A(u,z_0)$-, $\bar{A}(\bar{u},\bar{z}_0)$- and $Q$-operators
\begin{align*}
	& A(u,z_0)
	\Psi_{\bm{x}_n}(\bm z_n) = (u-x_1)\cdots(u-x_n)
	\Psi_{\bm{x}_n}(\bm z_n) , \\
	& \bar{A}(\bar{u},\bar{z}_0)
	\Psi_{\bm{x}_n}(\bm z_n) = (\bar{u}-\bar{x}_1)\cdots(\bar{u}-\bar{x}_n)
	\Psi_{\bm{x}_n}(\bm z_n) , \\
	& Q_n(u) \Psi_{\bm x_n}(\bm z_n)
	= q(u ,\bm x_{n}) \Psi_{\bm x_n}(\bm z_n)
\end{align*}
are constructed in an iterative way using raising
operators $\Lambda_k$ or dual raising operators $\hat{\Lambda}_k$
\begin{align*}
	&\Psi_{\bm{x}_n}(\bm z_n)=
	\Lambda_n( x_n)
	\Lambda_{n-1}( x_{n-1})\cdots
	\Lambda_{2}( x_{2}) [z_{10}]^{-s-x_1} , \\
	&\Psi_{\bm{x}_n}(\bm z_n)=
	\hat{\Lambda}_n( z_n)
	\hat{\Lambda}_{n-1}( z_{n-1})\cdots
	\hat{\Lambda}_{2}(z_{2}) [z_{10}]^{-s-x_1} .
\end{align*}
Using the $Q$-operator method and the complex generalization of A-type Gustafson integral, we shown in Section~\ref{sect:MB} that these two methods of constructing eigenfunctions lead to the same result. However, the mechanisms are different. When the function $\Psi_{\bm{x}_{k}}(\bm{z}_k)$ is obtained from $\Psi_{\bm{x}_{k-1}}(\bm{w}_{k-1})$ by application of the integral operator $\Lambda_k(x_k)$, one integrates $\Psi_{\bm{x}_{k-1}}(\bm{w}_{k-1})$ with respect to coordinate variables $\bm{w}_{k-1}$
\begin{equation*}
	\Psi_{\bm{x}_{k}}(\bm{z}_k)=[\Lambda_k(x_k) \Psi_{\bm{x}_{k-1}}](\bm z_k) = \int \mathrm{d}^{2} \bm w_{k-1}
	\Lambda(\bm z_k,\bm w_{k-1};x_k) \Psi_{\bm{x}_{k-1}}(\bm w_{k-1}) ,
\end{equation*}
and the spectral variables stay fixed.
On the other hand, when we obtain $\Psi_{\bm{x}_{k}}(\bm{z}_k)$ applying the integral operator $\hat{\Lambda}_k(z_k)$ to the function $\Psi_{\bm{y}_{k-1}}(\bm{z}_{k-1})$ we treat the latter as the function of spectral variables $\bm{y}_{k-1}$, so the integration is performed with respect to $\bm{y}_{k-1}$
\begin{equation*}
	\Psi_{\bm{x}_{k}}(\bm{z}_{k}) = \hat{\Lambda}_k(z_k) \Psi_{\bm{y}_{k-1}}(\bm{z}_{k-1}) = \int \mathcal{D}\bm{y}_{k-1} \mu(\bm{y}_{k-1}) \hat{\Lambda}(\bm{x}_k,\bm{y}_{k-1};z_k) \Psi_{\bm{y}_{k-1}}(\bm{z}_{k-1}) ,
\end{equation*}
where the Sklyanin's measure $\mu(\bm{y}_{k-1})$ is given by expression~\eqref{mu}.
In this situation, the variables~$\bm{z}_{k-1}$ are regarded as fixed parameters, and after the integration this set includes one more ``parameter'' $z_k$.

The expression for $\Lambda_k(x)$ in the operator form reads
\begin{align*}
	\Lambda_k(x) & = \lambda_k(x) \mathcal{R}_{12}(x)\mathcal{R}_{23}(x)\cdots
	\mathcal{R}_{k-1 k}(x) [z_{k0}]^{-s-x} ,
\end{align*}
where $z_{k0}=z_k-z_0$ and $\lambda_k(x)$ is the normalization constant~\eqref{lambda}. The kernel $\hat{\Lambda}(\bm{x}_k,\bm{y}_{k-1};z_k)$ of the dual raising operator is given by the formula~\eqref{dual}.

The eigenfunctions are symmetric under the action of Weyl group of type $A_n$ (the symmetric group $\mathfrak{S}_n$) in the space of spectral variables. That is, for any permutation $\tau\in\mathfrak{S}_n$ we have
\begin{equation*}
	\Psi_{\tau \bm x_n}(\bm z_n) = \Psi_{\bm x_n}(\bm z_n) ,
\end{equation*}
where $\tau \bm x_n = \bigl(x_{\tau(1)},\bar{x}_{\tau(1)}, \dots , x_{\tau(n)},\bar{x}_{\tau(n)}\bigr)$. This property agrees with the same symmetry of $A$-operator's and $Q$-operator's eigenvalues. It is the consequence of the following commutation relation for raising operators:
\begin{equation*}
	\Lambda_n(x) \Lambda_{n-1}(y) =
	\Lambda_n(y) \Lambda_{n-1}(x) .
\end{equation*}
It is demonstrated in Section~\ref{sect:PermSym} that the latter identity can be derived from the commutation relation for $Q$-operators using the reduction~\eqref{RedQL} from $Q$-operator to $\Lambda$-operator.

The constructed set of eigenfunctions is orthogonal
\begin{align*}
\langle\Psi_{\bm{y}_n}|\Psi_{\bm{x}_n}\rangle = \mu^{-1}(\bm{x}_n)
\delta^{(2)}\bigl(\bm{x}_n,\bm{y}_n \bigr) ,
\end{align*}
and complete
\begin{equation*}
\int \mathcal{D}\bm{x}_n \mu(\bm{x}_n)
|\Psi_{\bm{x}_n}\rangle \langle\Psi_{\bm{x}_n}| = \II .
\end{equation*}
Here $\delta^{(2)}\bigl(\bm{x}_n,\bm{y}_n \bigr)$ is the delta function~\eqref{deltasym} of spectral variables $\bm{y}_n$ and $\bm{x}_n$, which possesses the same symmetry under permutations of these variables as the eigenfunctions. And $\II$ is the identity operator in the coordinate space.

Principal series representations characterized by the
parameters $s$ and $1-s$ are equivalent and the manifestation
of this symmetry $s \rightleftarrows 1-s$ on the level of eigenfunctions
has the form
\begin{align*}
\Psi_{\bm x_n}(\bm z_n;1-s) = c(\bm{x}_n)
[z_{12}]^{2s-1} [z_{23}]^{2s-1}
\cdots [z_{n0}]^{2s-1} \Psi_{\bm x_n}(\bm z_n;s) ,
\end{align*}
where the coefficient reads
\begin{equation*}
	c(\bm{x}_n) = [-1]^{n(n-1)s} \prod\limits_{k=1}^n \frac{\mathbf{\Gamma}^{n-1}(s+x_k)}{\mathbf{\Gamma}^{n-1}(1-s+x_k)} .
\end{equation*}
Note that $c(\bm{x}_n)$ differs from the coefficient~\eqref{cx}, which was calculated in the non-symmetric normalization.

In Section~\ref{sect:Basic}, we found closed form expressions for the scalar products
\smash{$\langle\Psi_{\bm{y}_n}|[z_{n0}]^{x} \Psi_{\bm{x}_{n}}\rangle$},
\smash{$\langle\Psi_{\bm{y}_n}|[z_{n0}]^{x} \Psi_{\bm{x}_{n-1}}\rangle$} and \smash{$\bigl\langle\Psi^{0^{\prime}}_{\bm{y}_n} | \Psi^{0}_{\bm{x}_{n}}\bigr\rangle$},
where \smash{$\Psi^{0^{\prime}}_{\bm{y}_n}$} and \smash{$\Psi^{0}_{\bm{x}_{n}}$} are eigenfunctions of $A(u,z_0)$ and $A(u,z_{0'})$, correspondingly. These expressions are given by formulas~\eqref{In1}, \eqref{Jn1} and~\eqref{In0}, which follow from the
recurrence relations~\eqref{In1}, \eqref{Jn1} and~\eqref{I0}. The latter relations were derived with the help of the algebraic approach based on the following formulas:
\begin{itemize}\itemsep=0pt
	\item Yang--Baxter equation~\eqref{YB} for $\mathcal{R}$-operator,
	\item the star-triangle relation~\eqref{star-tr},
	\item eigenvalue equation~\eqref{QPsi} for $Q$-operator, i.e., the fact that eigenfunctions of the model diagonalize the $Q$-operator,
	\item hermitian conjugation rules~\eqref{conj} and~\eqref{QdagQinv} for operators $[\hat{p}]^\alpha$ and $Q_n(x)$,
	\item formulas~\eqref{Chain1} and~\eqref{Star1} for action of the operator $[\hat{p}]^\alpha$ on power functions.
\end{itemize}
The algebraic technique turned out to be an effective tool for calculation of scalar products. Derivation of all above mentioned relations uses one and the same trick. We apply the recurrent formula for one of the eigenfunctions and rewrite the higher rank $\Lambda_n$-operator in terms of $Q$-operator, such that the second function in the scalar product is its eigenfunction.
The $Q$-operator disappears leaving the corresponding eigenvalue, and since the expressions for $Q$- and $\Lambda_n$-operators almost coincide, all that remains from the latter is an operator acting nontrivially in the $n$-th space. These compact algebraic manipulations are much simpler than repeated application of the diagrammatic star-triangle relation. They also involve a rather limited number of coefficients, in contrast to the use of the diagrammatic technique when one has to take into account coefficients appearing after application of every graphical transformation.

The similar algebraic technique also proved to be useful in the derivation of operator relations. For example, in Section~\ref{sect:SpinSym} we easily reduced with its help the formulas connecting $Q$- and $\Lambda$-operators for spins $s$ and $1-s$ to simple local relations for $\mathcal{R}$-operator which is easily verified by the use of operator form~\eqref{star-tr} of star-triangle relation.

We should note that all relations on the level of
$\Lambda$-, $\hat{\Lambda}$- and $Q$-operators itemized in this section
have analog in the case of Ruijsenaars hyperbolic system
\cite{BDKK1,BDKK2,BDKK3,BDKK4}, but there is not any analog
of the local $\mathcal{R}$-operator and the
corresponding Yang--Baxter algebra.

The constructed set of eigenfunctions was used for the calculation of the two-dimensional Basso--Dixon diagram in \cite{DKO} and has direct analogue in the case of higher dimensions \cite{DFO,DO1,DO3,DO2}.
After that significant progress has been made in understanding the
multi-leg fishnet integrals in two dimensions and its connection to the geometry
of Calabi--Yau varieties~\mbox{\cite{Duhr1,Duhr2}}.
The recent work \cite{LS} shows intriguing similarities of the conformal integrals in different dimensions so that there exists some hope to connect two-dimensional results of this paper, \cite{Duhr1,Duhr2} and \cite{DKO} with higher-dimensional analogs \cite{BD,BD1,DFO,DO1,DO2,DO3}.

In recent years, significant progress has been made in understanding the role of the Gustafson integrals \cite{G2,G1,G3} in connection with integrable spin chains
\cite{DM,DMV2,DMV1,M} and its generalization to the complex field \cite{DM,DM3,DMV2,DMV1,MN,N,SS0,SS1,SS2}.
We have demonstrated in this paper the Gustafson integrals
at work and hope that it can help to perform calculations in the program of the
hexagonalization of Fishnet integrals \cite{AO, O1,O2} or in more complicated
fishnet model \cite{AFKO}.

\appendix

\section{Useful formulas and diagrammatic rules}
\label{App-DiagTech}

In this appendix, we itemize the useful formulas and
formulate the diagrammatic rules.

\subsection{Gamma function of the complex field}

The gamma function of the complex field~\cite[Section~1.4]{GGR} and~\cite[Section 1.3]{N}
is the ratio of standard Euler gamma functions and is defined as follows:
\begin{equation*}
	\bm{\Gamma}(a) = \frac{\Gamma(a)}{\Gamma(1 - \bar{a})}.
\end{equation*}
We denote its products as
\begin{align*}
	\bm{\Gamma}(a,b) = \bm{\Gamma}(a) \bm{\Gamma}(b).
\end{align*}
Notice that $\bm{\Gamma}(a)$ depends on two parameters $(a, \bar{a}) \in \mathbb{C}^2$ such that $a - \bar{a} \in \mathbb{Z}$, but for brevity we display only the first one. Moreover, for $\rho \in \mathbb{R}$ we write
\begin{align*}
	\bm{\Gamma}(a + \rho) \equiv \frac{\Gamma(a + \rho)}{\Gamma(1 - \bar{a} - \rho)}.
\end{align*}
From the well-known properties of the ordinary gamma function, it is easy to prove the following relations:
\begin{align}\label{gamma-diff}
	& \bm{\Gamma}(a + 1) = - a \bar{a} \bm{\Gamma}(a), \\
	& \bm{\Gamma}(a) = [-1]^{a} \bm{\Gamma}(\bar{a}), \\ \label{gamma-refl}
	& \bm{\Gamma}(a) \bm{\Gamma}(1 - a) = [-1]^a.
\end{align}
and the complex conjugation rule
\begin{equation*}
	\mathbf{\Gamma}(\rho+a)^\ast = \bm{\Gamma}(\rho - \bar{a}) = [-1]^a \mathbf{\Gamma}(\rho-a),
	\qquad \rho \in \mathbb{R},
\end{equation*}
where $a$, $\bar{a}$ are of the form
\begin{equation*}
	a = \frac{n}{2} + \imath \nu, \qquad
	\bar{a} = -\frac{n}{2} + \imath \nu,
	\qquad n \in \mathbb{Z}, \quad \nu \in \mathbb{R} .
\end{equation*}

\subsection{Chain rule and star-triangle relation}

\begin{figure}[t]
	\centering
	\begin{tikzpicture}[thick, line cap = round, scale=1]
		\draw[->-] (0,0) to node[midway,above]{$a$} (2,0);
		\node[left] at (0,0) {$z$}; \node[right] at (2,0) {$w$};
	\end{tikzpicture}
	\caption{Diagrammatic representation of $[z-w]^{-a}$.}	\label{propag}
\end{figure}
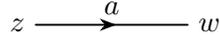

The directed line (Figure \ref{propag}) corresponds to the function
\begin{equation*}
	\frac{1}{[z-w]^a}\equiv\frac{1}{(z-w)^a (\bar{z}-\bar{w})^{\bar a}}=
	\frac{(\bar{z}-\bar{w})^{a-\bar a}}{|z-w|^{2a}},
\end{equation*}
where we always assume $a-\bar a \in \mathbb{Z}$.
The flip of the arrow gives an additional sign factor
\begin{equation*}
	\frac{1}{[z-w]^a}=
	\frac{(-1)^{a-\bar a}}{[w-z]^{a}}= \frac{[-1]^{a}}{[w-z]^{a}}.
\end{equation*}
There are two useful integral identities which can be rewritten in a different equivalent forms and can be depicted diagrammatically -- chain relation and star-triangle relation.
The proofs can be found in \cite[Appendix~A]{DKM}.
The first identity is the \textit{chain relation} depicted in Figure \ref{Rules1}.
The bold vertex in the figure corresponds to the integration over $\mathbb{C}$.

\begin{figure}[t]
	\centering
	\begin{tikzpicture}[thick, line cap = round, scale=1.2]
		\def\rad{0.08} 
		\draw[->-] (0,0) to node[midway,above]{$a$} (2,0);
		\draw[->-] (2,0) to node[midway,above]{$b$} (4,0);
		\draw[fill = black] (2,0) circle (\rad);
		\node at (6,0) {$=\pi (-1)^{c-\bar{c}} \mathbf{\Gamma}^{-1}(a,b,c)$};
		\draw[->-] (8,0) to node[midway,above]{$a+b-1$} (11,0);
	\end{tikzpicture}
	\caption{The chain relation, $a+b+c=2$.}
	\label{Rules1}
\end{figure}

It can be represented in two equivalent forms.
The first ones is an integral identity
\begin{align}\label{Chain}
	\int \mathrm{d}^2 w \frac{1}{[z-w]^a [w-z_0]^{b}}=
	\frac{\pi [-1]^c}{\mathbf{\Gamma}(a,b,c)}
	\frac{1}{[z-z_0]^{a+b-1}} ,
\end{align}
where $c=2-a-b$, $\bar c=2-\bar a-\bar b$.
This formula can be rewritten in a form which is very closed
to the usual Euler beta integral \cite{N}
\begin{align*}
\int \mathrm{d}^2 w [z-w]^{a-1} [w-z_0]^{b-1} =
\pi \mathbf{B}(a,b) [z-z_0]^{a+b-1},
\end{align*}
where
\begin{align*}
\mathbf{B}(a,b) =
\frac{\mathbf{\Gamma}(a) \mathbf{\Gamma}(b)}{\mathbf{\Gamma}(a+b)}.
\end{align*}
The equivalent formula represents this integral identity as the result
of application of the operator~$[\hat{p}]^{a}$ to the particular
function $[z-z_0]^{b-1}$
\begin{align}\label{Chain1}
[\hat{p}]^{a} [z-z_0]^{b-1} =
\frac{[\imath]^{1-a} \mathbf{\Gamma}(b)}
{\mathbf{\Gamma}(b-a)}
[z-z_0]^{b-a-1},
\end{align}
where we used~\eqref{d} and~\eqref{c}.

\begin{figure}[h]
	\centering
	 \begin{tikzpicture}[thick, line cap = round, scale=1.5]
		\def\rad{0.08} 
		\def\si{0.866} 
		\def\yc{0.577} 
		\def\yt{1.732} 
		\def\shi{3.35} 
		\def\sh{5} 
		\draw[->-] (1,\yc) to node[midway,above=1]{$c$} (0,0);
		\draw[->-] (1,\yc) to node[midway,above=1]{$b$} (2,0);
		\draw[->-] (1,\yc) to node[midway,right]{$a$} (1,\yt);
		\draw[fill = black] (1,\yc) circle (\rad);
		\node at (\shi,\yc) {$=\pi \mathbf{\Gamma}^{-1}(a,b,c)$};
		\draw[->-] (\sh,0) to node[midway,above]{$1-b\qquad$} (\sh+1,\yt);
		\draw[->-] (\sh+1,\yt) to node[midway,above]{$\qquad\; 1-c$} (\sh+2,0);
		\draw[->-] (\sh+2,0) to node[midway,below]{$1-a$} (\sh,0);
	\end{tikzpicture}
	\caption{Star-triangle relation, $a+b+c=2$.}
	\label{Rules}
\end{figure}
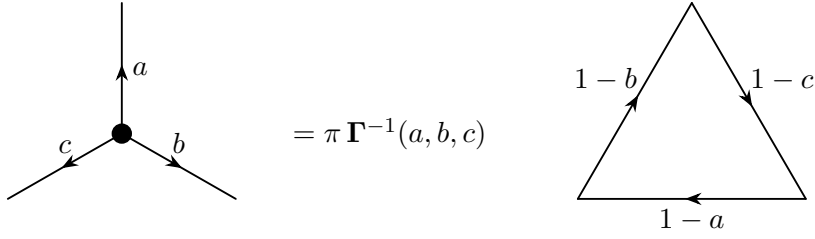

The second identity is the \textit{star-triangle relation} shown in Figure~\ref{Rules}.

This identity can be represented in a three equivalent forms.
The first ones is the integral identity
\begin{align}\label{Star}
\int \mathrm{d}^2 w \frac{1}{[w-z_1]^a[w-z_2]^b [w-z_3]^c}
 = \frac{\pi}{\mathbf{\Gamma}(a,b,c)}
\frac{1}{[z_{12}]^{1-c}[z_{31}]^{1-b}[z_{23}]^{1-a}} ,
\end{align}
where again $a+b+c=2$, $\bar a+\bar b+\bar c=2$.
This integral identity can be represented as the result of application
of operator $[\hat{p}_1]^{a-1}$ to the particular function
$[z_{12}]^{-b} [z_{13}]^{-c}$
\begin{align}\label{Star1}
[\hat{p}_1]^{a-1} [z_{12}]^{-b} [z_{13}]^{-c}
= \frac{[\imath]^{-a}}
{\mathbf{\Gamma}(b,c)}
[z_{12}]^{c-1} [z_{31}]^{b-1} [z_{23}]^{a-1}.
\end{align}
The last and most compact form is the operator form of the star-triangle relation \cite{DM1,DM2, Isa}
\begin{align}\label{star-tr}
[\hat{p}]^{a} [z-z_0]^{a+b} [\hat{p}]^{b} =
[z-z_0]^{b} [\hat{p}]^{a+b} [z-z_0]^{a}.
\end{align}
In the main text, we need formula for the application of considered operator to the constant function $\psi(z) = 1$. Using the chain rule
\eqref{Chain1}, we obtain
\begin{align*}
[\hat{p}]^{a} [z -z_0]^{a+b} [\hat{p}]^{b}\cdot 1 =
[z-z_0]^{b} [\hat{p}]^{a+b} [z-z_0]^{a}\cdot 1 =
\frac{[\imath]^{a+b+1} [-1]^{-b}}{\mathbf{\Gamma}(-a,1-b)}.
\end{align*}
Finally, we give two representations for the $\delta$ function.
The first one is
\begin{equation}\label{delta1}
	\delta^2(z)=\lim_{\varepsilon\to 0}\frac{\varepsilon}{\pi} \frac{1}{[z]^{1-\varepsilon}}
\end{equation}
and the second relation
\begin{equation*}
	\int \mathrm{d}^2 w \frac{1}{[z_1-w]^{2-\alpha}[w-z_2]^\alpha} =
	\frac{\pi^2}{\mathbf{\Gamma}(\alpha,2-\alpha)} \delta^2(z_1-z_2)
\end{equation*}
results from the chain relation~(\ref{Chain}) and (\ref{delta1}).

\subsection{Two identities}

In the main text, we have used identities
\begin{align}\label{del}
\lim\limits_{\alpha\to y}
q^{-1}(y,\alpha)
[z_{n0}]^{s+y} \mathcal{R}_{n 0}(y) [z_{n0}]^{-s-\alpha} \Phi(z_n) =
\Phi(z_0)
\end{align}
and
\begin{align}
& \nonumber
\lim\limits_{\alpha\to x}
q^{-1}(x,\alpha)
[z_{n0}]^{s+x} \mathcal{R}_{n 0}(x) [z_{n0}]^{-s-y}
\mathcal{R}_{n-1 n}(y) [z_{n-1 0}]^{-s-\alpha} \\
&\label{del1}
\qquad =
[\imath]^{s-y} \frac{\mathbf{\Gamma}(x-y,s+y)}
{\mathbf{\Gamma}(s-y,s+x)} [z_{n-1 0}]^{-s-y}
\end{align}
in order to perform various reductions of the commutation
relations for $Q$-operators.

The derivation of these identities is based on the use
of all relations itemized in previous section.
Let us start from~\eqref{del}.
We choose $\alpha = y-\varepsilon$ so that
\begin{align*}
q(y ,y-\varepsilon) =
[\imath]^{s-y} \frac{\mathbf{\Gamma}(\varepsilon)}{\mathbf{\Gamma}(s-y)}, \qquad
\mathbf{\Gamma}(\varepsilon) =
\frac{\Gamma(\varepsilon)}{\Gamma(1 - \varepsilon)} =
\varepsilon \frac{\Gamma(1+\varepsilon)}{\Gamma(1 - \varepsilon)}
\end{align*}
and calculate the corresponding limit $\varepsilon\to 0$ using explicit
representation for $\mathcal{R}$-operator, formula~\eqref{d} and
then representation~\eqref{delta1} for delta function
\begin{align*}
&
\lim\limits_{\varepsilon\to 0}
q^{-1}(y,y-\varepsilon)
[z_{n0}]^{s+y} \mathcal{R}_{n 0}(y) [z_{n0}]^{-s-y+\varepsilon} \Phi(z_n) \\
&\quad =
[z_{n0}]^{1+y-s} [\imath]^{y-s} \mathbf{\Gamma}(s-y)
\lim\limits_{\varepsilon\to 0}
\frac{1}
{\mathbf{\Gamma}(\varepsilon)}
[\hat{p}_n]^{y-s} [z_{n0}]^{-1+\varepsilon} \Phi(z_n) \\
&\quad =
\lim\limits_{\varepsilon\to o} \frac{\varepsilon}{\pi}
\int \mathrm{d}^2 w \frac{[z_{n0}]^{1+y-s} \Phi(w)}{[z_n-w]^{1+y-s} [w-z_0]^{1-\varepsilon}} =
\int \mathrm{d}^2 w \frac{[z_{n0}]^{1+y-s} \Phi(w)}{[z_n-w]^{1+y-s}}
\delta^2(w-z_0) = \Phi(z_0) .
\end{align*}
The derivation of~\eqref{del1} is similar but more involved.
First of all, we use~\eqref{Chain1}
\begin{align*}
\mathcal{R}_{n-1 n}(y) [z_{n-1 0}]^{-s-\alpha} & = [\hat{p}_{n-1}]^{s+y-1} [z_{n-1 n}]^{y-s}
[\hat{p}_{n-1}]^{1-2s} [z_{n-1 0}]^{-s-\alpha} \\
& = 	
\frac{[-1]^{-\alpha}}{\mathbf{\Gamma}(s+\alpha,s-\alpha)}
[\hat{p}_{n-1}]^{s+y-1} [z_{n-1 n}]^{y-s}
[z_{n-1 0}]^{s-\alpha-1}
\end{align*}
and then~\eqref{Star1}
\begin{align*}
[z_{n0}]^{s+x} \mathcal{R}_{n 0}(x) [z_{n0}]^{-s-y} [z_{n-1 n}]^{y-s} & =
[-1]^{y-s}
[z_{n 0}]^{1-s+x} [p_n]^{x-s} [z_{n 0}]^{x-y-1} [z_{n n-1}]^{y-s} \\
& =
\frac{[-1]^{x+y-2s} [\imath]^{x-s+1}}
{\mathbf{\Gamma}(1+y-x,s-y)}
[z_{n0}]^{x-y} [z_{n-1 n}]^{y-x} [z_{0 n-1}]^{x-s} .
\end{align*}
Next, we choose $\alpha = x-\varepsilon$
and calculate the corresponding limit $\varepsilon\to 0$
\begin{align*}
&
\lim\limits_{\varepsilon\to 0}
q^{-1}(x,x-\varepsilon)
[z_{n0}]^{s+x} \mathcal{R}_{n 0}(x) [z_{n0}]^{-s-y}
\mathcal{R}_{n-1 n}(y) [z_{n-1 0}]^{-s-\alpha} \\
&\qquad =
\frac{[-1]^{y} [z_{n0}]^{x-y}}{\mathbf{\Gamma}(1+y-x,s-y,s+x)}
\lim\limits_{\varepsilon\to 0}
\frac{1}
{\mathbf{\Gamma}(\varepsilon)}
[\hat{p}_{n-1}]^{s+y-1} [z_{n-1 n}]^{y-x} [z_{n-1 0}]^{x-\alpha-1} \\
&\qquad =
\frac{[-1]^{y} [z_{n0}]^{x-y}}
{\mathbf{\Gamma}(1+y-x,s-y,s+x)}
[\imath]^{s+y} \bm{\Gamma}(s+y)
\lim\limits_{\varepsilon\to 0} \frac{\varepsilon}{\pi}
\int \mathrm{d}^2 w \frac{[w-z_{n}]^{y-x}}
{[z_{n-1}-w]^{s+y} [w-z_0]^{1-\varepsilon}} \\
&\qquad =
\frac{[-1]^{y} [z_{n0}]^{x-y}}
{\mathbf{\Gamma}(1+y-x,s-y,s+x)}
[\imath]^{s+y} \bm{\Gamma}(s+y)
\int \mathrm{d}^2 w \frac{[w-z_{n}]^{y-x}}
{[z_{n-1}-w]^{s+y}}
\delta^2(w-z_0) \\
&\qquad =
\frac{[-1]^{y} [z_{n0}]^{x-y}}
{\mathbf{\Gamma}(1+y-x,s-y,s+x)}
[\imath]^{s+y} \bm{\Gamma}(s+y) \frac{[z_{0 n}]^{y-x}}
{[z_{n-1 0}]^{s+y}} \\
&\qquad =
\frac{[-1]^{x} [\imath]^{s+y} \mathbf{\Gamma}(s+y)}
{\mathbf{\Gamma}(1+y-x,s-y,s+x)} [z_{n-1 0}]^{-s-y} .
\end{align*}

\section{Formula for the multi-dimensional delta function}

In this appendix, we are going to prove
the key formula
\begin{align}\label{deltan}
\lim\limits_{\varepsilon\to 0_+}
\frac{n \varepsilon}
{\bigl(\underline{\bm{y}_{n-1}}-\underline{\bm{x}_{n-1}}+\varepsilon\bigr)
\prod_{k=1}^{n-1}(x_k-y_k+\varepsilon)} =
(2\pi)^{n-1} \prod\limits_{k=1}^{n-1}
\delta^{(2)}(x_k-y_k) .
\end{align}
For $n=2$, the main formula states
\begin{align*}
\lim\limits_{\varepsilon\to 0_+}
\frac{2\varepsilon}
{(y_1-x_1+\varepsilon)(x_1-y_1+\varepsilon)} =
2\pi \delta^{(2)}(x_1-y_1)
\end{align*}
and can be proven in a following way.
We have $x_1 = \tfrac{k}{2}+\imath\nu$,
$y_1 = \tfrac{p}{2}+\imath\mu$,
\begin{align*}
&
\lim\limits_{\varepsilon\to 0_+}
\frac{2\varepsilon}
{(y_1-x_1+\varepsilon)(x_1-y_1+\varepsilon)} =
\lim\limits_{\varepsilon\to 0_+}
\biggl[\frac{1}
{y_1-x_1+\varepsilon} +
\frac{1}
{x_1-y_1+\varepsilon}\biggr] \\
&\qquad =
\lim\limits_{\varepsilon\to 0_+}\biggl[
\frac{1}
{\tfrac{k-p}{2}+\imath(\nu-\mu)+\varepsilon}+
\frac{1}
{\tfrac{p-k}{2}+\imath(\mu-\nu)+\varepsilon}
\biggr] \\
&\qquad =
\delta_{p k}\lim\limits_{\varepsilon\to 0_+}\biggl[
\frac{1}
{\imath(\nu-\mu)+\varepsilon}+
\frac{1}
{\imath(\mu-\nu)+\varepsilon}
\biggr] =
2\pi \delta_{p k} \delta(\mu-\nu) =
2\pi \delta^{(2)}(x_1-y_1) .
\end{align*}
In the case $p \neq k$, there is not any singularity for $\mu,\nu \in \mathbb{R}$
and everything cancel for $\varepsilon \to 0$ -- it is the origin of the Kronecker symbol $\delta_{p k}$. After that, everything is reduced to the Sokhotski--Plemelj formula for $x\in \mathbb{R}$
\begin{equation*}
	\lim\limits_{\varepsilon\to 0_+}
	\biggl(\frac{1}{x -\imath\varepsilon}
	- \frac{1}{x +\imath\varepsilon}\biggr)
	= 2\pi\imath \delta(x) .
\end{equation*}
Let us consider the next example $n=3$ to clarify the origin
of the appearance of the factor $n$ in the numerator
\begin{align*}
&
\lim\limits_{\varepsilon\to 0_+}
\frac{3\varepsilon}
{(y_1+y_2-x_1-x_2+\varepsilon)(x_1-y_1+\varepsilon)(x_2-y_2+\varepsilon)} \\
&\qquad =
\lim\limits_{\varepsilon\to 0_+}
\frac{3\varepsilon}
{(x_1-y_1+\varepsilon)(y_1-x_1+2\varepsilon)}
\biggl[\frac{1}{x_2-y_2+\varepsilon}+\frac{1}{y_1+y_2-x_1-x_2+\varepsilon}
\biggr] \\
&\qquad =
\lim\limits_{\varepsilon\to 0_+}
\biggl[
\frac{1}
{x_1-y_1+\varepsilon}+
\frac{1}
{y_1-x_1+2\varepsilon}\biggr]
\biggl[\frac{1}{x_2-y_2+\varepsilon}+
\frac{1}{y_1+y_2-x_1-x_2+\varepsilon}
\biggr] .
\end{align*}
Repeating all previous steps, it is easy to derive natural
generalization of $n=2$ formula
\begin{align*}
\lim\limits_{\varepsilon\to 0_+}
\biggl[
\frac{1}
{x_1-y_1+\varepsilon}+
\frac{1}
{y_1-x_1+a\varepsilon}\biggr] = 2\pi \delta^{(2)}(x_1-y_1),
\end{align*}
which is based on generalized Sokhotski–Plemelj formula
\begin{equation*}
	\lim\limits_{\varepsilon\to 0_+}
	\biggl(\frac{1}{x -\imath\varepsilon}
	- \frac{1}{x +\imath a \varepsilon}\biggr)
	= 2\pi\imath \delta(x)
\end{equation*}
valid for positive $a$.
We have
\begin{align*}
&
\lim\limits_{\varepsilon\to 0_+}
\biggl[
\frac{1}
{x_1-y_1+\varepsilon}+
\frac{1}
{y_1-x_1+2\varepsilon}\biggr]
\biggl[\frac{1}{x_2-y_2+\varepsilon}+
\frac{1}{y_1+y_2-x_1-x_2+\varepsilon}
\biggr] \\
& \quad =
2\pi \delta^{(2)}(x_1-y_1)
\lim\limits_{\varepsilon\to 0_+}
\biggl[
\frac{1}{y_2-x_2+\varepsilon}+\frac{1}{x_2-y_2+\varepsilon}
\biggr] = (2\pi)^2 \delta^{(2)}(x_1-y_1) \delta^{(2)}(x_2-y_2) .
\end{align*}
In the general case, we have very similar representation
of initial expression in the form of the product of simplest terms
\begin{align*}
&
\lim\limits_{\varepsilon\to 0_+}
\frac{n \varepsilon}
{\bigl(\underline{\bm{y}_{n-1}}-\underline{\bm{x}_{n-1}}+\varepsilon\bigr)
\prod_{k=1}^{n-1}(x_k-y_k+\varepsilon)} = \lim\limits_{\varepsilon\to 0_+}
\biggl(\frac{1}
{x_1-y_1+\varepsilon}+
\frac{1}
{y_1-x_1+(n-1)\varepsilon}\biggr) \\
&\qquad{} \times \biggl(\frac{1}{x_2-y_2+\varepsilon}+
\frac{1}{y_1+y_2-x_1-x_2+(n-2)\varepsilon}
\biggr)
\cdots \\
&\qquad{} \times
\biggl(\frac{1}{\underline{\bm{y}_{n-1}}-\underline{\bm{x}_{n-1}}+\varepsilon}+
\frac{1}{x_{n-1}-y_{n-1}+\varepsilon}\biggr).
\end{align*}
For $\varepsilon \to 0$, we have avalanche of simplifications --
the leftmost bracket produces $2\pi \delta^{(2)}(x_1-y_1)$,
then next bracket gives $2\pi \delta^{(2)}(x_2-y_2)$ and similarly till the last bracket which gives $2\pi \delta^{(2)}(x_{n-1}-y_{n-1})$.

\subsection*{Acknowledgements}

We are grateful to N.~Belousov, S.~Khoroshkin and A.~Manashov for fruitful discussions.
The work was supported by the
Theoretical Physics and Mathematics Advancement Foundation BASIS (S.D. and P.A.) and by the Ministry of Science and Higher Education of the Russian Federation (P.A.), agreement 075-15-2025-344 dated 29/04/2025 for Saint Petersburg Leonhard Euler International Mathematical Institute at PDMI RAS.

\pdfbookmark[1]{References}{ref}
\LastPageEnding

\end{document}